\documentclass[aps,pre,10pt,final,showpacs]{revtex4}
\usepackage{amssymb}
\usepackage{amsmath}
\usepackage{graphicx}
\usepackage{color}
\usepackage{hyperref}
\usepackage[normalsize]{subfigure}
\usepackage{amssymb}
\usepackage{enumerate}
\usepackage{amsthm,amsfonts,amsmath}
\usepackage{bm}
\usepackage{graphicx}
\usepackage{graphics,epsf,bbold}
\usepackage{siunitx}
\usepackage{color}
\usepackage{xspace}
\usepackage{accents}

\usepackage[T1]{fontenc}
\usepackage[english]{babel}
\usepackage{verbatim}
\usepackage{siunitx}
\usepackage{tensor}
\usepackage{xspace}
\usepackage{bm,bbm}
\usepackage{color}
\usepackage{accents}
\usepackage{setspace}
\usepackage{mathtools}
\usepackage[normalem]{ulem}
\usepackage[capitalize]{cleveref}

\definecolor{dBgreen}{rgb}{.3,.5,.45}

\newcommand{\rev}[1]{\textcolor{black}{#1}}

\def\Tcal{\mathcal{S}}
\def\TcalY{\mathcal{S}}
\def\sTcal{\sqrt{\Tcal}}
\def\sTcalY{\sqrt{\TcalY}}

\newcommand{\bitem}{\begin{itemize}}
	\newcommand{\eitem}{\end{itemize}}
\newcommand{\benum}{\begin{enumerate}}
	\newcommand{\eenum}{\end{enumerate}}
\newcommand{\beq}{\begin{equation}}
	\newcommand{\eeq}{\end{equation}}
\newcommand{\bal}{\begin{align}}
	\newcommand{\eal}{\end{align}}
\newcommand{\baln}{\begin{align*}}
	\newcommand{\ealn}{\end{align*}}
\newcommand{\beqn}{\begin{equation*}}
	\newcommand{\eeqn}{\end{equation*}}

\DeclareMathOperator\sgn{sgn}
\DeclareMathOperator\erf{erf}
\DeclareMathOperator\erfc{erfc}

\newcommand*{\vcenteredhbox}[1]{\begingroup
\setbox0=\hbox{#1}\parbox{\wd0}{\box0}\endgroup}

\newcommand\st[1]{\textup{\tiny{#1}}}
\newcommand\pt{\partial}
\newcommand\Round[1]{{\left(#1\right)}}
\newcommand\Square[1]{{\left[#1\right]}}
\newcommand\Angle[1]{{\left<#1\right>}}

\def\Acal{\mathcal{A}}
\def\Dcal{\mathcal{D}}

\def\Fcal{\mathcal{F}}

\def\Hcal{\mathcal{H}}

\def\e{\epsilon}
\def\kB{k_{\st{B}}}
\def\R{R}

\def\l{\ell}
\def\out{\alpha}
\def\Young{\theta_{\st{Y}}}

\begin{document}

\title{Thermal fluctuations of an interface near a contact line}

\author{D. Belardinelli}
\email{belardinelli@roma2.infn.it}
\affiliation{Department of Physics \& INFN, University of Rome ``Tor Vergata'', Via della Ricerca Scientifica 1, 00133, Rome, Italy.}

\author{M. Sbragaglia}
\email{sbragaglia@roma2.infn.it}
\affiliation{Department of Physics \& INFN, University of Rome ``Tor Vergata'', Via della Ricerca Scientifica 1, 00133, Rome, Italy.}

\author{M. Gross}
\email{gross@is.mpg.de}
\affiliation{Max-Planck-Institut f\"{u}r Intelligente Systeme, Heisenbergstr.\ 3, 70569 Stuttgart, Germany, and\\
IV.\ Institut f\"{u}r Theoretische Physik, Universit\"{a}t Stuttgart, Pfaffenwaldring 57, 70569 Stuttgart, Germany}

\author{B. Andreotti}
\affiliation{
Physique et M\'ecanique des Milieux H\'et\'erog\`enes, UMR 7636 ESPCI -CNRS, Univ. Paris-Diderot, 10 rue Vauquelin, 75005, Paris, France}

\date{\today}

\begin{abstract}
The effect of thermal fluctuations near a contact line of a liquid interface partially wetting an impenetrable substrate is studied analytically and numerically. Promoting both the interface profile and the contact line position to random variables, we explore the equilibrium properties of the corresponding fluctuating contact line problem based on an interfacial Hamiltonian involving a ``contact'' binding potential. To facilitate an analytical treatment we consider the case of a one-dimensional interface. The effective boundary condition at the contact line is determined by a dimensionless parameter that encodes the relative importance of thermal energy and substrate energy at the microscopic scale. We find that this parameter controls the transition from a partially wetting to a pseudo-partial wetting state, the latter being characterized by a thin prewetting film of fixed thickness. In the partial wetting regime, instead, the profile typically approaches the substrate via an exponentially thinning prewetting film. We show that, independently of the physics at the microscopic scale, Young's angle is recovered sufficiently far from the substrate. The fluctuations of the interface and of the contact line give rise to an effective disjoining pressure, exponentially decreasing with height. Fluctuations therefore provide a regularization of the singular contact forces occurring in the corresponding deterministic problem.
\end{abstract}

\maketitle

\section{Introduction}\label{sec:Intro}
From the macroscopic point of view, the shape of a droplet deposited on a non-wetting surface is determined by the the excess free energy associated with the interfaces, namely the surface tensions of the liquid-vapor ($\gamma$), solid-liquid ($\gamma_{\st{sl}}$), and solid-vapor interfaces ($\gamma_{\st{sv}}$). The contact angle $\Young$ made by the liquid-vapour interface with respect to the solid is then controlled by Young's law: 
\beq\label{eq_Young}
\cos \Young = \frac{\gamma_{\st{sv}}-\gamma_{\st{sl}}}{\gamma}.
\eeq
The understanding and modeling of the statics and dynamics of wetting, such as spreading or motion of a droplet on a solid surface, is a subject at the forefront of physics, chemistry, and engineering \cite{bonn_wetting_2009,snoeijer_moving_2013}. Compared to the static case, the dynamical problem is intrinsically out of equilibrium down to the molecular scale and must be described using a multi-scale analysis. In particular, contact line motion is associated with a viscous stress which, under the assumption of no slip, is singular at the contact line and leads to a dissipation of energy at all length scales between the molecular scale and the size of the drop \cite{huh_scriven_1971}. One possibility to regularize this singularity is to assume the presence of a thin precursor film, which results, for instance, from an interface binding potential that has a local minimum at a finite height above the substrate \cite{eggers_contact_2005,pismen_solvability_2008}. Another approach is based on a microscopic consideration of the contact line motion: indeed, for sufficiently small contact line velocity, dynamics becomes dominated by physico-chemical heterogeneities, with a thermally activated dynamics across defects \cite{BH69,Prevost99, Blake93, Petrovforced_1992, Petrovcomb_1992,GiaPNAS16}. This small velocity regime has been proven to be a temperature-dependent activation process over nano-scale defects by~\cite{Prevost99} and has been modeled using Kramers reaction path theory based on a macroscopic description of the energy landscape \cite{JdG84, rolley2007, Davitt_2013a}. Alternatively, one may also consider thermal fluctuations within a continuum description of the contact line problem, based on the equations of fluctuating hydrodynamics \cite{landau_fluidmech_book,flekkoy_fluctuating_1996} \footnote{For simple liquids, the Navier-Stokes equations often turn out to be valid  down to the nanometer scale \cite{bocquet_nanofluidics_2010}}. In the limit of small slopes and small curvatures, these reduce to the stochastic lubrication equations \cite{davidovitch_spreading_2005,gruen_thin-film_2006,rauscher_wetting_2008}. Within this approach, typically, a thin precursor film \cite{popescu_precursor_2012} is assumed to be present. In numerical solutions, the impenetrable nature of the wall is taken into account, for instance, by rejecting algorithmically negative configurations \cite{nesic_dynamics_2015,nesic_nonlinear_2015}.  The corresponding \emph{static} problem, however, admits a general no-flux boundary condition at the wall \cite{burkhardt_propagator_1989} and, as we shall discuss in this study, a prewetting film is not a necessity. In view of the singularity problem of the moving contact line \cite{huh_scriven_1971}, it is furthermore tempting to ask whether thermal fluctuations alone can provide any form of {\it microscopic} regularization within a purely local fluctuating hydrodynamic framework. The first step is then to investigate \emph{equilibrium} properties of a contact line on a homogeneous substrate in the presence of thermal fluctuations and to explore the effects of such fluctuations on the scale of the thermal length, where their strength is comparable to surface tension.

Various approximations have been proposed in the literature for the microscopic description of capillarity in equilibrium. The most rigorous approach is density functional theory (DFT), in which the grand potential is minimized with respect to the density field. The interface elevation profile $h(x)$ can then be deduced from the Gibbs dividing surface. A description of the density profile across the interface is particularly important in the presence of long-range van der Waals interactions \cite{deGennesRMP, dietrich_wetting_1988}. The effect of long-range interactions has been assessed, for instance, in Refs.\ \cite{dietrich_microscopic_1991,napiorkowski_structure_1993, mecke_effective_1999} for free interfaces,  and in Refs.\ \cite{getta_line_1998, bauer_quantitative_1999} for an inhomogeneous wetting film. Whenever the radius of curvature of the interface is much larger than the thickness of the interface, a further simplification, known as the sharp interface approximation \cite{getta_line_1998,MerchantPFA1992,snoeijer_microscopic_2008,MWSA11,Weijs2014}, can be used, in which the grand potential is computed assuming that the two phases have a homogeneous density. The grand potential, which remains a non-local functional of $h(x)$, is then minimized with respect to the interface profile $h(x)$. This description captures the microscopic properties such as the stress-anisotropy near the interface, the disjoining pressure and the line tension and is consistent with macroscopic thermodynamics in the form of Laplace pressure and Young's law \cite{schimmele_linetension_2007,MWSA11,weijs_origin_2011,MDSA12b}. It is expected to be more accurate for fluids consisting of particles that interact via short-ranged forces, such as colloid-polymer mixtures \cite{lekkerkerker_life_2008, vandecan_theoretical_2008}. The Derjaguin approximation is an uncontrolled approximation of the DFT in the sharp interface approximation. The interaction between the liquid-vapor interface and the substrate is described by a local binding potential associated with the effect of disjoining pressure \cite{safran_statistical_1994, white_deryaguin_1983, solomentsev_microscopic_1999}. Along with this approximation, the effect of interface curvature (Laplace pressure) is captured by a local additive gradient contribution, leading to the classical square-gradient Hamiltonian for the interface profile $h(x)$ \cite{sekimoto_morphological_1987,indekeu_line_1992,dobbs_line_1993}. A different local approximation, uncontrolled as well, and based on a generalization of the disjoining pressure was introduced in Ref.\ \cite{snoeijer_microscopic_2008} and further studied in Ref.\ \cite{Pahlavan}. In practice, these local approximations often turn out to be quite accurate \cite{bauer_quantitative_1999}. They have the practical advantage that the equilibrium condition reduces to a second order differential equation and a non-local integral equation is avoided.  DFT can be used to predict the small scale structure close to the contact line, using an expansion around a purely repulsive system. For a typical Lennard-Jones interaction, it predicts the presence of a nanoscopic precursor film in front of the three-phase contact line \cite{getta_line_1998}. The thickness of this film is the relevant order parameter of the wetting transition. Upon crossing the wetting temperature $T_\st{w}$ from below, along the liquid-vapor coexistence, the film becomes macroscopically thick \cite{dietrich_wetting_1988}. Depending on the shape of the binding potential, the nature of the wetting transition can be first-order, for which the film thickness jumps abruptly from a finite to an infinite (macroscopic) value, or continuous, in which case the film thickness increases smoothly upon crossing $T_\st{w}$ \cite{dietrich_wetting_1988}. 
\rev{However, at least far from $T_\st{w}$, thin precursor films are often not observed in molecular dynamics simulations of partially wetting liquids having Lennard-Jones interactions \cite{lundgren_waterMD_2002, weijs_origin_2011, tretyakov_parameter_2013}.} 
\rev{In fact, the existence and size of these precursor films depend on the characteristics of the intermolecular interactions \cite{iseleholder_precursors_langm2016, iseleholder_precursors_pre2016} and other material properties \cite{diaz_hysteresis_2010, diaz_contactangle_pre2016}} \footnote{\rev{Precursor films typically occur for volatile liquids that have a large vapor pressure and a strong adsorption preference for the substrate \cite{diaz_hysteresis_2010, diaz_contactangle_pre2016}}}. Furthermore, the theoretically predicted thickness of these films is often smaller than the particle diameter, except very close to the wetting transition \cite{berim_nanodropDFT_2008, nold_fluid_2014, nold_nanoscale_2015, hughes_liquid_2015}. 

Within all the above descriptions, the physics near the contact line is oversimplified since thermal fluctuations are ignored; only their average effect is possibly included in the form of effective model parameters. However, thermal noise actually excites capillary waves at liquid interfaces \cite{mandelstam_uber_1913,buff_interfacial_1965,evans_nature_1979,rowlinson_molecular_1982}. In the case of a free (Gaussian) interface, this leads to an interfacial roughness that grows linearly with the extent of the interface in one dimension (fluctuating line) and logarithmically in two dimensions (fluctuating surface) \cite{rowlinson_molecular_1982, safran_statistical_1994, flekkoy_fluctuating_1996}.  Near an impenetrable boundary, fluctuation modes are restricted, giving rise to an ``entropic repulsion'' of the interface from the wall \cite{fisher_walks_1984, bricmont_random_1986, lebowitz_effect_1987}. In the context of critical wetting transitions, interfacial fluctuations have been rationalized in terms of a renormalization of the binding potential \cite{brezin_critical_1983, kroll_universality_1983, fisher_wetting_1985, lipowsky_scaling_1987, huse_comment_1987, juelicher_functionalRG_1990, spohn_fixedpoints_1991, indekeu_thermal_2010}. The effect of fluctuations on wetting transitions and on the interface morphology has been extensively studied \cite{vallade_transition_1981,burkhardt_propagator_1989,forgacs_review_1991,fisher_effective_1991,jin_effective_1993, parry_fluctuations_1993, burkhardt_two-dimensional_1998, parry_universality_1999, parry_droplet_2001, romero-enrique_interfacial_2004,parry_derivation_2006}. In the case of a one-dimensional interface, the solution can be obtained via a mapping to a quantum mechanical eigenvalue problem, whereas, for two-dimensional interfaces, field theoretical methods have to be employed. In many analytical investigations of the shapes of wedges of droplets, however, contact lines are either taken to be fixed \cite{vallade_transition_1981, burkhardt_propagator_1989, de_coninck_microscopic_1989, de_coninck_contact_1994, ben_arous_construction_1996, de_coninck_random_2009}, or a mesoscopic precursor film is supposed in front of the wedge \cite{getta_line_1998, bauer_quantitative_1999, romero-enrique_interfacial_2004, nesic_dynamics_2015}. Due to the entropic repulsion effect, such a film would necessarily have a finite thickness. In view of the fact that such films are \rev{often} not observed in molecular dynamics simulations of partially wetting liquids \cite{lundgren_waterMD_2002, weijs_origin_2011, tretyakov_parameter_2013} -- and they are far too thin to be detected experimentally -- it is natural to assume, instead, that the interface touches the substrate at a well-defined location. This location should not be considered as fixed, but as a fluctuating quantity. Indeed, in the so-called molecular kinetic theory \cite{BH69, Blake93,blake_dynamics_review_2011} as well as in mesoscopic approaches to dynamic wetting \cite{colosqui_crossover_2013, colosqui_terraced_wetting_2015}, a fluctuating contact line is naturally present.
It should be emphasized that, far away from the contact line, the contact angle is selected by the forces exerted on the three corners of the liquid wedge and, as a consequence of the homogeneity of the substrate \cite{snoeijer_microscopic_2008}, one expects to recover Young's law after averaging. Fluctuations of a contact line have been previously studied within various analytical models \cite{pomeau_thermal_1983, clarke_thermal_1992, abraham_divergence_1993, jakubczyk_point_2006}. 
In Refs.\ \cite{pomeau_thermal_1983, clarke_thermal_1992}, however, the effect of the impenetrability of the substrate on the fluctuations of the interface has been disregarded. In Ref.\ \cite{abraham_divergence_1993}, the concept of point tension (which is the one-dimensional analogue of the line tension) was generalized by taking into account fluctuations of the contact point within the two-dimensional Ising model. Noteworthy, in \cite{de_coninck_spreading_1993}, based on a ``solid-on-solid'' model, the spreading of a droplet was investigated, taking into account the entropic repulsion and allowing both interface and contact line to fluctuate.

In this study, we investigate the effect of thermal fluctuations on the morphology of a one-dimensional interface near a contact line within an exactly solvable model, taking fully into account the impenetrability of the substrate. The fluctuating interface is modeled based on a path integral representation of the partition function of a standard (Gaussian) capillary wave Hamiltonian for a short-range binding potential \cite{vallade_transition_1981,kroll_universality_1983, burkhardt_propagator_1989,forgacs_review_1991, indekeu_line_1992}.
Crucially, each realization of the interface $h(x)$ is supposed to have a well-defined position of the contact line. In order to facilitate an analytical treatment, we consider the limiting case of a pure ``contact potential'' \cite{burkhardt_propagator_1989}, which gives rise to an effective boundary condition for the interfacial propagator at the wall, characterized by a single dimensionless parameter. 
While the identification of the boundary condition parameter by asymptotic matching with an ``inner layer'' description including the details of intermolecular interactions remains out of the scope of the present paper, we explore different scenarios, from cases where interfaces are microscopically bound to the wall, to cases where the interface unbinds. We find that the fluctuating nature of the contact line gives rise to an exponentially decaying precursor film in front of the liquid wedge. Far from the surface, the behavior of the average profile is linear and Young's law is recovered analytically. The effect of fluctuations on the profile is captured in terms of an effective disjoining pressure. The associated binding potential is of finite range and therefore regularizes the singular contact force appearing in the deterministic case.

The paper is organized as follows: in Section \ref{sec:deterministic} we recall the derivation (similar to Refs. \cite{snoeijer_microscopic_2008,sekimoto_morphological_1987,safran_statistical_1994}) of the equilibrium properties of sharp partially wetting interfaces both in the macroscopic theory (\cref{sec:deterministic1}) as well as when a disjoining pressure is included (\cref{sec:disjoining}). The results in \cref{sec:disjoining} will be useful to describe the effects of thermal fluctuations within the framework of an effective deterministic theory. In \cref{sec:contact}, via a straightforward extension of the macroscopic theory of \cref{sec:deterministic1}, we incorporate thermal fluctuations in a model based on contact interactions. Specifically, we discuss in \cref{sec:fixed} the case of a {\it pinned contact line} and -- more extensively -- in \cref{sec:hxR} the analogous case of a {\it fluctuating contact line}. A summary of our central findings is given in \cref{sec:conclusions}. Technical details are reported in the appendices \ref{app:av_prop}-\ref{app:squarewell}.

\section{Partial wetting for a deterministic interface}\label{sec:deterministic}
\subsection{General framework}

\begin{figure}[t!]\centering
	\includegraphics[height=0.17\textheight]{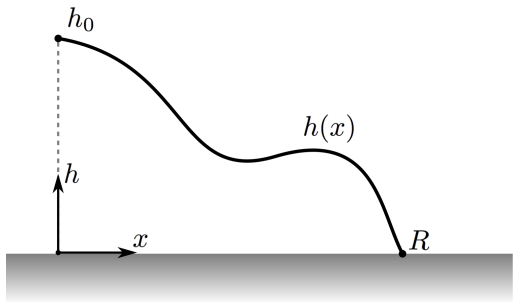}\hfill
	\caption{Sketch of a one-dimensional interface described by an elevation profile $h=h(x)$, where $x$ is the coordinate along the substrate. The position of the contact line, at which $h=0$, is denoted by $x=\R$, while $h_0$ plays the role of a macroscopic reference cut-off length (e.g., the capillary length \cite{rowlinson_molecular_1982}).}
	\label{fig_wedge_sketch_gen}
\end{figure}


We consider an interface that is invariant along the transverse direction (\cref{fig_wedge_sketch_gen}). The interface is described by a single elevation profile $h=h(x)$, where $x$ is the coordinate along the substrate. It joins the substrate at the contact line, located at $x=\R$ and at which $h=0$, i.e.,
\beq\label{eq:hvar(R)=0}
h(\R)=0.
\eeq
We generally require that $h(x)$ identically vanishes for $x\geq\R$, i.e., we do not assume the presence of a thin wetting film in front of the macroscopic profile. As the problem will be treated in two dimensions, the contact line is in fact a contact \emph{point}. However, for simplicity we shall keep the name ``contact line''. We briefly turn to the three-dimensional problem in the conclusion (\cref{sec:conclusions}). 

We assume that the fluid is limited to the region $x>0$. The height at the origin $x=0$ is denoted by $h_0=h(0)$, which is assumed to be imposed. The volume of liquid, conversely, is not imposed. This hypothesis is introduced for simplicity: it allows us to recover a straight wedge shape far from the contact line. As discussed later on, in presence of thermal fluctuations, $h_0$ plays the role of a macroscopic cut-off length and is a relevant parameter of the problem. We then introduce a free energy functional as (prime $'$ means here derivative with respect to $x$)
\beq\label{eq:Action}
\Acal_\R[h] \equiv \int_0^\R dx\, \Gamma\Round{h(x),h'(x)},
\eeq
where $\Gamma(h,h')$ is a suitable free energy density. To avoid possible confusion with the Helmholtz free energy obtained after integrating over all degrees of freedom, we will henceforth refer to the above free energy functional $\Acal_\R[h]$ as the \emph{action}. In the conventional macroscopic approach to capillarity, $\Gamma(h,h')$ can be expressed as a function of the surface tension coefficient $\gamma$ and Young's angle $\Young$, defined by \cref{eq_Young}: 
\beq\label{eq:Gamma0}
\Gamma(h,h')=\gamma\Round{\sqrt{1+h'^2}-\cos\Young}.
\eeq
In the next sections we consider this model -- and refined versions of it including details at a microscopic scale -- in the small slope approximation and take it as a stepping stone for the treatment of fluctuations. 
For the remaining part of this section we shall not assume a particular form for $\Gamma$.

In order to obtain the equilibrium (stationary) profile, the action $\Acal_\R[h]$ has to be minimized under the constraints $h(0)=h_0$ and $h(\R)=0$, with $\R$ variable. Let us call the stationary profile the ``classical'' profile and denote it by $h_\st{cl}(x)$ (the reason of this notion is explained below). We shall assume that $h_\st{cl}(x)$ identically vanishes for $x\geq\R_\st{cl}$. Then, for small variations of the profile and of the contact line position over their stationary values, i.e., for $h(x)=h_\st{cl}(x)+\delta h(x)$ and $\R=\R_\st{cl}+d\R$, the first variation of the action $\delta\Acal=\Acal_{\R}[h]-\Acal_{\R_\st{cl}}[h_\st{cl}]$ reads:
\beq\label{eq:var_E}
\delta\Acal = \int_0^\R dx\, \Round{\frac{\pt\Gamma}{\pt h_\st{cl}} - \frac{d}{dx} \frac{\pt\Gamma}{\pt h_\st{cl}'}} \delta h(x) + \left.\frac{\pt\Gamma}{\pt h_\st{cl}'}\right|_{x=\R_\st{cl}} \delta h(\R_\st{cl}) + \Gamma\Round{0,h_\st{cl}'(\R_\st{cl})} d\R,
\eeq
where $h_\st{cl}(0)=h_0$ (i.e., $\delta h(0)=0$) and $h_\st{cl}(\R_\st{cl})=0$ have been used. The term $\delta h(\R_\st{cl})$ can be interpreted as a variation of the profile $h=h(x)$ at $x=\R_\st{cl}$ stemming from a sole variation of the contact line. In fact, a virtual displacement of the contact line induces a non vanishing $h(\R_\st{cl})\equiv\delta h(\R_\st{cl})$ \footnote{Note that $\delta h(\R_\st{cl})$ can be negative, depending on the sign of $d\R$, while $h(\R_\st{cl})$ is non-negative by construction. A more rigorous interpretation is given by \cref{eq_dh_dR_coupling}.}. By Taylor expanding \cref{eq:hvar(R)=0} and using $h_{\st{cl}}(\R_{\st{cl}})=0$ we obtain, at first order,
\beq\label{eq_dh_dR_coupling}
\delta h(\R_{\st{cl}}) + h_{\st{cl}}'(\R_{\st{cl}}) d\R = 0.
\eeq
The last two terms in \cref{eq:var_E} can then be combined such that
\beq\label{eq:total_variation}
\delta\Acal = \int_0^\R dx\, \Round{\frac{\pt\Gamma}{\pt h_\st{cl}} - \frac{d}{dx} \frac{\pt\Gamma}{\pt h_\st{cl}'}} \delta h(x) + \Round{\Gamma\Round{0,h_\st{cl}'(\R_\st{cl})} - h_{\st{cl}}'(\R_{\st{cl}}) \left.\frac{\pt\Gamma}{\pt h_\st{cl}'}\right|_{x=\R_\st{cl}}} d\R.
\eeq
The same result is obtained when imposing the constrains using Lagrange multipliers \cite{sekimoto_morphological_1987}. Stationarity of the action at equilibrium requires $\delta\Acal=0$ with respect to independent (and arbitrary) variations of the profile and contact line, implying
\beq\label{eq:bulk}
\frac{\pt\Gamma}{\pt h_\st{cl}} - \frac{d}{dx} \frac{\pt\Gamma}{\pt h_\st{cl}'} = 0
\eeq
and
\beq\label{eq:bound}
\Gamma\Round{0,h_\st{cl}'(\R_\st{cl})} - h_{\st{cl}}'(\R_{\st{cl}}) \left.\frac{\pt\Gamma}{\pt h_\st{cl}'}\right|_{x=\R_\st{cl}} = 0,
\eeq
respectively. Multiplying \cref{eq:bulk} by $h_\st{cl}'(x)$ and integrating over $x$, we find that
\beq\label{eq:Ginvar}
G(h,h') \equiv \Gamma(h,h') - h'\frac{\pt\Gamma}{\pt h'}
\eeq
must be constant for a stationary solution, i.e., $G(h_\st{cl}(x),h_\st{cl}'(x))=\textup{const}$. Physically, this invariant expresses the horizontal balance of the forces acting on the liquid wedge extending from the contact line to an arbitrary cross section at position $x$. The actual value of the constant can be computed at the contact line, where it follows from \cref{eq:bound} immediately as $G\Round{0,h_\st{cl}'(\R)} = 0.$ Accordingly, we get the stationarity condition
\beq\label{eq:gmacro0}
G(h_\st{cl}(x),h_\st{cl}'(x)) = 0,
\eeq
which is a first order differential equation encapsulating both \cref{eq:bulk} and the boundary condition~\eqref{eq:bound}. To find $h_\st{cl}(x)$ one solves \cref{eq:gmacro0} under the constrains $h_\st{cl}(0)=h_0$ and $h_{\st{cl}}(\R_{\st{cl}})=0$ to determine the stationary contact line location $\R_{\st{cl}}$. 
Recall that the stationary profile is understood to be identically vanishing for $x\geq\R_{\st{cl}}$.\\
Throughout this work we shall use the term ``classical'' synonymous for ``deterministic'', and thereby distinguish the stochastic solution considered in \cref{sec:contact} below.
There, we shall work with the Euclidean counterpart of quantum mechanics wherein the role of the Planck constant is played by the temperature. When introducing thermal fluctuations in the theory, the limit of small temperature is expected to reproduce the deterministic (stationary) solution. We shall refer to this procedure as taking the ``classical limit''. 

\subsection{Classical macroscopic description}\label{sec:deterministic1}


\begin{figure}[t!]\centering
	\includegraphics[height=0.17\textheight]{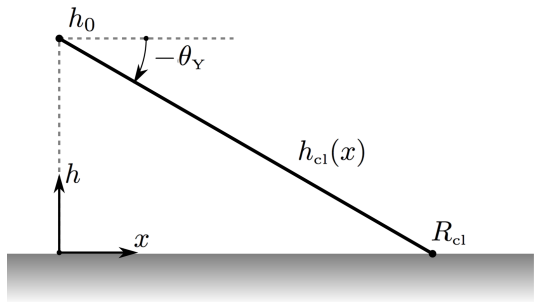}\hfill
	\caption{Macroscopic description of an interface partially wetting a substrate (see \cref{sec:deterministic1}): The classical (i.e.\ stationary) profile $h_\st{cl}(x)$ (solid line) resulting from the minimization of the action in \cref{eq:ActionMacro} is a straight wedge with slope angle $-\Young$ [see \cref{eq:classic}]. The position $\R_\st{cl}$ of the contact line follows from the boundary condition $h_\st{cl}(0)=h_0$ and the value of the macroscopic contact angle $\Young$. The shaded area represents the (impenetrable) wall.}
	\label{fig_wedge_sketch}
\end{figure}


The previous results can be easily specialized to the free energy density in \cref{eq:Gamma0}, resulting in the action
\beq\label{eq:ActionMacro}
\Acal_\R[h] = \gamma \int_0^\R dx\, \Round{\sqrt{1+h'(x)^2}-\cos\Young}.
\eeq
In order to facilitate the study of thermal fluctuations later, we apply the {\it gentle-slope approximation}, i.e., we perform in \cref{eq:ActionMacro} an expansion with respect to the slope $h'(x)$. Accordingly, also the angle $\Young$ is assumed to be small. The action in \cref{eq:ActionMacro} then simplifies to
\beq\label{eq:ActionMacroG-S}
\Acal_\R[h] = \frac{\gamma}{2} \int_0^\R dx\, \Round{h'(x)^2+\Young^2},
\eeq
and the free energy density, correspondingly, to
\beq\label{eq:Gamma0G-S}
\Gamma(h,h')=\frac{\gamma}{2}\Round{h'^2+\Young^2}.
\eeq
The variation of $\Gamma$ with respect to the profile [\cref{eq:bulk}] leads to
\beq\label{eq:of_motion}
h_\st{cl}''(x)=0
\eeq
showing that from the bulk equation one can only infer that the profile is linear, but not obtain any information on the value of the contact angle. The mechanism of contact angle selection comes from the variation with respect to the contact line position. Specifically, the invariant follows from \cref{eq:Ginvar} as
\beq\label{inv_GSapp}
G(h,h')=\frac{\gamma}{2}\Round{\Young^2-h'^2},
\eeq
which is the unbalanced Young's force per unit line. The stationarity condition \eqref{eq:gmacro0} then gives Young's law in the form
\beq\label{eq:stationary}
h_\st{cl}'(x)^2=\Young^2
\eeq
and results in a classical profile having the form of a straight wedge (see \cref{fig_wedge_sketch}):
\begin{align}\label{eq:classic}
&h_{\st{cl}}(x) = h_0 -\Young x, &&\R_{\st{cl}} = \frac{h_0}{\Young}.
\end{align}
Note that in \cref{eq:classic} the location $\R_\st{cl}$ of the contact line follows from the externally imposed values of $h_0$ and the contact angle $\Young$. The stationarity condition in \cref{eq:stationary} is formally equivalent to
\beq\label{eq:normal}
h_\st{cl}''(x) - \Young \delta(x-R_\st{cl}) = 0,
\eeq
revealing the presence of a singular term (i.e., Dirac's delta function) in addition to the usual interface curvature term $h''_{\st{cl}}(x)$. If Dirac's delta function $\delta(x-R_\st{cl})$ is replaced by a disjoining pressure term, the outer angle is selected ``internally'' and the singularity in \cref{eq:normal} is {\it regularized} \cite{snoeijer_microscopic_2008}. For the contact line problem considered in the present study, we will investigate the effects of thermal fluctuations close to the contact line, and show that they produce a regularization similar to a disjoining pressure. In the next section we therefore discuss the basic features of such a regularized microscopic description within a deterministic framework.

\subsection{Regularization by a disjoining pressure}\label{sec:disjoining}


\begin{figure}[t!]\centering
	\includegraphics[height=0.17\textheight]{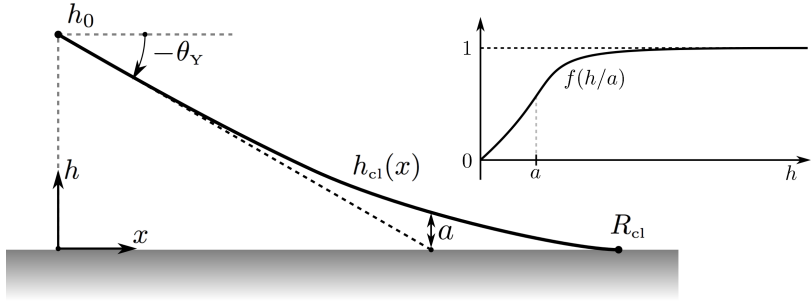}\hfill
	\caption{Interface partially wetting a substrate in the presence of a disjoining pressure (see \cref{sec:disjoining}): the minimization of the action [\cref{freeenergy:disjoining}] yields the profile $h_\st{cl}(x)$ and the location of the contact line $\R_\st{cl}$. The parameter $a$ is a regularization length and describes the range of the binding potential $f(h/a)$, such that $a\to 0$ corresponds to the purely macroscopic case considered in \cref{fig_wedge_sketch}. The typical behavior of $f$ is sketched on the upper right side. The value of the macroscopic contact angle $\Young$ is determined by the binding potential [see \cref{eq_dhdx_lim_bulk}] and not imposed externally. While the classical profile has vanishing slope at the contact line [see \cref{eq_dhdx_lim_wall}], we do \emph{not} assume a precursor film to be present in front [i.e., right to the point $(\R_\st{cl},0)$] of the profile. The dashed line represents the extrapolation of a profile with constant slope angle $-\Young$ towards the substrate (represented by the shaded area).}
	\label{fig_wedge_sketch_disj}
\end{figure}


Let us introduce a regularization of the action at a microscopic scale $a$, based on a {\it binding potential} whose derivative represents the (negative of a) {\it disjoining pressure} \cite{deGennesRMP,safran_statistical_1994}. In the presence of such a binding potential the entire classical solution follows from a pressure balance, without the necessity of imposing an external condition for the contact angle. In fact, the contact angle is selected internally, i.e., it is determined by the (asymptotic value of the) binding potential and, therefore, ultimately by the molecular interactions. We write the action in the form:
\beq\label{freeenergy:disjoining}
\Acal_\R[h]  = \gamma\int_0^\R dx\, \Square{\sqrt{1+h'(x)^2}-1+(1-\cos\Young)f(h(x)/a)},
\eeq
where the dimensionless binding potential $f(h/a)$ is a regular monotonic function obeying $f(0)=0$, $f(\infty)=1$.
\rev{The prefactor multiplying $f$ is the negative of the spreading coefficient $S=\gamma_\st{sv}-\gamma_\st{sl} - \gamma = \gamma(\cos\Young-1)$.}
For a two-dimensional fluid consisting of particles that interact by van der Waals forces, the binding potential tends to $1$ algebraically, $\sim 1/h^3$ \cite{kroll_universality_1983} \footnote{In the analogous three-dimensional case, the algebraic decay is $\sim 1/h^2$ \cite{deGennesRMP}.}. The molecular scale $a$ can be related to the Hamacker constant and the surface tension $\gamma$ \cite{deGennes_book,deGennesRMP,Israelachvili}. 
\rev{The typical shape of the interface potential is illustrated in the inset to \cref{fig_wedge_sketch_disj}. We remark that alternative forms of $f$ at intermediate $h$ are possible, see, e.g., Refs.\ \cite{BrochardWyart_spreading_1991, deGennes_book}.
As motivated in the introduction, we focus here on true partial wetting states, for which no precursor film exists in front of the contact line. As demonstrated below, this requires $f(0)=0$.}
In the gentle-slope approximation \cref{freeenergy:disjoining} simplifies to
\beq
\Acal_\R[h]  = \frac{\gamma}{2}\int_{0}^{\R}dx\, \Square{h'(x)^2  + \Young^2 f\Round{h(x)/a}},
\label{eq_fEfunc_reg}\eeq
corresponding to the free energy density
\beq\label{eq:GammaaG-S}
\Gamma(h,h')=\frac{\gamma}{2}\Square{h'^2+\Young^2f(h/a)}.
\eeq
The variation of $\Gamma$ with respect to the profile [\cref{eq:bulk}] leads to 
\beq\label{eq:eqnofmotion}
h_\st{cl}''(x)=\frac{\Young^2}{2a}f'(h_\st{cl}(x)/a),
\eeq
which is compatible with many solutions, depending on the form of the binding potential at $h=0$. The mechanism of contact angle selection can be easily inferred by computing from \cref{eq:Ginvar} the invariant associated with \cref{eq:GammaaG-S}:
\beq\label{eq:GammaMeso}
G(h,h')=\frac{\gamma}{2}\Square{\Young^2f(h/a)-h'^2}.
\eeq
According to \cref{eq:gmacro0}, the stationary condition for the regularized profile reads
\beq\label{dhdxfull}
h_\st{cl}'(x)^2=\Young^2 f\Round{h_\st{cl}(x)/a}.
\eeq
\rev{Since we require $f(0)=0$, \cref{dhdxfull} automatically leads to a vanishing slope at the contact line:}
\beq\label{eq_dhdx_lim_wall}
h_\st{cl}'(\R_\st{cl}) = 0,
\eeq
which can be interpreted as a film of vanishing thickness extending for $x>\R_\st{cl}$. On the other hand, in the limit $h/a\to\infty$, i.e., far from the substrate, we have $f(h/a)\sim f(\infty)=1$, and hence
\beq\label{eq_dhdx_lim_bulk}
h_\st{cl}'(0)^2 = \Young^2f(h_0/a) \overset{h_0/a\to\infty}{\sim} \Young^2.
\eeq
This shows explicitly that Young's contact angle is obtained as an outer asymptotics in the limit of a height much larger than $a$. This selection is independent of the detailed shape  of the function $f(h/a)$. Thus, $a$ can be considered to be the length scale at which the (outer) contact angle is selected.
\rev{The preceding analysis is summarized in \cref{fig_wedge_sketch_disj}, where the main panel displays the typical shape of the profile resulting from an interface potential as sketched in the inset \footnote{\rev{Crucially, the requirement that $h'(\R_\st{cl})=0$ at a finite contact point $\R_\st{cl}$ restricts the possible behaviors of $f(h/a)$ for $h\to 0$. As can be easily inferred from \cref{dhdxfull}, for $R_\st{cl}$ to be finite an algebraic behavior $f\sim h^p$ with $0<p<2$ is required. Other behaviors lead to $R_\st{cl}=\infty$ [cf.\ Ref.\ \cite{starov_wettingreview_2009}}.}}.

\section{Partial wetting for a fluctuating interface}\label{sec:contact}

\begin{figure}[t!]\centering
	\includegraphics[height=0.17\textheight]{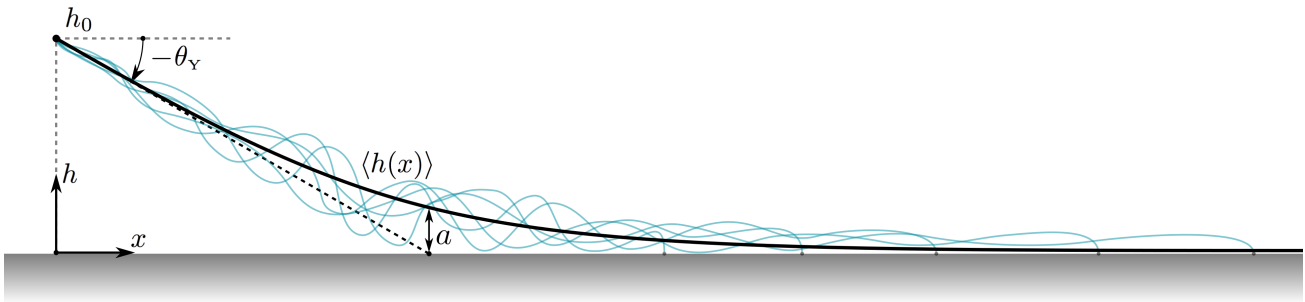}\hfill
	\caption{Thermally fluctuating interface wetting a substrate (see \cref{sec:contact} and, in particular, \cref{sec:hxR}). The thin blue lines represent different stochastic realizations of the interface profile, which are assumed to all start at a fixed ``outer scale'' $h_0$, and which touch the substrate at different random locations. The thick black line indicates the average profile $\Angle{h(x)}$ [see \cref{eq_avg_profile_fluctR}], which, compared to the classical solution [dotted line, \cref{eq:classic}], smoothly crosses over to a thin precursor film. In the outer region, both the classical and average profiles are expected to obey Young's law [\cref{eq_Young}]. The effect of the fluctuations of the interface and the contact line can be captured in terms of an effective binding potential with regularization length $a$ [cf.\ inset to \cref{fig_wedge_sketch_disj}].}
	\label{fig_wedge_sketch_fluct}
\end{figure}


In this section we introduce thermal fluctuations into the macroscopic approach described in \cref{sec:deterministic1}. We discuss the influence of thermal fluctuations on the structure of the profile around the contact line and show that they lead to a regularization of the problem in the form of an effective disjoining pressure. It is convenient to introduce the functional
\beq\label{eq:Ham}
\Hcal_\R[h] \equiv \frac{1}{2}\int_{0}^{\R}dx\, h'(x)^2,
\eeq
which allows us to write the macroscopic action in \cref{eq:ActionMacroG-S} as
\beq\label{eq:action}
\Acal_\R[h] = \gamma\Round{\frac{\Young^2}{2}\R + \Hcal_\R[h]}.
\eeq 
We promote now both the profile $h=h(x)$ and the contact line position $\R$ to random variables whose statistics are governed by the action $\Acal_\R[h]$ given in \eqref{eq:action}. Thermal fluctuations introduce a new length scale, the {\it thermal length}
\beq\label{eq:ell}
\l = \frac{\kB T}{\gamma},
\eeq
where $T$ is the temperature and $\kB$ the Boltzmann constant. The partition function $Z$ for our model is then obtained as a path integral \cite{feynman_book, kleinert_book, chaichian_book} by integrating the statistical weight $e^{-\Acal_\R[h]/\kB T}=e^{-\Young^2\R/2\l}e^{-\Hcal_\R[h]/\l}$ over all the possible realizations of $h=h(x)$ and $\R$ satisfying the conditions $h(0)=h_0$ and $h(\R)=0$, that is
\beq\label{eq:Z}
Z \equiv \int_0^\infty d\R\, \int_{h(0)=h_0}^{h(\R)=0} \Dcal_\theta h\, e^{-\Acal_\R[h]/\kB T} = \int_0^\infty d\R\, e^{-\Young^2\R/2\l} c(h_0,0;\R),
\eeq
where the propagator 
\beq\label{eq:fixedRpart}
c(h_0,0;\R) \equiv \int_{h(0)=h_0}^{h(\R)=0} \Dcal_\theta h\, e^{-\Hcal_\R[h]/\l} = \int_{h(0)=h_0}^{h(\R)=0} \Dcal_\theta h\, e^{-\int_{0}^\R dx\, h'(x)^2/2\l}
\eeq
has been introduced. This expression for the propagator can be understood as the (non-normalized) probability density associated with the set of all profiles connecting the height $h_0$ with the height $h=0$ over a distance $\R$ under the condition to remain non-negative in between.
The presence of an impenetrable wall at $h=0$ is embodied in the notation $\Dcal_\theta h$ for the measure, which ensures that only non-negative $h=h(x)$ contribute to the path integral (see, e.g., Refs.\ \cite{majumdar_brownian_2005,Rajabpour_2009,Farhi_1990,Clark_1980}). 
The precise nature of the wall (i.e., its attractive or repulsive character) is encoded in the real dimensionless parameter $\theta$, which will appear in a boundary condition at $h=0$, as explained below [see \cref{eq:prop_Robin}]. Restricted path integrals of the type appearing in \cref{eq:fixedRpart} are well known in the study of first-passage phenomena \cite{majumdar_brownian_2005, majumdar_spatial_2006} and have a quantum mechanical analogue in a particle confined to the half line \cite{ohya_running_2011,ohya_path_2012,bonneau_self-adjoint_2001,berman_boundary_1991}, where the thermal length $\l$ plays the role of the Planck constant $\hbar$. 
We remark that, for fixed $\R$, the propagator $c(h_0,0;\R)$ itself plays a role of a partition function, and the functional $\Hcal_\R[h]$ the role of the associated action (see Appendix~\ref{app:fixedR}). The ensemble defined by $c(h_0,0;\R)$ in \cref{eq:fixedRpart} corresponds to the \rev{classical ``contact potential'' model introduced in Ref.\ \cite{burkhardt_propagator_1989}. 
We shall henceforth call it the {\it pinned contact line} model and study it further in \cref{sec:fixed}.} 

We have seen (\cref{sec:deterministic1}) that, in the macroscopic approach, the selection of Young's angle results from the variations with respect to the contact line position $R$. We therefore expect, once fluctuations are taken into account, that the appropriate ensemble is the one defined by $Z$ in \cref{eq:Z}, where variations of the contact line are indeed allowed. By construction, $Z$ does not depend on the variable $\R$, which has been integrated out, but it depends on the variable $\Young$ appearing in the exponential weight $e^{-\Young^2\R/2\l}$. In the remaining part of our study we shall refer to $Z$ as the {\it fluctuating contact line ensemble}. The first obvious difference between the pinned contact line problem and the fluctuating contact line problem is that, in the latter, the contact line fluctuates around the mean value
\beq\label{eq:Rav_general}
\Angle{\R} = \frac{1}{Z} \int_0^\infty d\R\, \R e^{-\Young^2\R/2\l} c(h_0,0;\R) = -\frac{\l}{\Young} \frac{\pt \ln Z}{\pt\Young}.
\eeq
According to \cref{eq:Z}, the fluctuating contact line partition function $Z$ is the integral over all the possible $\R$ of the propagator $c(h_0,0;\R)$ with a weight $e^{-\Young^2\R/2\l} \simeq e^{S\R/\kB T}$, where the prefactor of $\R$ in the exponential is essentially the spreading coefficient $S=\gamma\,(\cos\Young-1) \simeq -\gamma \, \Young^2/2$ normalized by the thermal energy $\kB T$. In the language of statistical mechanics, such a weighted integral can be interpreted as a way to establish a contact with a ``reservoir'' characterized by Young's angle $\Young$. Accordingly, the scale $h_0$ represents the scale at which Young's angle is recovered -- independently from the properties of the model (such as the specific boundary conditions) near the contact line.

Another interesting point to be investigated concerns role played by the invariant $G(h,h')$ defined in \cref{inv_GSapp} and, more generally, the averaged equations of motion [\cref{eq:of_motion}]. In the macroscopic approach, the stationary profile has zero curvature, $h_\st{cl}''(x)=0$. Due to the presence of a contact line and the topological constraints imposed by the impenetrable wall, we will see that the same result does not hold in the presence of thermal fluctuations, and the average profile develops a non zero curvature, $\Angle{h''(x)} \neq 0$. As a consequence, Ehrenfest's theorem \cite{shankar_QM_book,maeda_unitary_2015} is violated. This fact can be interpreted as a regularization effect of the classical problem induced by thermal fluctuations. 

In the fluctuating contact line ensemble, the average profile is given by (see \cref{app:fluctuatingR})
\beq\label{eq_avg_profile_fluctR}
\Angle{h(x)} = \frac{1}{Z} \int_x^\infty d\R\, \Angle{h(x)}_\R e^{-\Young^2\R/2\l} c(h_0,0;\R),
\eeq
where
\beq\label{eq:profileR}
\Angle{h(x)}_\R = \frac{1}{c(h_0,0;\R)} \int_{0}^{\infty} dh\, h\, c(h_0,h;x) c(h,0;\R-x),\qquad 0<x<\R,
\eeq
is the average profile for fixed $\R$, which follows straightforwardly from the Chapman-Kolmogorov equation [see \cref{eq:C-K_eq}]. Note that, by construction, $\Angle{h(x)}_\R$ vanishes for $x\geq\R$. 
The propagator $c(h_0,h;x)$ in \cref{eq:profileR} can be expressed in terms of a path integral, as discussed in \cref{app:av_prop}.
As is well known \cite{feynman_book, kleinert_book, chaichian_book}, in the domain $h_0,h\geq 0$ and for any $X>0$, this propagator can be equivalently obtained as a solution to a diffusion equation:
\beq\label{eq:diff}
\frac{\pt c}{\pt X} = \frac{\l}{2}\frac{\pt^2c}{\pt h^2}, \hspace{.4in} c(h_0,h;0) = \delta_{h_0}(h),
\eeq
where the second equation serves as an initial condition for the first order differential equation in $X$ \footnote{We have chosen the notation $\delta_{h_0}(h)$ for Dirac's delta function here, as explained in Appendix~\ref{app:av_prop}.}. 
Since \cref{eq:diff} is a second order differential equation in $h$, two boundary conditions are required. A physically meaningful propagator must obey $c(h_0,h;X)\to0$ for $h\to\infty$ \footnote{In fact, $c(h_0,h;X)$ must vanish for $h\to\infty$ more rapidly that $1/h$, in such a way that the integral of $c(h_0,h;X)$ over all $h\geq0$ is finite.}. This boundary condition characterizes the behavior of the system far from the solid substrate. The other boundary condition must embody the presence of an impenetrable wall at $h=0$. To this aim, we \emph{impose} that the conditional probability density which can be associated with \cref{eq:profileR} has vanishing flux at $h=0$. As detailed in Appendix~\ref{app:fixedR}, this is ensured by the boundary condition
\beq\label{eq:prop_Robin}
\l \left.\frac{\partial\ln c}{\partial h}\right|_{h=0} = - \theta.
\eeq
Here, $\theta$ is a dimensionless parameter independent of $X$ and $h_0$, which can interpreted as an effective (coarse grained) boundary condition parameter characterizing the importance of thermal fluctuations relative to the attractive interaction with the wall \cite{burkhardt_propagator_1989}. If one interprets $-\l\ln c$ as a dimensionless free energy, the parameter $\theta$ in \cref{eq:prop_Robin} can be understood as the dimensionless energy per unit height (i.e., a normal force) needed to detach the interface from the wall: a positive $\theta$ indicates binding of the interface, while a negative $\theta$ indicates repulsion of the interface from the wall. Note that \cref{eq:prop_Robin} reduces to Dirichlet (absorbing) and Neumann (reflecting) boundary conditions in the limits $\theta\to -\infty$ and $\theta=0$, respectively. However, \cref{eq:prop_Robin}, which is known as Robin boundary condition, is more general and covers all boundary conditions compatible with a zero conditional probability flux at the wall. We return to a discussion of the meaning of $\theta$ below. 

The sign of $\theta$ is intimately connected to the presence of bound states in the quantum mechanical problem described by the Schr\"{o}dinger-like equation \eqref{eq:diff}. To see this, let us write the formal expansion of the propagator 
\beq\label{eq_prop_eigen_decomp}
c(h_0,h;X) = \sum_\e \psi_\e^*(h_0) \psi_\e(h) e^{-\e X/\l}
\eeq
in terms of orthonormal solutions $\psi=\psi_\e(h)$ to the stationary problem (prime $'$ means here derivative with respect to the variable $h$)
\beq\label{eq:eigen}
-\frac{\l^2}{2}\psi'' = \e \psi,
\eeq
where $\e$ is a dimensionless energy eigenvalue [the sum over $\e$ in \cref{eq_prop_eigen_decomp} is assumed to take into account both the discrete and continuous parts of the spectrum]. Note that the reversibility condition $c(h,h_0;X)=c(h_0,h;X)$ [see \cref{eq:h-h0_symm}] implies that the energy eigenvalues are real. The boundary condition in \cref{eq:prop_Robin} requires
\beq\label{eq:Robin}
\theta \psi(0) + \l \psi'(0) = 0.
\eeq
It can be shown \cite{bonneau_self-adjoint_2001} that this condition ensures the self-adjointness of the differential operator appearing on the l.h.s.\ of \cref{eq:eigen}, which in turn is consistent with the requirement that all the $\e$ are real. As discussed in \cref{app:cont_prop}, for $\theta\leq0$ only positive energy eigenstates (scattering states) exist, while for $\theta>0$ one additional eigenstate with energy $\e=-\theta^2/2$ (bound state) arises. 
As is furthermore shown in Appendix~\ref{app:cont_prop}, the propagator resulting from \cref{eq_prop_eigen_decomp} is given by
\beq\label{eq:propagator}
c(h_0,h;X) = \frac{e^{-(h-h_0)^2/2\l X}+e^{-(h+h_0)^2/2\l X}}{\sqrt{2\pi \l X}} + \frac{\theta}{\l} e^{-\theta(h+h_0)/\l}e^{\theta^2X/2\l}\erfc\Round{\frac{h+h_0-\theta X}{\sqrt{2\l X}}},
\eeq
where
\beq\label{eq:erfc}
\erfc(z) = \frac{2}{\sqrt{\pi}} \int_z^\infty dt\, e^{-t^2} \overset{z\to\pm\infty}{\sim} 1\mp1 + \frac{e^{-z^2}}{\sqrt{\pi}z} \Round{1 - \frac{1}{2z^2}}.
\eeq
Note the asymptotic result, which will be useful in the following. 

\rev{We close this section by a few comments.
It has been shown in Refs.\ \cite{upton_interfacemodel_1999, upton_interfacemodel_2002} that, at large scales, a wetting interface in the two-dimensional Ising model is described by a contact potential model, with the boundary condition parameter $\theta$ being related to the distance to the wetting transition, $\theta\propto T_\st{w}-T$. The macroscopic contact angle is in this case given by $\theta$ [see \cref{eq_fixedR_angle}] and it turns out that a thin precursor film in front of the effective contact line is always present [see \cref{eq:filmRfixed_rescaled}].
Here, instead, we approach the fluctuating contact point problem from a macroscopic point of view, directly based on the classical action in \cref{eq:action}. As discussed in \cref{sec:Intro}, this is motivated by the fact precursor films are not necessarily observed in experiments or simulations.
Indeed, it turns out that the fluctuating contact point ensemble [\cref{eq:Z}] does not involve a precursor film (except if $\theta=\Young$), see \cref{sec:hxR}. This ensemble is characterized by two parameters, $\Young$ and $\theta$, the latter arising from the general requirement of a vanishing probability flux through the wall [see \cref{eq:prop_Robin}].}

Generally, the boundary condition in \cref{eq:prop_Robin} can be understood as a parametrization of the short-distance physics that would otherwise be represented by an interface binding potential $u(h/\sigma)$ \cite{belchev_robinbc_2010, ohya_running_2011} \footnote{The intermolecular regularization length scale is denoted here by $\sigma$ and should be distinguished from the regularization length scale induced by thermal fluctuations (later denoted by $a$).}. Indeed, such an approach is routinely followed in the construction of effective field theories (see, e.g., Ref.\ \cite{gopalakrishnan2006self} and references therein) \rev{and can be rationalized based on the renormalization group} \cite{mueller_renormalization_2004, kolomeisky_universality_1992}. 
For a sufficiently short-ranged $u(h/\sigma)$, one expects that the effect of the binding potential can be fully captured by a \rev{``contact potential'', i.e.,} boundary condition of the form of \cref{eq:prop_Robin} \cite{burkhardt_propagator_1989, WoodParry01}. 
\rev{While the identification of the boundary condition parameter $\theta$} is beyond the scope of the present study, we note that, for the analytically solvable and paradigmatic problem of a square well potential (see Appendix \ref{app:squarewell}), bound states are present if the range $\sigma$ is larger than the thermal lengthscale $\l$ \footnote{This is true up to a rescaling by a dimensionless factor.}.  Phenomenologically, therefore, the case $\theta<0$ corresponds to $\sigma < \l$, resulting in a pure continuum spectrum, whereas $\theta>0$ corresponds to the situation $\sigma > \l$. For the remaining part of this study, we therefore simply assume $\theta$ to be given and take advantage of the fact that the contact potential model admits an analytical solution of the fluctuating contact line problem. The physically acceptable range of $\theta$ \rev{and its relation to $\Young$} will be delineated in the course of the study.

Before turning to the fluctuating contact line problem in \cref{sec:hxR}, which contains our main results, we discuss the pinned contact line problem [\cref{sec:fixed}] . This enables us to clearly exhibit the new features that arise when the contact line is allowed to fluctuate and, furthermore, to connect to previous literature on this problem.
\begin{figure}[t!]\centering
\subfigure[]{\includegraphics[width=0.49\linewidth]{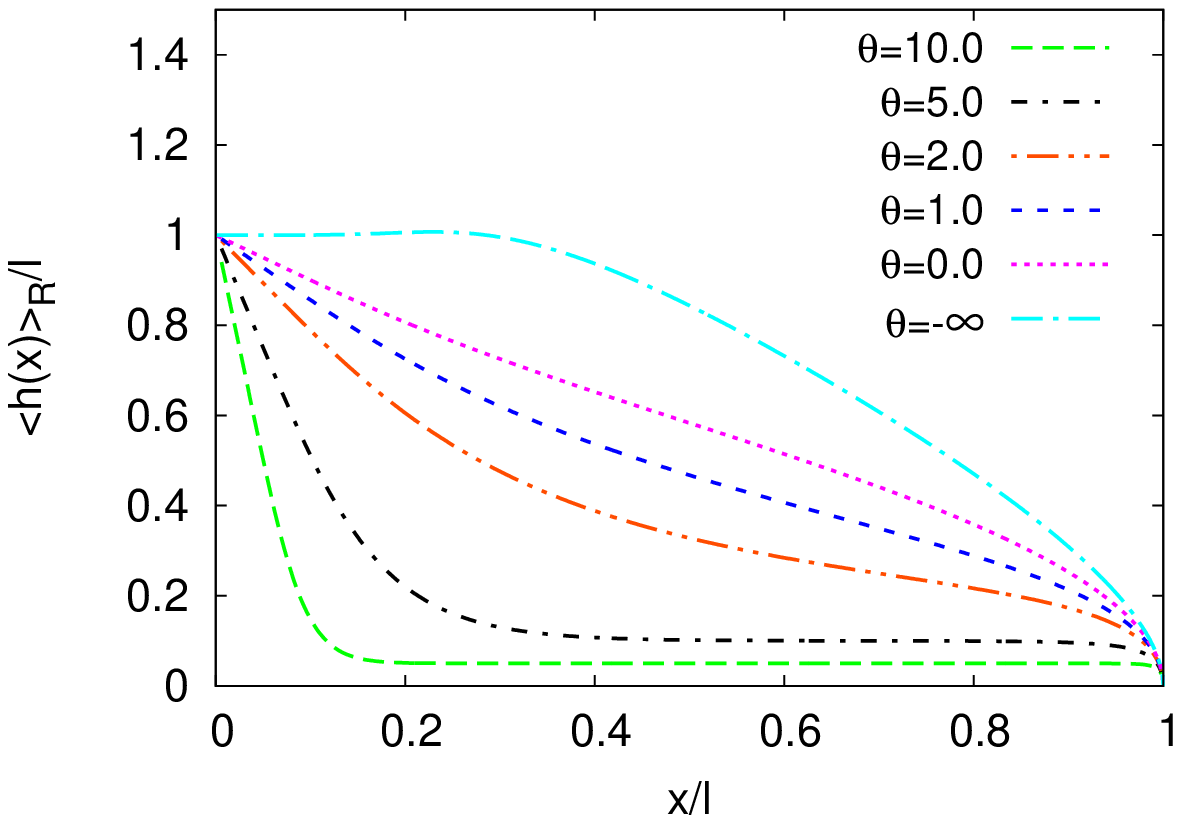}}
\subfigure[]{\includegraphics[width=0.49\linewidth]{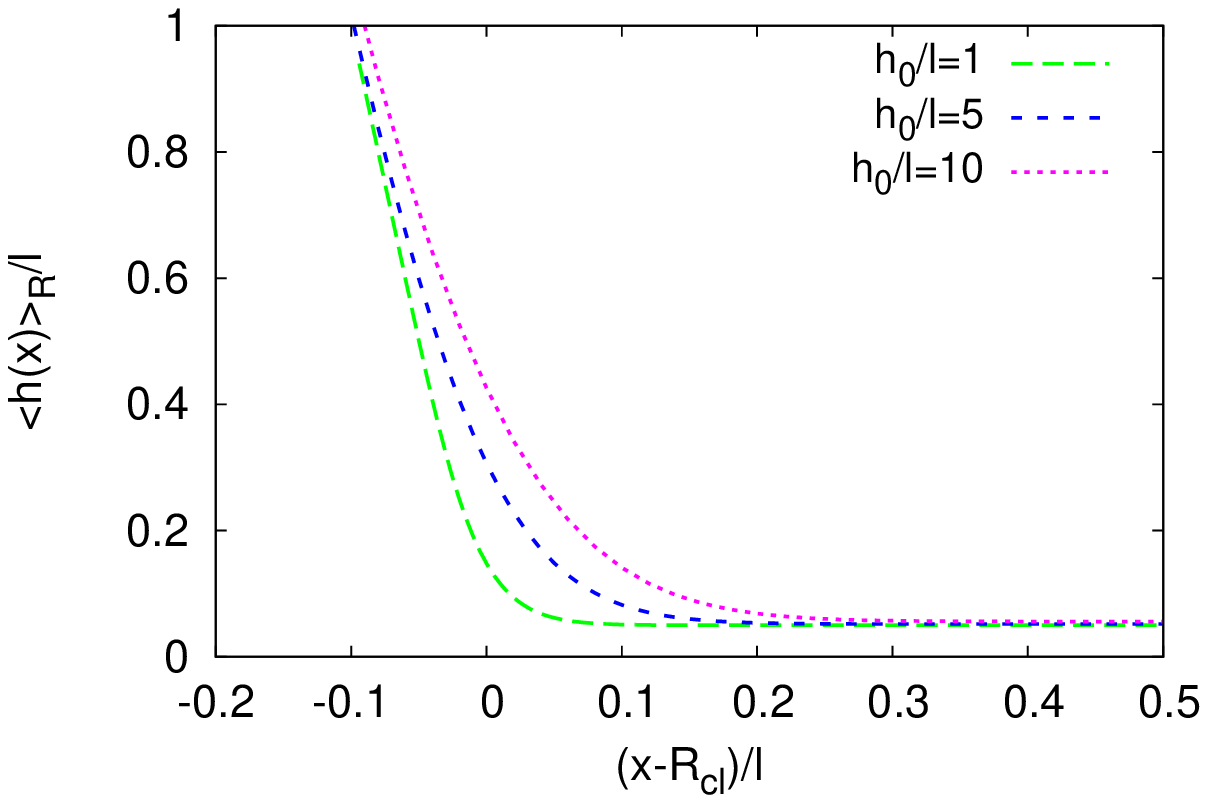}}
	\caption{Average profiles for a fixed contact line position $\R$, obtained by integrating \cref{eq:profileR}, using the propagator \cref{eq:propagator} and the boundary condition in \cref{eq:prop_Robin}. Both the height and the coordinate along the substrate are made dimensionless with respect to the thermal length scale $\l$ [\cref{eq:ell}]. (a) Typical average profiles, for fixed $R/\l=h_0/\l=1$ and various values of the boundary condition parameter $\theta$. Recall that $\theta=0$ corresponds to totally reflecting and $\theta=-\infty$ to totally absorbing boundary conditions. For large positive $\theta$, the interface develops a thin prewetting film with a thickness that increases as $\theta$ decreases. For $\theta\lesssim 0$, the film disappears, and the profile assumes a non-monotonous shape with a negative outer angle [see \cref{eq:dh_R/dx}] in the limit of $\theta=-\infty$. (b) Average profiles for fixed $\theta=10$ and various values of $h_0/\l$ and $\R/\l$, keeping $\l$ fixed and $h_0/R=1$. To highlight the interface shape close to the contact line region, we have subtracted the extrapolated position ($\R_{\st{cl}}=h_0/\theta$) of the contact line from the horizontal coordinate. \label{fig:fixedR}}
\end{figure}

\section{The pinned contact line problem}\label{sec:fixed}

In the previous section, we have determined in \cref{eq:propagator} the general expression for the propagator, which can be used to compute the profile $\Angle{h(x)}_\R$ by numerical integration of \cref{eq:profileR}, keeping $R$ fixed. This condition corresponds to a contact line anchored on a strong defect. Such a situation is essentially of pedagogical interest here and serves to highlight the differences that emerge when the contact line is free to move along the solid. In fact, previous studies of the model defined by \cref{eq:diff,eq:prop_Robin} have typically considered $\R\to\infty$ (or, alternatively, employed periodic boundary conditions), thereby disregarding the behavior near the contact line \cite{ParryWoodRascon00,WoodParry01,romero-enrique_interfacial_2004, jakubczyk_point_2006}.

The characteristic behavior of the average profile $\Angle{h(x)}_\R$ is illustrated in \cref{fig:fixedR}(a). We here use the thermal length $\l$ as a common length to rescale both coordinates. The profile then depends on three parameters: $h_0/\l$, $R/\l$ and $\theta$. One observes that $\theta$ controls the outer angle of the profile [i.e., its average slope at $x=0$, see \cref{eq:dh_R/dx} below], which is therefore \emph{not} Young's angle. We emphasize again that the selection of Young's angle results from the variations with respect to the contact line position $R$, which will be treated in \cref{sec:hxR}. 
Also the structure of the interface close to the contact line is strongly influenced by $\theta$. In particular, for $\theta>0$, the interface is bound to the substrate and develops a thin prewetting film. 
The nonzero thickness of the film is a manifestation of the fact that fluctuation modes can not cross the impenetrable wall, leading to an ``entropic repulsion'' of the interface from the substrate \cite{fisher_walks_1984} (see also \cref{sec:hxR}). The film thickness grows as $\theta$ decreases until for $\theta\simeq 0$ the outer wedge-like part of the profile and the prewetting film can not be clearly distinguished anymore. As $\theta$ becomes negative and the repulsive nature of the substrate is accordingly enhanced, the interface adopts a striking non-monotonous shape associated with a negative outer angle. We remark that, for $h_0\to 0$, we recover the one-dimensional ``droplet'' shapes studied in \cite{burkhardt_propagator_1989}. 

In \cref{fig:fixedR}(b), $\theta$ is fixed to a large positive value and the geometry is progressively inflated by changing $h_0$ and $R$ by the same proportions at constant $\l$. One observes that the asymptotic thickness of the prewetting film is independent of the endpoints of the profile. The thickness is determined by the decay length $\l/\theta$ of the bound state of the analogous quantum mechanical problem (see previous section). On a qualitative level, this bound state reflects the fact that the interface is bound to the wall for $\theta>0$. Due to the fact that the profile cannot intersect the wall, most trajectories actually stay close to it, within a distance governed by the length  $\l/\theta$. Indeed, in the case $\R\to\infty$, the asymptotic (large $x$) thickness of the prewetting film can be exactly computed to be $\Angle{h(x)}_\infty\sim \l/2\theta$ [see \cref{eq:filmRfixed_rescaled}]. This effect is in strict analogy with quantum mechanics, were the probability density function (the square of the wave function) is peaked close to the wall $h=0$ when a bound state exists. Furthermore, the curvature in the region of cross-over from the outer wedge to the prewetting film is dependent on the scale $h_0$, as the interface is more curved at increasing separation of scales.

A heuristic way to understand the effects of thermal fluctuations is based on capillary waves (see, e.g., Ref.\ \cite{forgacs_review_1991}). This reasoning will also provide us with natural rescaling factors for the dimensional variables of the problem and thereby enable us to identify a universal profile shape. Consider the case $\theta>0$, in the limit where the pinning point is sent to infinity ($R \to \infty$). As shown below, in such a case the parameter $\theta$ itself plays the role of the outer angle. In a first approximation, the interface profile is a wedge of width $\lambda_\st{M}=h_0/\theta$, which determines the largest possible wavelength of a standing capillary wave. The smallest wavelength $\lambda_\st{m}$ is set by a microscopic scale, which is typically the particle size of the fluid. For molecular and colloidal fluids it turns out that, in fact, the thermal length scale $\l$ can be used as a reasonable approximation to the particle size, hence $\lambda_\st{m}$ \cite{lekkerkerker_life_2008}. A characteristic scaling length for the height fluctuations of the profile is given by the square-root of the roughness of an equilibrated bulk interface of size $\lambda_\st{M}$ \cite{safran_statistical_1994, flekkoy_fluctuating_1996}, 
\beq
\delta h = \sqrt{\frac{\l h_0}{\theta}},
\eeq 
where we neglected any numerical prefactor. Note that, since capillary waves are not damped in our case, the characteristic interfacial width $\delta h$ grows with $\sqrt{h_0}$. For a one-dimensional interface, the average height does not depend on the microscopic cutoff length $\lambda_\st{m}$. This is different for a two-dimensional interface, where $(\delta h)_{\text{2D}} \simeq \l \sqrt{\ln (h_0/\theta \lambda_\st{m}) }$ \cite{safran_statistical_1994,flekkoy_fluctuating_1996}, which is only weakly dependent on the outer length-scale $h_0$. The characteristic scale $\delta h$ can be used to define a dimensionless height  
\beq 
\bar{h}\equiv \frac{h}{\delta h} = h \sqrt{\frac{\theta}{\l h_0}}.
\eeq 
Once this vertical rescaling is set, from the spatial diffusion equation \eqref{eq:diff} one deduces the characteristic horizontal lengthscale 
\beq
\delta x= \frac{(\delta h)^2}{\l} = \frac{h_0}{\theta},
\eeq 
which is simply the coordinate of the contact line defined by extrapolating the wedge down towards the substrate [cf.\ \cref{eq:classic}]. 
A rescaled coordinate can therefore be defined as: 
\beq
\bar{x} \equiv \frac{x}{\delta x}=x\frac{ \theta}{h_0}.
\eeq 
Finally, from the boundary condition in \cref{eq:prop_Robin}, one infers the  dimensionless {\it scale separation} parameter
\beq\label{eq:singleentryparam}
\Tcal \equiv \frac{\theta h_0}{\l} \simeq \frac{ 2\gamma (1-\cos \theta) \, \delta x }{\kB T}.
\eeq
This parameter compares the macroscopic scale $h_0$ to a microscopic thickness $\l/\theta$. The angle $\theta$ stands here for the contact angle at scale $h_0$ and will therefore be replaced in the fluctuating contact line problem [\cref{sec:hxR}] by Young's angle $\Young$. Upon identifying $\theta$ with $\Young$, $\gamma(\cos\theta-1)$ would become a spreading coefficient. In that case, the parameter $\Tcal$ then represents the ratio of the surface energy of the wedge and the thermal energy. For large $\Tcal$, surface tension effects at the outer scale dominates over thermal fluctuations. In other words, the separation of scales is so large that the thermal energy cannot trigger events with an energy cost $\gamma (1-\cos \theta) \, \delta x$. Hence, in this case, the ``distortion'' of the interface near the contact line region due to thermal fluctuations is expected to be relatively weak compared to the outer region where the profile is a straight wedge. The limit $\Tcal\to \infty$ can therefore be identified with the classical limit.  In contrast, at moderate $\Tcal$, temperature can provide sufficient energy to induce noticeable distortions of the interface not only close to the contact line region, but also further in the bulk. Indeed, in the extreme case of $\Tcal\to 0$, the profile is influenced at all scales (cf.\ Ref.\ \cite{burkhardt_propagator_1989}).

\begin{figure}[t]
\subfigure[]{\includegraphics[width=0.49\linewidth]{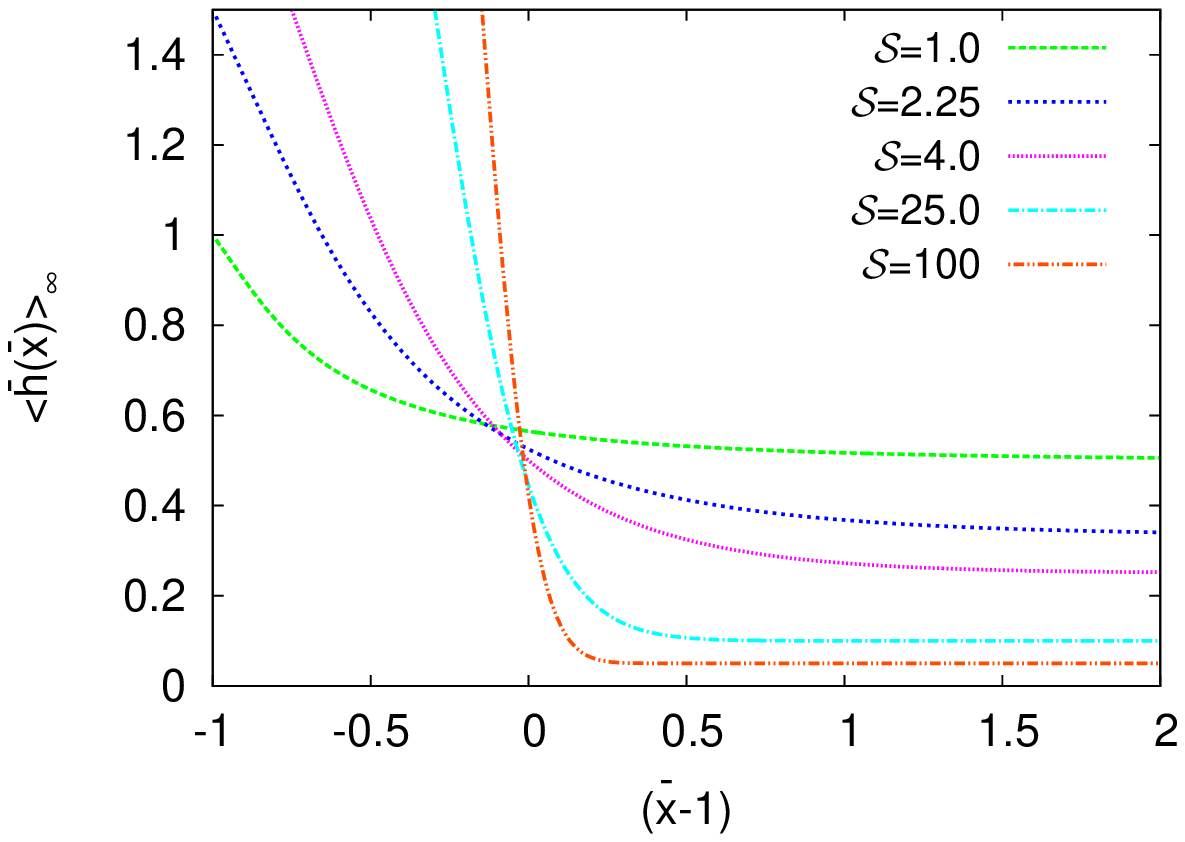}}
\subfigure[]{\includegraphics[width=0.49\linewidth]{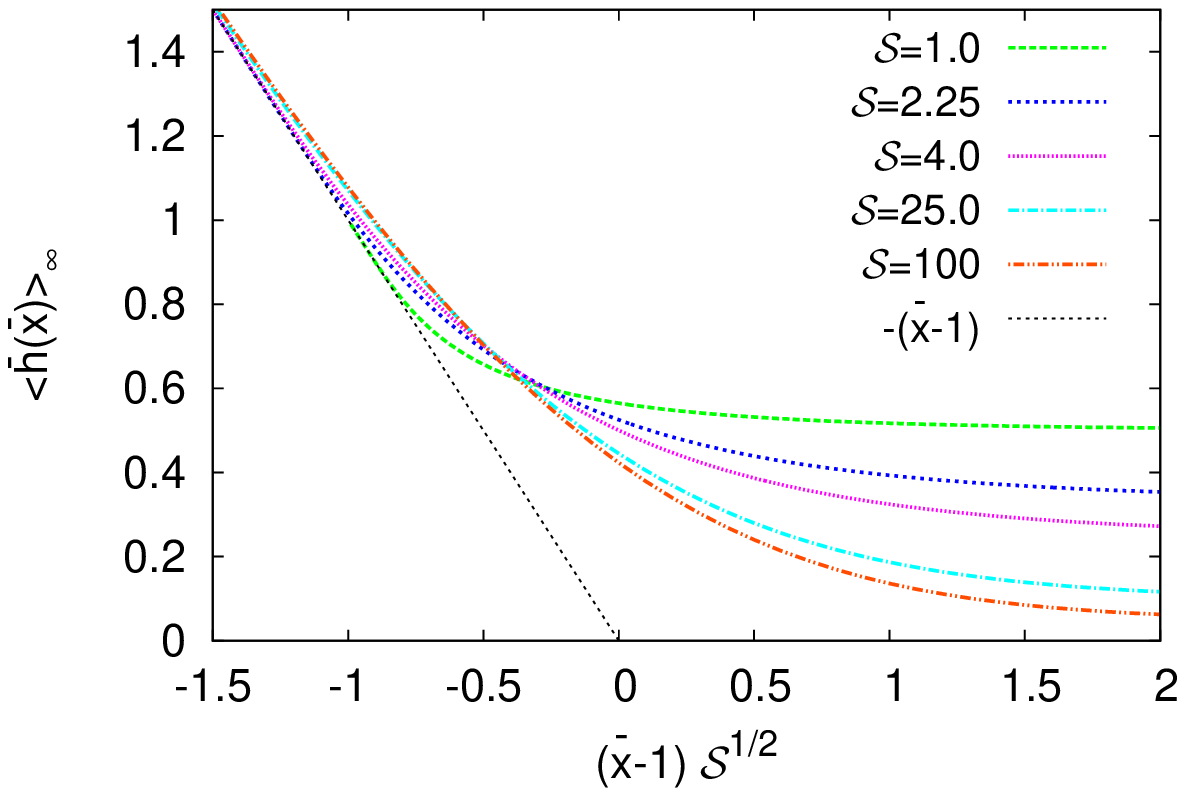}}
\caption{Average profiles for a fixed contact line position $\R$ in the limit $R \to \infty$, obtained by integrating \cref{eq:profileR}, using the propagator in \cref{eq:propagator} and the boundary condition in \cref{eq:prop_Robin}. In order to obtain a well defined outer angle, we keep $\theta>0$ (see also \cref{fig:fixedR}). The height and the horizontal lengthscale are made dimensionless as $\bar{h}=h/\delta h$ and $\bar{x}=x/\delta x$, with $\delta h=\sqrt{h_0\l/\theta}$ and $\delta x=h_0/\theta$ (see text for details). Additionally, the horizontal coordinate is shifted by $h_0/\theta \delta x = 1$ in order to translate the expected classical contact line position to the origin [cf.\ \cref{eq:angleRfixed_rescaled}]. The so obtained universal profiles are solely controlled by the dimensionless parameter $\Tcal$ in \cref{eq:singleentryparam}: if $\Tcal$ is large, the interface is significantly curved only near the contact line (classical limit). At moderate $\Tcal$, temperature can provide sufficient energy to induce a noticeable  ``repulsion'' of capillary waves from the impenetrable substrate. In all cases a thin prewetting film is formed, whose thickness approaches a finite value as $\bar x\to\infty$, depending on $\Tcal$  [see \cref{eq:filmRfixed_rescaled}]. In panel (b) the same data as in (a) is shown, but the horizontal coordinates is further rescaled by $\sTcal$, in order to highlight that $\sTcal$ represents the outer angle in dimensionless variables [see \cref{eq_outerangle_pinned}].} \label{fig:fixedR_rescaled}
\end{figure}

In the pinned contact line ensemble, the interface morphology is governed by the partition function in \cref{eq:fixedRpart}, which, with the aid of \cref{eq:propagator}, follows as:
\beq\label{eq:fixedRpart_expl}
c(h_0,0;\R) = \frac{2e^{-h_0^2/2\l\R}}{\sqrt{2\pi\l\R}} + \frac{\theta}{\l} e^{-\theta h_0/\l}e^{\theta^2\R/2\l}\erfc\Round{\frac{h_0-\theta\R}{\sqrt{2\l\R}}}.
\eeq
Analytically, the outer angle can be computed in terms of the averaged slope of the profile (see \cref{app:fixedR}): 
\beq\label{eq:dh_R/dx}
\Angle{h'(0)}_\R \equiv \left.\frac{d\Angle{h(x)}_\R}{dx}\right|_{x=0} = \l\frac{\pt\ln c(h_0,0;\R)}{\pt h_0} = -\theta - \frac{h_0}{\R} \frac{c^\st{ref}(h_0,0;\R)}{c(h_0,0;\R)},
\eeq
with $c^\st{ref}(h_0,0;\R)\equiv 2e^{-h_0^2/2\l\R}/\sqrt{2\pi\l\R}$ being the propagator for reflecting boundary conditions. Note that the first equality in \cref{eq:dh_R/dx} in fact serves to define the averaged slope. Such a definition is necessary, because, for a given stochastic realization of the profile $h(x)$, the quantity $h'(x)$ is a derivative of a Wiener path and therefore not strictly well-defined \cite{chaichian_book}. Based on \cref{eq:dh_R/dx} and the asymptotic expression in \cref{eq:erfc}, one infers that, in the limit $R \to \infty$ and for $\theta >0$, the outer angle is given by $\theta$, i.e.,
\beq \Angle{h'(0)}_\infty=-\theta.
\label{eq_fixedR_angle}\eeq 
In rescaled variables, this corresponds to
\beq \Angle{\bar h'(0)}_\infty \equiv \frac{d\Angle{\bar h(\bar x)}_\infty}{d\bar x}\Big|_{\bar x=0}=-\sTcalY.
\label{eq_outerangle_pinned}\eeq 
In the limit $\R\to\infty$, analytical calculations are in fact feasible for the entire average profile [\cref{eq:profileR}]. We obtain (see Appendix~\ref{app:cont_prop}): 
\beq\label{eq:hav_theta}
\Angle{h(x)}_\infty = \frac{\l}{2\theta} +\Round{\frac{h_0-\theta x}{2}-\frac{\l}{4\theta}}\erfc\Round{\frac{\theta x-h_0}{\sqrt{2\l x}}}- \frac{\l}{4\theta}e^{2\theta h_0/\l}\erfc\Round{\frac{h_0+\theta x}{\sqrt{2\l x}}}+\sqrt{\frac{\l x}{2\pi}}e^{-(h_0-\theta x)^2/2\l x}
\eeq
which, in rescaled variables, becomes   
\beq\label{eq:hav_theta_rescaled}
\Angle{\bar{h}(\bar{x})}_\infty = \frac{1}{2\sqrt{\Tcal}}-\Round{\sqrt{\Tcal}\frac{\bar{x}-1}{2}+\frac{1}{4\sqrt{\Tcal}}}\erfc\Round{\sqrt{\Tcal}\frac{\bar{x}-1}{\sqrt{2 \bar{x}}}}- \frac{1}{4\sqrt{\Tcal}}e^{2\Tcal}\erfc\Round{\sqrt{\Tcal}\frac{\bar{x}+1}{\sqrt{2 \bar{x}}}}+\sqrt{\frac{\bar{x}}{2\pi}}e^{-\Tcal(\bar{x}-1)^2/2\bar{x}}.
\eeq
Indeed, the scale separation $\Tcal$ is seen to be the only control parameter governing the behavior of the average profile $\Angle{\bar h(\bar x)}_\infty$. The presence of the precursor film can be inferred from the asymptotics at large $\bar{x}$: Employing the asymptotic relation in \cref{eq:erfc}, \cref{eq:hav_theta_rescaled} becomes 
\beq\label{eq:filmRfixed_rescaled}
\Angle{\bar{h}(\bar{x})}_\infty \overset{\bar{x}\to\infty}{\sim} \frac{1}{2\sTcal} + \frac{\Tcal-1}{\Tcal^2}\sqrt{\frac{2}{\pi\bar{x}^3}}e^{\Tcal}e^{-\Tcal\bar{x}/2},
\eeq
revealing the emergence of a film whose dimensionless thickness is $\Angle{\bar{h}(\bar{x})}_\infty \sim 1/2\sqrt{\Tcal}$ in the limit $\bar{x}\to\infty$. This corresponds, in unscaled variables, to $\Angle{h(x)}_\infty\sim \l/2\theta$. In the limit of small $\bar{x}$, we obtain:
\beq\label{eq:angleRfixed_rescaled}
\Angle{\bar{h}(\bar{x})}_\infty \overset{\bar{x}\to0}{\sim} \sTcal(1 - \bar{x}),
\eeq
confirming that the quantity $\sTcal$ plays the role of an outer angle in the rescaled profile [see \cref{eq_outerangle_pinned}]. In unscaled variables, \cref{eq:angleRfixed_rescaled} becomes 
\beq \Angle{h(x)}_\infty \overset{x\to0}{\sim} h_0 -\theta x.
\eeq 
The properties analytically derived above are illustrated in \cref{fig:fixedR_rescaled}, where profiles for $\theta>0$ and $\R\to\infty$, expressed in rescaled variables, are displayed for different values of $\Tcal$. In rescaled variables, the scale separation $\Tcal$ controls the rescaled thickness of the prewetting film [see \cref{eq:filmRfixed_rescaled}], which is independent of any macroscopic length. $\Tcal$ furthermore controls the curvature in the region of cross-over from the outer wedge to the prewetting film. 

\begin{figure}[t]\centering	
\subfigure[]{\vcenteredhbox{\includegraphics[width=0.49\linewidth]{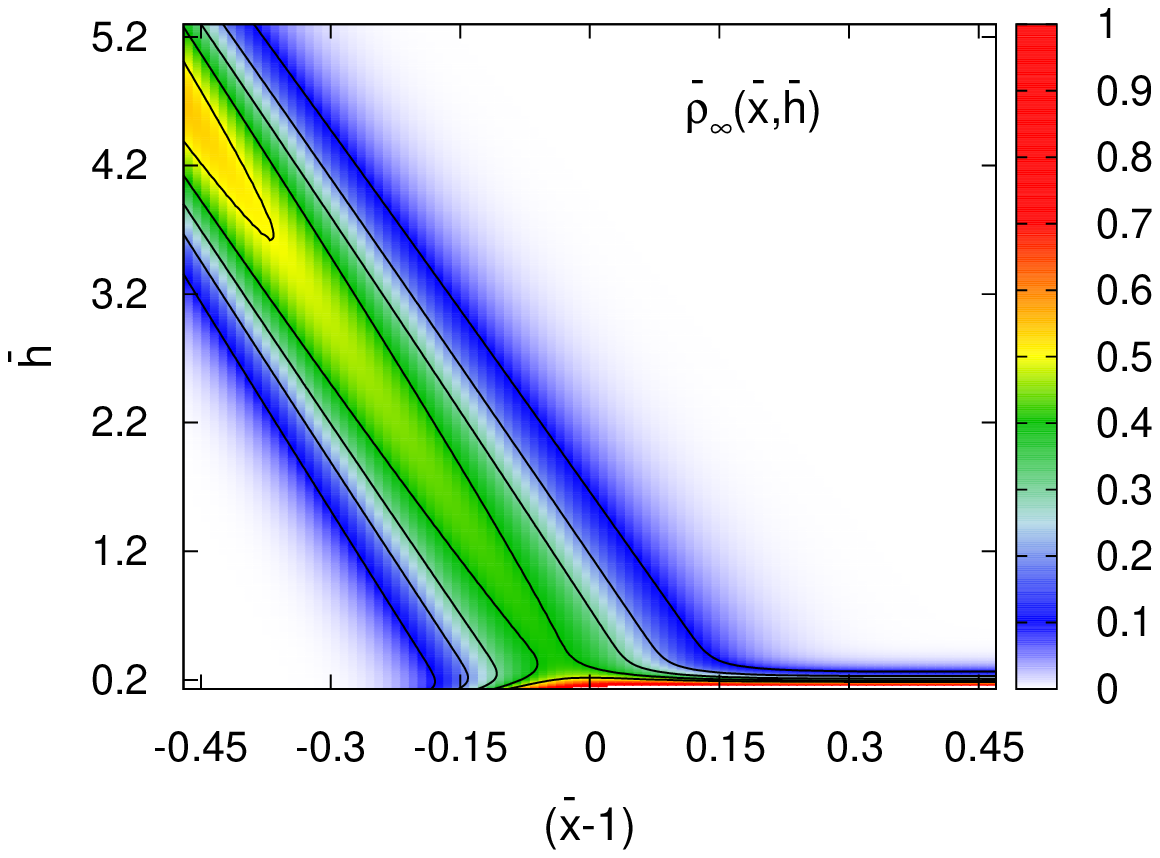}}}
\subfigure[]{\vcenteredhbox{\includegraphics[width=0.49\linewidth]{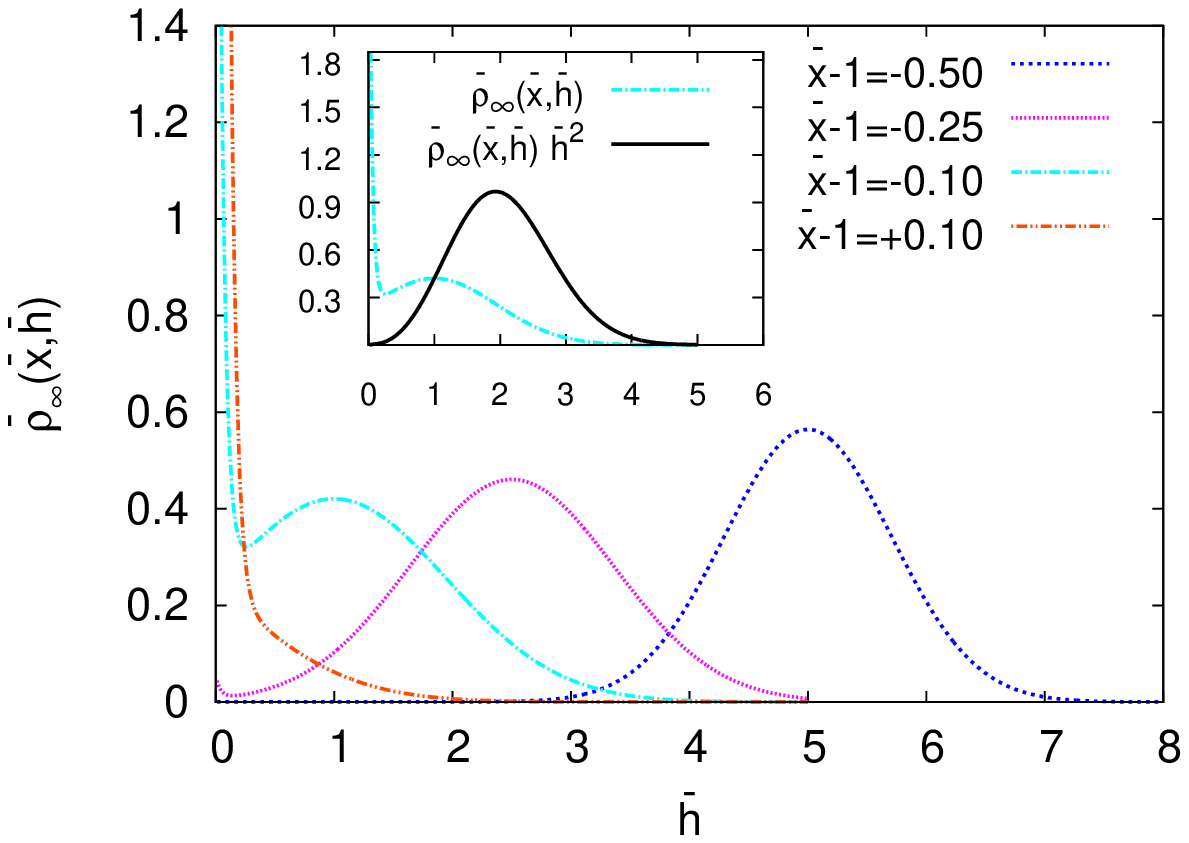}}}
	\caption{Conditional probability distribution function $\bar{\rho}_{\infty}(\bar{x},\bar{h})$ [\cref{eq:Phx}] for $\Tcal=100$. (a) Intensity plot of $\bar{\rho}_{\infty}(\bar{x},\bar{h})$ near the location of the contact line ($\bar x\simeq 1$). Contour levels (black solid lines) are equispaced by $0.1$ between $0.0$ and $0.5$. (b) $\bar{\rho}_{\infty}(\bar{x},\bar{h})$ {\it vs.} $\bar{h}$ for various $\bar{x}$. To highlight the asymmetry of $\bar{\rho}_{\infty}(\bar{x},\bar{h})$ for $\bar{x} \simeq 1$ the inset reports both $\bar{\rho}_{\infty}(\bar{x},\bar{h})$ and $\bar{\rho}_{\infty}(\bar{x},\bar{h}) \bar{h}^2$ {\it vs.} $\bar{h}$ for a fixed $\bar{x}-1=-0.1$.  \label{fig:P}}
\end{figure}

The results for the average profiles are finally complemented with an investigation of the conditional probability distribution function $\rho_{R}(x,h)$, which expresses the probability to find the interface with a height $h$ at the coordinate $x$, normalized such that $\int_{0}^{\infty} \rho_R(x,h)\, d h=1$ (independently of $x$). $\rho_R(x,h)$ can be computed exactly and in the limit $R \to \infty$ we obtain (for $\theta>0$) a well defined limiting probability distribution $\rho_\infty(x,h)$ (see \cref{app:cont_prop}). In rescaled variables, $\bar{\rho}_{\infty}(\bar{x},\bar{h})\equiv \sqrt{\l h_0/\theta}\rho_{\infty}(x,h)$ is given by \footnote{The prefactor $\sqrt{\l h_0/\theta}$ comes from the requirement $\bar{\rho}_\infty(\bar{x},\bar{h})d\bar{h}=\rho_\infty(x,h)dh$.}
\beq\label{eq:Phx}
\bar{\rho}_{\infty}(\bar{x},\bar{h}) = \frac{1+e^{-2\sTcal\bar{h}/\bar{x}}}{\sqrt{2\pi\bar{x}}}e^{-[\bar{h}-\sTcal(1-\bar{x})]^2/2\bar{x}} + \sTcal e^{-2\sTcal \bar{h}}\erfc\Round{\frac{\bar{h}+\sTcal (1-\bar{x})}{\sqrt{2\bar{x}}}}.
\eeq
The characteristic properties of $\bar\rho_\infty$ are illustrated in \cref{fig:P}. \cref{fig:P}(a) reports a plot of $\bar{\rho}_{\infty}(\bar{x},\bar{h})$, with both $\bar{x}$ and $\bar{h}$ that lie close to the contact line region. We can clearly identify two distinct behaviors for $\bar{x} \ll 1$ and for $\bar{x} \gg 1$. The transition between the two regions occurs around the classical contact line location $\bar{x} \simeq 1$. In this region, due to the entropic repulsion of the fluctuations, the probability distribution is asymmetric [see inset in \cref{fig:P}(b)]. Due to the presence of the $\erfc$ function, it is difficult to expose these facts analytically in \cref{eq:Phx}. However, based on \cref{eq:erfc}, in the limit of large $\Tcal$ one finds the tractable expression
\beq\label{eq:prob_smallT}
\bar{\rho}_{\infty}(\bar{x},\bar{h}) \overset{\Tcal \to \infty}{\sim} \Square{1+\frac{\bar{h}+\sTcal (1+\bar{x})}{\bar{h}+\sTcal (1-\bar{x})}e^{-2 \sTcal \bar{h}/\bar{x}}}\frac{e^{-[\bar{h}-\sTcal (1-\bar{x})]^2/2\bar{x}}}{\sqrt{2\pi\bar{x}}} + \frac{1-\sgn(\bar{h}+\sTcal(1-\bar{x}))}{2}2\sTcal e^{-2 \sTcal \bar{h}},
\eeq
where $\sgn(z)$ is the sign of $z$. We infer from \cref{eq:prob_smallT} that if $\bar{x}\ll1$ then
$$
\bar{\rho}_{\infty}(\bar{x},\bar{h}) \overset{\Tcal \to \infty}{\sim} \frac{1}{\sqrt{2\pi\bar{x}}}e^{-[\bar{h}-\sTcal (1-\bar{x})]^2/2\bar{x}},
$$
and $\bar\rho_\infty$ is basically the conditional probability of a free interface fluctuating around the straight wedge profile in \cref{eq:angleRfixed_rescaled}. For $\bar{x} \gg 1$, instead, the probability becomes 
$$
\bar{\rho}_{\infty}(\bar{x},\bar{h}) \overset{\Tcal \to \infty}{\sim} 2\sTcal e^{-2 \sTcal \bar{h}},
$$
that is a distribution function completely unrelated to the outer problem \cite{romero-enrique_interfacial_2004}. Consequently, for $\bar{x}\gg 1$ the profile is localized near the wall, as a manifestation of the presence of the precursor film.  

All the above results focus on the case $\theta>0$, for which in the limit $R \to \infty$ a well defined wedge structure emerges. 
In contrast, if $\theta \le 0$, no wedge structure survives in the limit $R \to \infty$, because the interface is not anymore bound to the surface (see also \cref{fig:fixedR}). Hence the characteristic outer scale $\lambda_\st{M}$ used for the rescaling grows with $R$. This yields a rescaling that differs from the one adopted with $\theta<0$. These properties, together with some relevant analytical examples useful to connect to previous literature \cite{burkhardt_propagator_1989}, are discussed in \cref{app:cont_prop}.

In conclusion, for the case of a pinned contact line, a prewetting film emerges, whose thickness is controlled by the balance between thermal fluctuations and surface energy. Due to the presence of undamped capillary waves, the overall shape of the interface depends on the macroscopic cut-off length $h_0$. We shall now investigate how these results are modified in the physical case of a contact line let free to move.

\section{The fluctuating contact line problem}\label{sec:hxR}

In the previous section we have examined the situation where the contact line is pinned, in which case the outer angle is constrained by the \rev{boundary condition}, and not selected by surface tensions. Here, we discuss the structure of an unpinned interface close to the contact line on a homogeneous substrate. Accordingly, we assume each realization of the fluctuating profile to have a well defined contact point $\R$, but perform a weighted average over all possible values of $\R$ [see \cref{eq:Z} as well as \cref{fig_wedge_sketch_fluct}]. We consider the ensemble in which both $h(x)$ and $\R$ can fluctuate as the proper one for an interface that partially wets a homogeneous substrate. We shall see that the assumption of a fluctuating $\R$ gives rise to a fundamentally different phenomenology from the pinned contact line problem considered in the previous section: in particular, it turns out that Young's contact angle is automatically selected at the outer scale ($h_0$) -- a property that is directly associated with the fact that the contact line can move freely. 
We study in the following both the morphology of the average profiles and the statistics of $\R$, and their dependence on the \rev{boundary condition parameter $\theta$}. In particular, We show that $\theta$ controls a transition from partial wetting to pseudo-partial wetting, the latter being a state in which the outer contact angle is finite but where the solid gets covered by a microscopic flat liquid film \cite{BrochardWyart_spreading_1991, deGennes_book}. The order parameter of the transition is the average position of the contact line, $\Angle{R}$, which diverges when the inner parameter $\theta$ tends to Young's angle $\Young$. Finally, we investigate the regularization induced by thermal fluctuations: this effect can be captured in terms of an effective disjoining pressure contribution, which turns out to decay exponentially (instead of algebraically, as for van der Waals interactions \cite{kroll_universality_1983}, for instance).

As a preparatory step, the relevant dimensionless parameters in the fluctuating contact line problem are constructed. The Young's angle $\Young$ is now expected to be the outer angle at the scale $h_0$. Thus, following the ideas put forward in the previous section, we define the dimensionless height as
\beq \label{eq_hbar_Y}
\bar{h}\equiv h \sqrt{\frac{\Young}{\l h_0}}.
\eeq 
Once this vertical rescaling is set, invariance of the spatial diffusion equation \eqref{eq:diff} fixes the rescaled horizontal coordinate to 
\beq \label{eq_xbar_Y} \bar{x}\equiv x\frac{\Young}{h_0},
\eeq 
implying that the classical contact line position is $\bar{x}=\bar{R}_\st{cl}=1$. Analogously to \cref{eq:singleentryparam}, the scale separation parameter is again defined in terms of the contact angle, which is now Young's angle $\Young$:
\beq\label{eq:singleentryparam_Y}
\TcalY\equiv \frac{\Young h_0}{\l} \simeq \frac{2 \gamma (1-\cos \Young) \, \delta x }{\kB T}.
\eeq
Due to the presence of the boundary condition parameter $\theta$ [see \cref{eq:prop_Robin}], a further dimensionless number in addition to $\TcalY$ plays an important role, namely the ratio  $\theta/\Young$. In summary, for the fluctuating contact line problem, we expect the invariant rescaling to be characterized by two dimensionless numbers, the scale separation $\TcalY$, expressing the importance of the surface energy of the Young wedge with respect to the thermal energy, and $\theta/\Young$, which controls the microscopic deviation from Young's angle. This will indeed be confirmed by analytical expressions for the rescaled profiles [see, e.g., \cref{eq:hav_theta<alpha_rescaled} below].


\begin{figure}[t]\centering
        \subfigure[]{\includegraphics[width=0.49\linewidth]{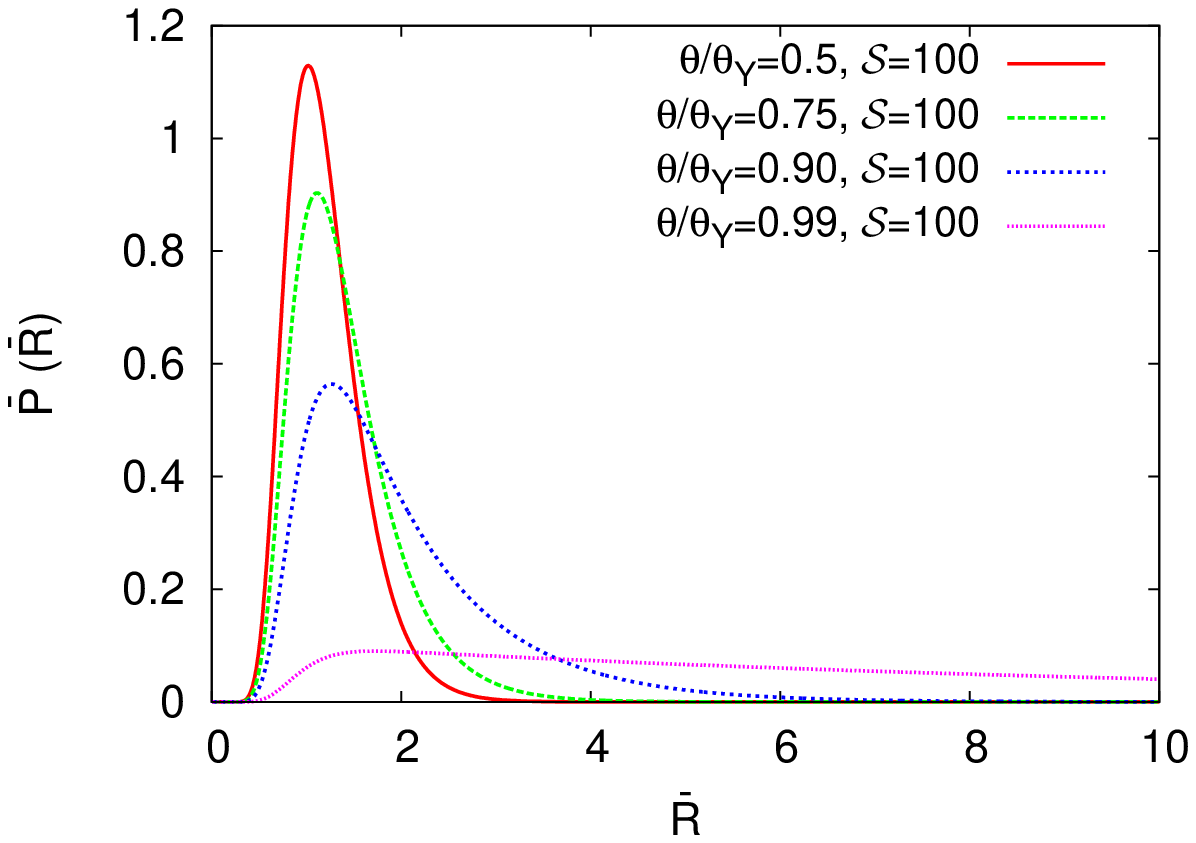}}
	\subfigure[]{\includegraphics[width=0.49\linewidth]{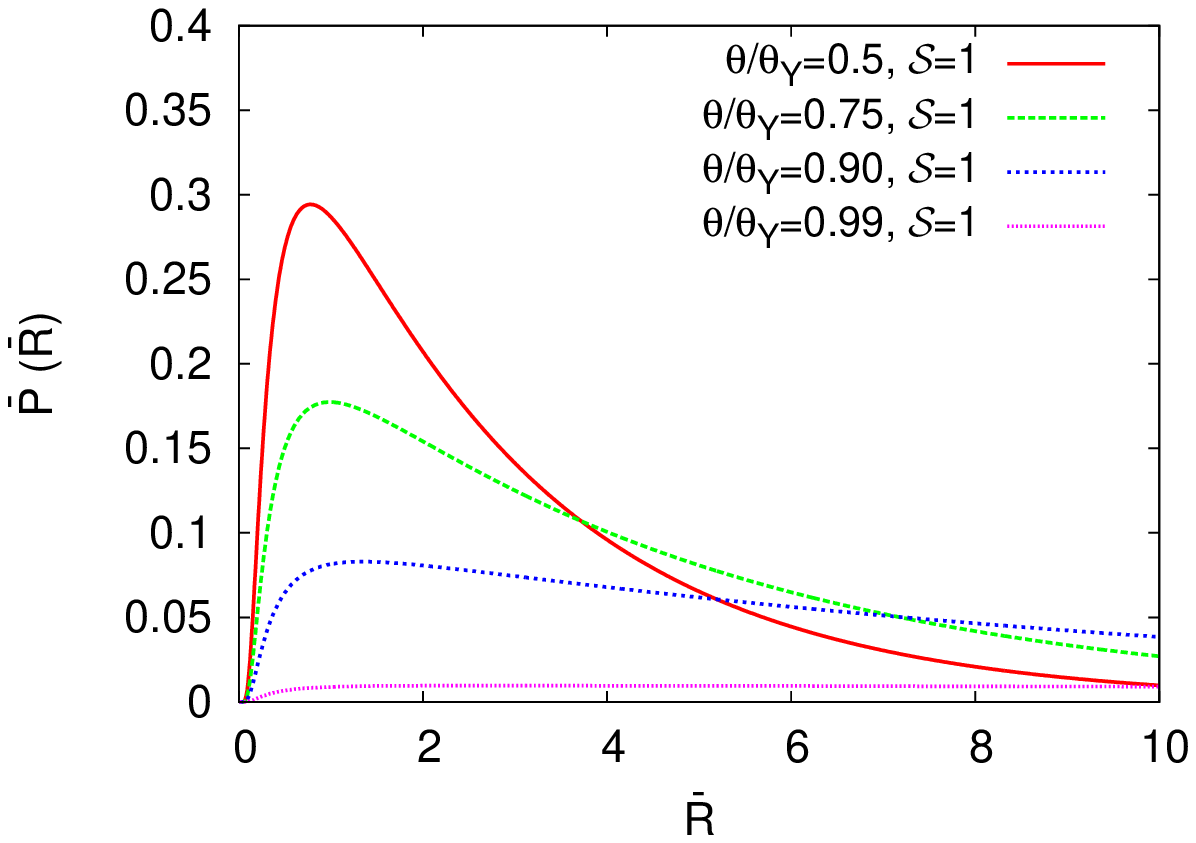}}
	\caption{Probability distribution of the (rescaled) contact line position $\bar\R$, \cref{eq:P(R)}, for various values of $\theta/\Young$ and scale separation $\TcalY=100$ (panel (a)) and $\TcalY=1$ (panel (b)) [\cref{eq:singleentryparam_Y}]. $\bar P(\bar\R)$ develops a fat tail as $\theta\nearrow\Young$, leading to a divergence of the mean contact line position $\Angle{\bar R}$. \label{fig:Rfluct}}
\end{figure}


Before embarking on the analysis of the interface morphology, we first characterize the statistical properties of the fluctuating contact line ensemble. For $\theta/\Young < 1$, the partition function in \cref{eq:Z} is finite and simplifies to
\beq\label{eq:partitionfunctionTEXT}
Z=\frac{2e^{-\Young h_0/\l}}{\Young-\theta}.
\eeq
For $\theta/\Young \ge 1$, instead, the partition function diverges (see \cref{app:part_hav}). 
\rev{The total free energy (apart from a constant and per $\kB T$) follows from \cref{eq:partitionfunctionTEXT} as $
-\ln Z = (\Young^2/\l)(h_0/\Young) + \ln(\Young-\theta)$ and decomposes into a ``surface'' (line) and an excess contribution. The former is the free energy of a straight interface of Young angle $\Young$ (in the gentle-slope approximation), while the latter can be identified with the \emph{point tension} $\tau$ \cite{schimmele_linetension_2007}. Since in our model the wetting transition occurs at $\Young=0$, this implies a logarithmic divergence, $\tau\sim \ln\Young$, of the point tension at the wetting transition provided the wall is reflective ($\theta=0$). A similar logarithmic divergence of the point tension has been found for an interface with a fluctuating contact point in the two-dimensional Ising model \cite{abraham_divergence_1993} \footnote{We recall that in a pure ``contact potential'' model, the boundary condition parameter $\theta\propto T_\st{w}-T$, implying that in this case the wall is indeed reflective at the wetting transition \cite{burkhardt_propagator_1989, upton_interfacemodel_1999, upton_interfacemodel_2002}.}.} To further characterize the transition at $\theta=\Young$, we investigate the statistics of $R$ and its dependence on $\TcalY$ and $\theta/\Young$. The probability distribution function of $R$ for $\theta<\Young$ is defined by the statistical weights characterizing the partition function in \cref{eq:Z}:
\beq\label{eq:P(R)_unscaled}
P(R)=\frac{c(h_0,0;\R) \,e^{-\Young^2 R/2 \l}}{Z}
\eeq
where $c(h_0,0;\R)$ is given by \cref{eq:fixedRpart_expl} and $Z$ by \cref{eq:partitionfunctionTEXT}. 
When expressed in dimensionless variables [\cref{eq_hbar_Y,eq_xbar_Y}], the rescaled probability distribution $\bar{P}(\bar{\R})\equiv P(R)h_0/\Young$ depends only on the parameters $\TcalY$ and $\theta/\Young$:
\beq\label{eq:P(R)}
\bar{P}(\bar{\R})=\frac{1-\theta/\Young}{2} \Square{\sqrt{\frac{2\TcalY}{\pi\bar{\R}}}e^{\TcalY[1-(\bar{\R}+1/\bar{\R})/2]} + \TcalY(\theta/\Young) e^{\TcalY(1-\theta/\Young)[1-(1+\theta/\Young)\bar{\R}/2]}\erfc\Round{\sTcalY \frac{1-(\theta/\Young)\bar{\R}}{\sqrt{2\bar{\R}}}}}.
\eeq
In \cref{fig:Rfluct}, $\bar{P}(\bar{\R})$ is plotted for different values of the scale separation $\TcalY$ and the boundary condition parameter $\theta/\Young$. 
Comparing both panels, we observe that, at fixed $\theta/\Young$, $\TcalY$ mainly governs the fluctuations around the classical contact line position $\bar{R}=1$; in particular, the variance of the distribution increases upon reducing $\TcalY$. For $\TcalY \to \infty$ the classical limit is recovered and thermal fluctuations are insufficient to induce significant displacements of the contact line. The parameter $\theta/\Young$, on the other hand, controls the tails of the distribution for large $\bar{R}$, which become fatter with increasing $\theta/\Young$. 
These insights are confirmed from an asymptotic expansion of $\bar P(\bar{R})$ for large $\bar \R$:
\beq\label{eq:P(R)asympt}
\bar{P}(\bar{R}) \overset{\bar{\R}\to\infty}{\sim} \begin{dcases}
	\frac{(1-\theta/\Young)[1-\TcalY (\theta/\Young)]}{\TcalY (\theta/\Young)^2}\sqrt{\frac{\TcalY}{2\pi\bar{\R}^3}}e^{\TcalY}e^{-\TcalY \bar{\R}/2} &\textup{if } \theta/\Young<0, \\
	\Round{1-\frac{\TcalY}{2\bar{\R}}}\sqrt{\frac{\TcalY}{2\pi\bar{\R}}}e^{\TcalY}e^{-\TcalY \bar{\R}/2} &\textup{if } \theta/\Young=0, \\
	(1-\theta/\Young) \TcalY (\theta/\Young) e^{\TcalY (1-\theta/\Young)}e^{-\TcalY [1-(\theta/\Young)^2]\bar{\R}/2} &\textup{if } 0< \theta/\Young<1
\end{dcases}
\eeq
where we made use of the expression in \cref{eq:erfc}. For $\theta/\Young<1$, the rescaled average contact point $\Angle{\bar{R}}\equiv\Angle{\R}\Young/h_0$ follows from \cref{eq:Rav_general} as
\beq\label{eq:Rav}
\Angle{\bar{\R}} = 1 + \frac{1}{\TcalY}\frac{1}{1-\theta/\Young},
\eeq
and the variance as \footnote{The variance is computed from $\Angle{\R^2} = \frac{1}{Z}\frac{\l^2}{\Young} \frac{\pt}{\pt\Young}\Round{\frac{1}{\Young}\frac{\pt Z}{\pt\Young}}$.}
\beq\label{eq:fluctR}
\delta\bar{\R}^2 \equiv \Angle{\bar{R}^2}-\Angle{\bar{R}}^2 = \frac{1}{\TcalY}\Square{1 + \frac{2-\theta/\Young}{\TcalY(1-\theta/\Young)^2}}.
\eeq
When $\theta/\Young \nearrow 1$, the tail of $\bar{P}(\bar{R})$ does not decay to zero at infinity, but rather saturates to a constant. The distribution therefore flattens and all contact line locations become equally probable.  Indeed, for $\theta/\Young\nearrow1$ both $\Angle{\bar{\R}}\to\infty$ and $\delta\bar{R} \to \infty$ diverge. When $\theta/\Young \ge 1$, we find an infinite $\Angle{\bar{\R}}$, {\it independently} of the value of the scale separation $\TcalY$. This is in striking analogy with the results discussed for the pinned contact line problem in the limit $R \to \infty$ (see \cref{fig:fixedR_rescaled}), where a flat film always covers the substrate. The emergence of such a film in the fluctuating contact line ensemble is essentially a consequence of the fact that, in presence of a negative binding energy (i.e., for $\theta>0$), most of the trajectories stay close to the wall, hence they can prolong to larger distances from $\R_\st{cl}$.

The (in-)finiteness of $\Angle{\R}$ is strictly connected to the value of the outer angle. To see this, we average \cref{eq:dh_R/dx} over $\R$, to get, with the aid of \cref{eq:Z,eq_avg_profile_fluctR}, 
\beq\label{eq:outerangle}
\Angle{h'(0)} \equiv \left.\frac{d\Angle{h(x)}}{dx}\right|_{x=0} = \l\frac{\pt\ln Z}{\pt h_0} = -\theta - \frac{2e^{-\Young h_0/\l}}{Z}= \begin{cases}
	-\Young &\textup{if } \theta/\Young<1\\
	-\theta \qquad &\textup{if } \theta/\Young\geq 1,
\end{cases} 
\eeq
in dimensional variables.
Accordingly, below the pseudo-partial wetting transition point ($\theta/\Young<1$), where the partition function \cref{eq:partitionfunctionTEXT} is finite, the outer angle (i.e., the averaged slope at the scale $h_0$) in \cref{eq:outerangle} turns out to be exactly equal to Young's angle $\Young$. Upon crossing the pseudo-partial wetting point, the partition function $Z$ diverges and the outer angle is instead given by $\theta$.

\begin{figure}[t!]\centering
         \subfigure[]{\includegraphics[width=0.49\linewidth]{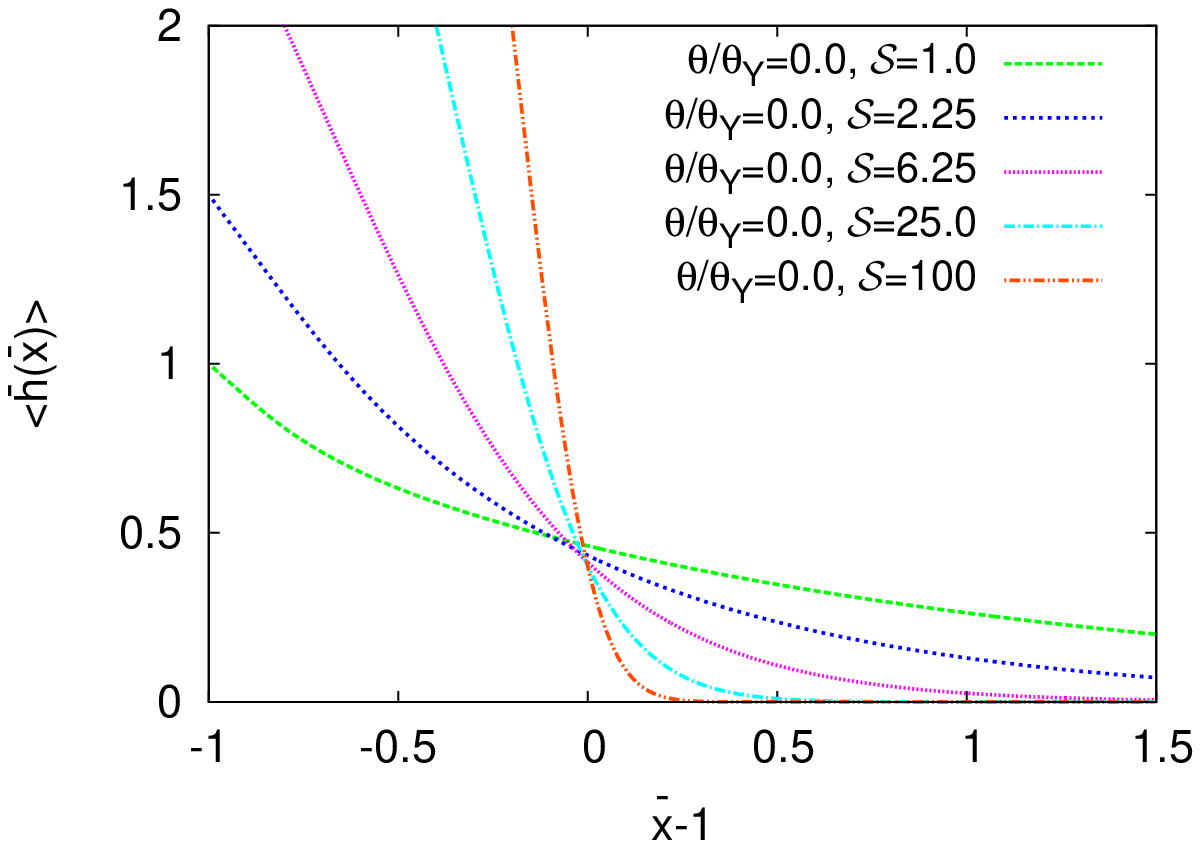}}
         \subfigure[]{\includegraphics[width=0.49\linewidth]{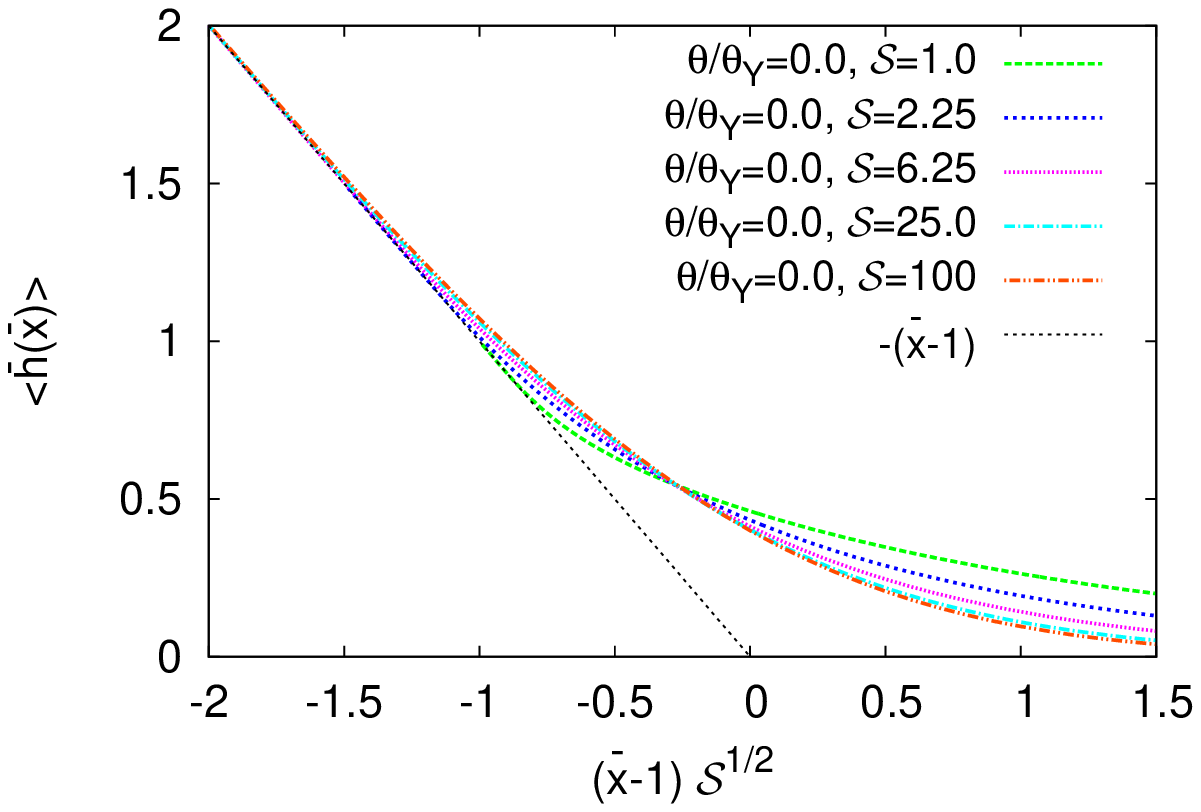}}
	\caption{(a) Average profiles [\cref{eq:hav_theta<alpha}] in the case of a fluctuating contact line, for various $\TcalY$ [\cref{eq:singleentryparam_Y}] and $\theta/\Young=0$ (reflecting wall). The dimensionless height and horizontal coordinate are defined as $\bar{h}=h/\sqrt{h_0\l/\Young}$ and $\bar{x}=x \Young/h_0$ [see \cref{eq_hbar_Y,eq_xbar_Y}]. In (b), the horizontal coordinate is further rescaled by $\sTcalY$ in order to emphasize that $\sTcalY$ plays the role of the outer angle in dimensionless variables [see also \cref{eq:outerangle}]. The average profile decays essentially exponentially for large $\bar x$ [\cref{eq:asymptotics_x_large}]. \label{fig:fixedYoung}}
\end{figure}

In the following, we focus mainly on those situations where a well defined (averaged) contact line position exists, i.e., we shall assume $\theta/\Young < 1$.
In this case, the average profile [\cref{eq_avg_profile_fluctR}] can be explicitly calculated (see \cref{app:part_hav}):
\beq\label{eq:hav_theta<alpha}
\begin{split}\Angle{h(x)} &= \frac{\theta \l}{(\Young+\theta)^2}e^{(\Young-\theta)h_0/\l}e^{(\theta^2-\Young^2)x/2\l}\erfc\Round{\frac{h_0-\theta x}{\sqrt{2\l x}}}+\frac{h_0-\Young x}{2}\erfc\Round{\frac{\Young x-h_0}{\sqrt{2\l x}}}\\
&\quad- \Square{\frac{\theta \l}{(\Young+\theta)^2}+\frac{\Young-\theta}{\Young+\theta}\frac{h_0+\Young x}{2}}e^{2\Young h_0/\l}\erfc\Round{\frac{h_0+\Young x}{\sqrt{2\l x}}}+\frac{2\Young}{\Young+\theta}\sqrt{\frac{\l x}{2\pi}}e^{-(h_0-\Young x)^2/2\l x}.\end{split}
\eeq
Expressed in rescaled variables, we correspondingly obtain
\beq\label{eq:hav_theta<alpha_rescaled}
\begin{split}\Angle{\bar{h}(\bar{x})} &= \frac{\theta/\Young}{(1+\theta/\Young)^2}\frac{1}{\sTcalY}e^{\TcalY (1-\theta/\Young)[1-(1+\theta/\Young)\bar{x}/2]}\erfc\Round{\sTcalY\frac{1-(\theta/\Young)\bar{x}}{\sqrt{2\bar{x}}}}-\sTcalY\frac{\bar{x}-1}{2}\erfc\Round{\sTcalY \frac{\bar{x}-1}{\sqrt{2 \bar{x}}}}\\
	&\quad- \Square{\frac{\theta/\Young}{(1+\theta/\Young)^2}\frac{1}{\sTcalY}+\sTcalY\frac{1-\theta/\Young}{1+\theta/\Young}\frac{\bar{x}+1}{2}}e^{2\TcalY}\erfc\Round{\sTcalY \frac{\bar{x}+1}{\sqrt{2 \bar{x}}}}+\frac{2}{1+\theta/\Young}\sqrt{\frac{\bar{x}}{2\pi}}e^{-\TcalY (\bar{x}-1)^2/2\bar{x}},\end{split}
\eeq
confirming that $\TcalY$ [\cref{eq:singleentryparam_Y}] and $\theta/\Young$ are the only remaining control parameters. The characteristic behavior of the profile is illustrated in \cref{fig:fixedYoung}, for a fixed value of $\theta/\Young$ and various values of $\TcalY$. The scale separation $\TcalY$ controls the curvature of $\Angle{\bar h(\bar x)}$ in the region of cross-over from the outer wedge down to zero height [see \cref{fig:fixedYoung}(a)]. Near the classical contact line location $\bar{x}=1$, the profiles approximately cross in a ``focal point'' at a characteristic height $\bar{h} \simeq 0.4$ --~we will show later on that the exact value is $1/\sqrt{2\pi}$ [see \cref{eq:asymptforce}]. This scale turns out to be relevant for the regularization induced by thermal fluctuations. In \cref{fig:fixedYoung}(b), we have rescaled the horizontal coordinate by $\sTcalY$ and observe that, far from the substrate, all profiles approach a straight wedge with a well defined contact angle.
According to \cref{eq:outerangle}, for $\theta/\Young<1$, this contact angle is simply Young's angle $\Young$. 
More generally, we define the averaged slope of the profile as the derivative of \cref{eq:hav_theta<alpha_rescaled} with respect to $\bar{x}$:
\beq\label{eq:hprimeav_theta_rescaled}
\begin{split}\Angle{\bar{h}'(\bar{x})} \equiv \frac{d\Angle{\bar{h}(\bar{x})}}{d\bar{x}} &= -\sTcalY\frac{1-\theta/\Young}{1+\theta/\Young}\frac{\theta/\Young}{2}e^{\TcalY (1-\theta/\Young)[1-(1+\theta/\Young)\bar{x}/2]}\erfc\Round{\sTcalY \frac{1-(\theta/\Young)\bar{x}}{\sqrt{2\bar{x}}}} -\frac{\sTcalY}{2} \erfc\Round{\sTcalY \frac{\bar{x}-1}{\sqrt{2 \bar{x}}}}\\
	&\quad-\frac{1-\theta/\Young}{1+\theta/\Young} \frac{1}{2}\sTcalY e^{2 \TcalY}\erfc\Round{\sTcalY \frac{\bar{x}+1}{\sqrt{2 \bar{x}}}}+ \frac{1}{\sqrt{2\pi\bar{x}}}e^{-\TcalY (\bar{x}-1)^2/2\bar{x}}.\end{split}
\eeq
Indeed the limit $\bar{x}\to0$ of \cref{eq:hprimeav_theta_rescaled} gives $\Angle{\bar{h}'(0)}=\sTcalY$ in rescaled variables, consistent with \cref{eq:outerangle}. 

\begin{figure}[t!]\centering
	\subfigure[]{\includegraphics[width=0.49\linewidth]{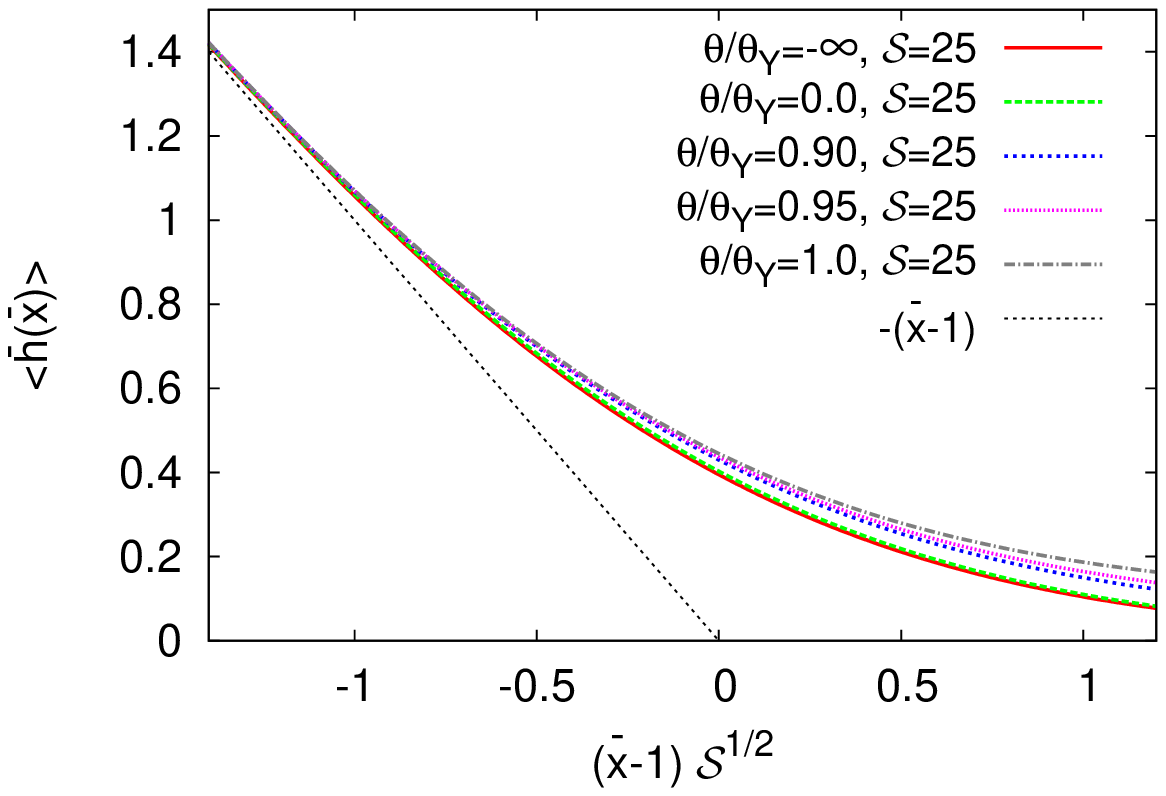}}
	\subfigure[]{\includegraphics[width=0.49\linewidth]{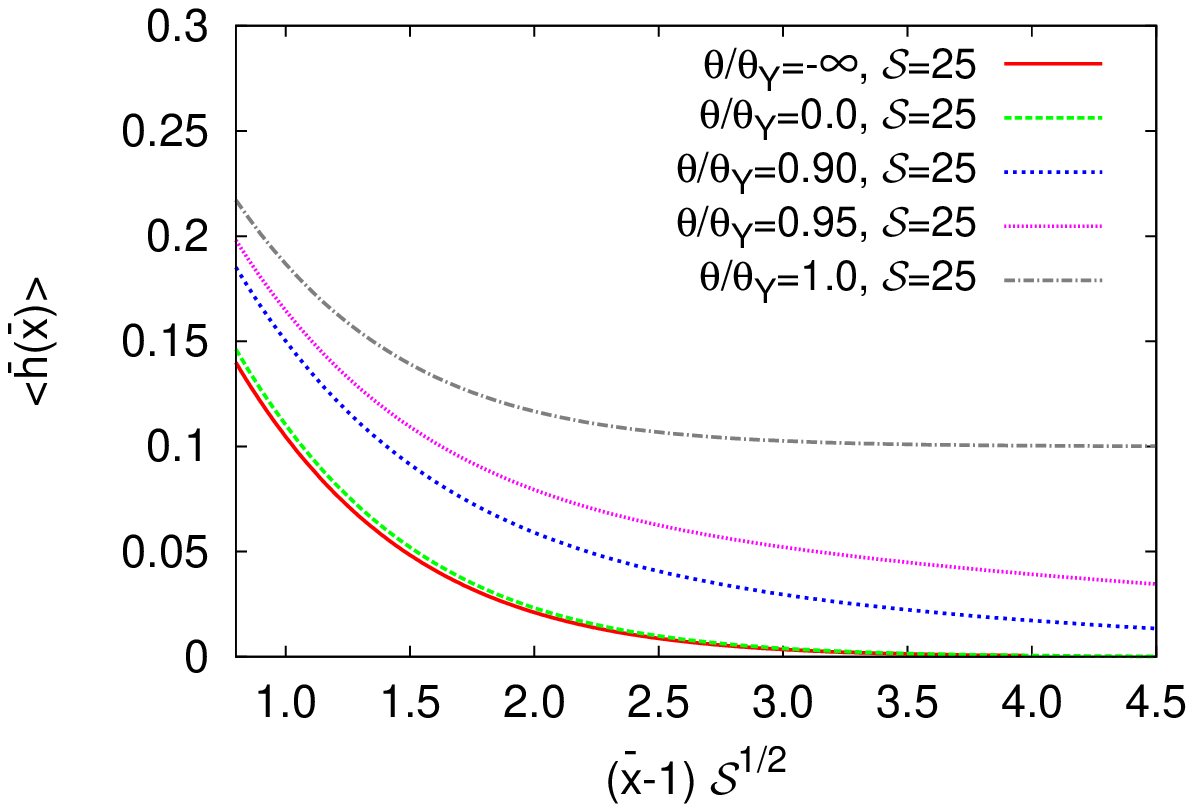}}
	\caption{Average profiles [\cref{eq:hav_theta<alpha}] in the case of a fluctuating contact line, for fixed scale separation $\TcalY=25$ \cref{eq:singleentryparam_Y} and various $\theta/\Young$. The dimensionless height and horizontal coordinate are defined as $\bar{h}=h/\sqrt{h_0\l/\Young}$ and $\bar{x}=x \Young/h_0$ [see \cref{eq_hbar_Y,eq_xbar_Y}]. (b) Magnification of the region near the solid substrate, for the same data as in (a). The shape of the profile is sensitive to the value of $\theta/\Young$ only near the substrate. When $\theta$ approaches $\Young$, the decay length of the exponentially thinning prewetting film increases, until at $\theta=\Young$ a film of constant thickness remains. In both panels, horizontal coordinates are scaled additionally by $\sTcalY$ in order to emphasize that $\sTcalY$ plays the role of the outer angle in dimensionless variables. \label{fig:fixedYoungb}}
\end{figure}

\cref{fig:fixedYoungb} illustrates the effect of the boundary condition parameter $\theta/\Young$ on the average profiles upon approaching the pseudo-partial wetting transition from below ($\theta\nearrow \Young$), for a fixed value of $\TcalY$. 
Far from the substrate, the profile shapes do not strongly depend on the value of $\theta/\Young$.  
The choice of $\theta/\Young$, however, does have a strong impact on the decay to zero of the average profiles. Indeed, a closer inspection of the region near $\bar{h}=0$ [\cref{fig:fixedYoungb}(b)] reveals that $\theta/\Young$ controls the properties on the average profile in the ``inner'' region: upon crossing the pseudo-partial wetting transition point, the profile does not decay anymore to zero, but rather forms a flat film, persisting at large $\bar{x}$ with a finite width. 
%
%



The impact of the scale separation $\TcalY$ as well as of the boundary condition parameter $\theta/\Young$ on the profile can be assessed based on the asymptotics of the latter. In the limit $\bar{x} \to 0$, \cref{eq:hav_theta<alpha_rescaled} reduces to
\beq\label{eq:asympt_xsmall}
\Angle{\bar{h}(\bar{x})} \overset{\bar{x}\to0}{\sim} \sTcalY(1-\bar{x}) + \frac{1}{\TcalY}\sqrt{\frac{2\bar{x}^3}{\pi}}e^{-\TcalY (\bar{x}-1)^2/2\bar{x}}.
\eeq
Accordingly, the (dimensionless) quantity $\TcalY$ controls the shape of the average profiles in the outer region [cf.\ \cref{fig:fixedYoungb}]. In particular, the way the profile crosses over to the inner region is manifestly independent of the parameter $\theta/\Young$. 
In contrast, for smaller $h$, the decay of the profile is affected by the parameter $\theta/\Young$. This becomes apparent from the asymptotics of \cref{eq:hav_theta<alpha_rescaled} for large $\bar{x}$:
\beq\label{eq:asymptotics_x_large}
\Angle{\bar{h}(\bar{x})} \overset{\bar{x}\to\infty}{\sim}
\begin{dcases}
\frac{(1-2\theta/\Young)(1-\TcalY \theta/\Young)}{(\theta/\Young)^2\TcalY^2}\sqrt{\frac{2}{\pi\bar{x}^3}}e^{\TcalY}e^{-\TcalY \bar{x}/2} &\textup{if } \theta/\Young<0, \\
\frac{1}{\TcalY}\sqrt{\frac{2}{\pi \bar{x}}}e^{\TcalY}e^{-\TcalY \bar{x}/2} &\textup{if } \theta/\Young=0, \\
\frac{1}{\sTcalY}\frac{2\theta/\Young}{(1+\theta/\Young)^2}e^{\TcalY (1-\theta/\Young)}e^{-\TcalY [1-(\theta/\Young)^2]\bar{x}/2} &\textup{if } 0< \theta/\Young<1.
\end{dcases}
\eeq
Note that the decay length scale of the average profiles diverges when $\theta/\Young \nearrow 1$, giving rise to a flat film that survives up to infinity. Specifically, we have $\Angle{\bar{h}(\bar{x})} \overset{\bar{x}\to\infty}{\sim} 1/2\sTcalY$ for $\theta/\Young=1$, which is in quantitative agreement with the result  of \cref{eq:filmRfixed_rescaled}.

In summary, while the scale separation $\TcalY$ controls the intensity of fluctuations of the rescaled contact point $\bar{R}$, the microscopic deviation from the Young's angle $\theta/\Young$ controls the ``inner'' properties of the model --- in particular, the asymptotic tails of the probability distribution of $\bar{R}$. These tails become long-ranged if $\theta/\Young \nearrow 1$. Crucially, if $\theta/\Young<1$, the outer angle is always selected in agreement with Young's law [see \cref{eq:outerangle}]. If $\theta/\Young \ge 1$, instead, the average position of the contact line diverges and a flat film persists to infinity; this changes the energy balance and yields an outer angle of $\theta$ instead of $\Young$. In this case, $\Angle{h}$ becomes identical to the the pseudo-partial wetting profile in \cref{eq:hav_theta}, studied for the pinned contact line problem in \cref{sec:fixed}. Thus, the parameter $\theta/\Young$ is an important characteristic of the fluctuating contact line ensemble, as it governs the transition from states with a finite $\Angle{R}$ to pseudo-partial wetting states ($\Angle{R} \to \infty$).


\begin{figure}[t!]\centering
	\subfigure[]{\includegraphics[width=0.49\linewidth]{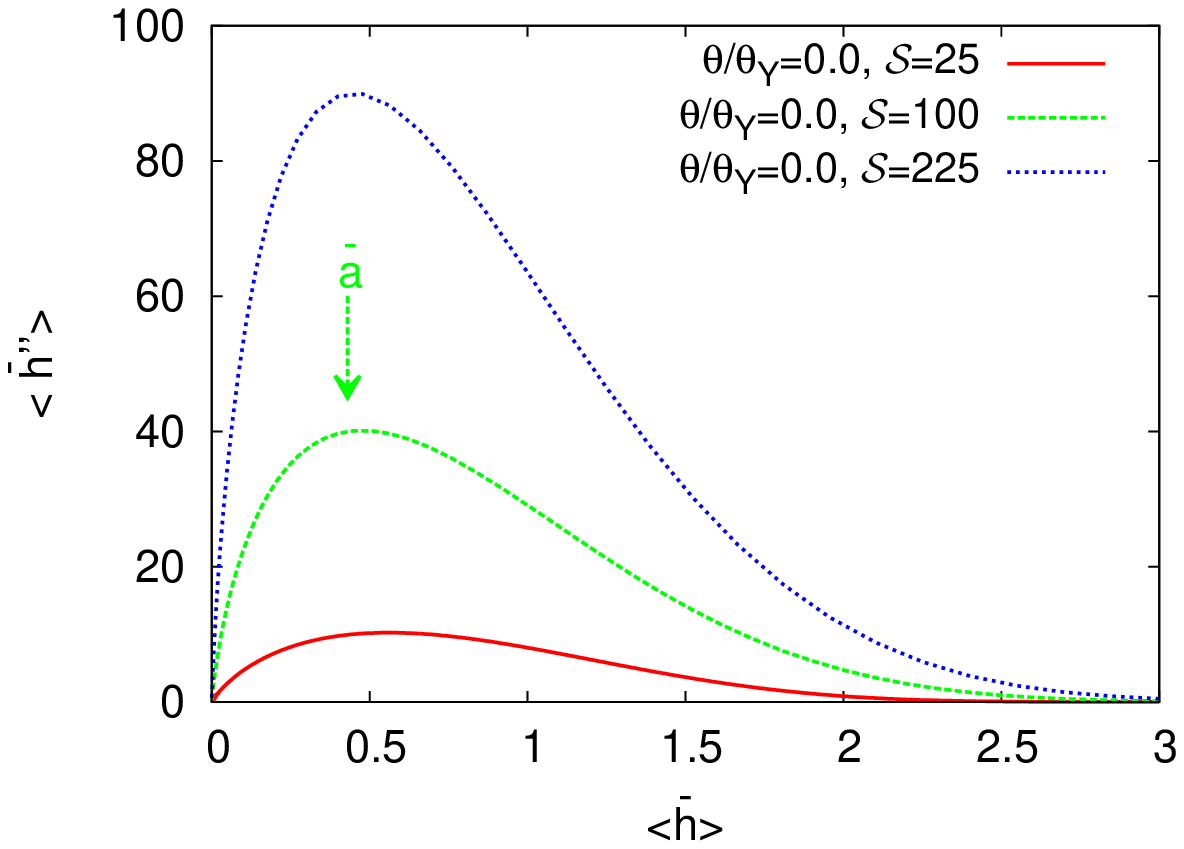}}
	\subfigure[]{\includegraphics[width=0.49\linewidth]{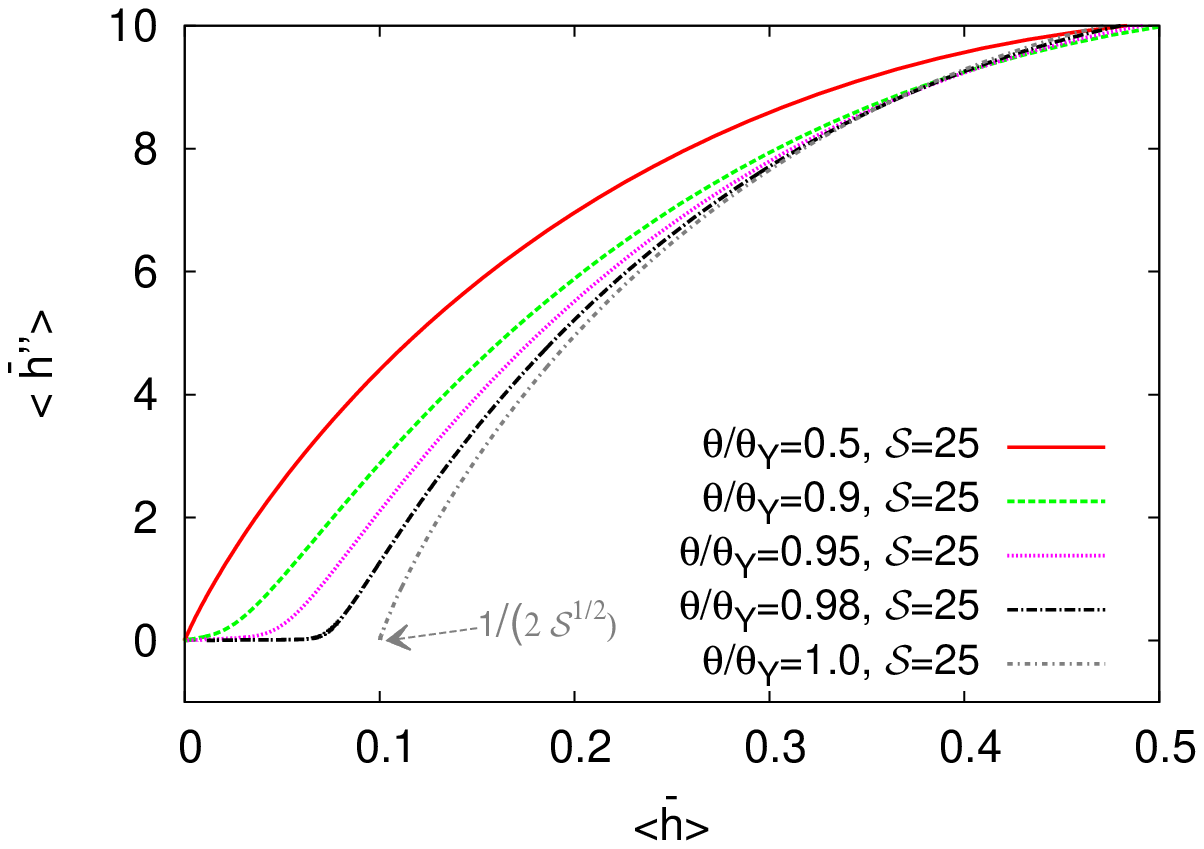}}\\
	\subfigure[]{\includegraphics[width=0.49\linewidth]{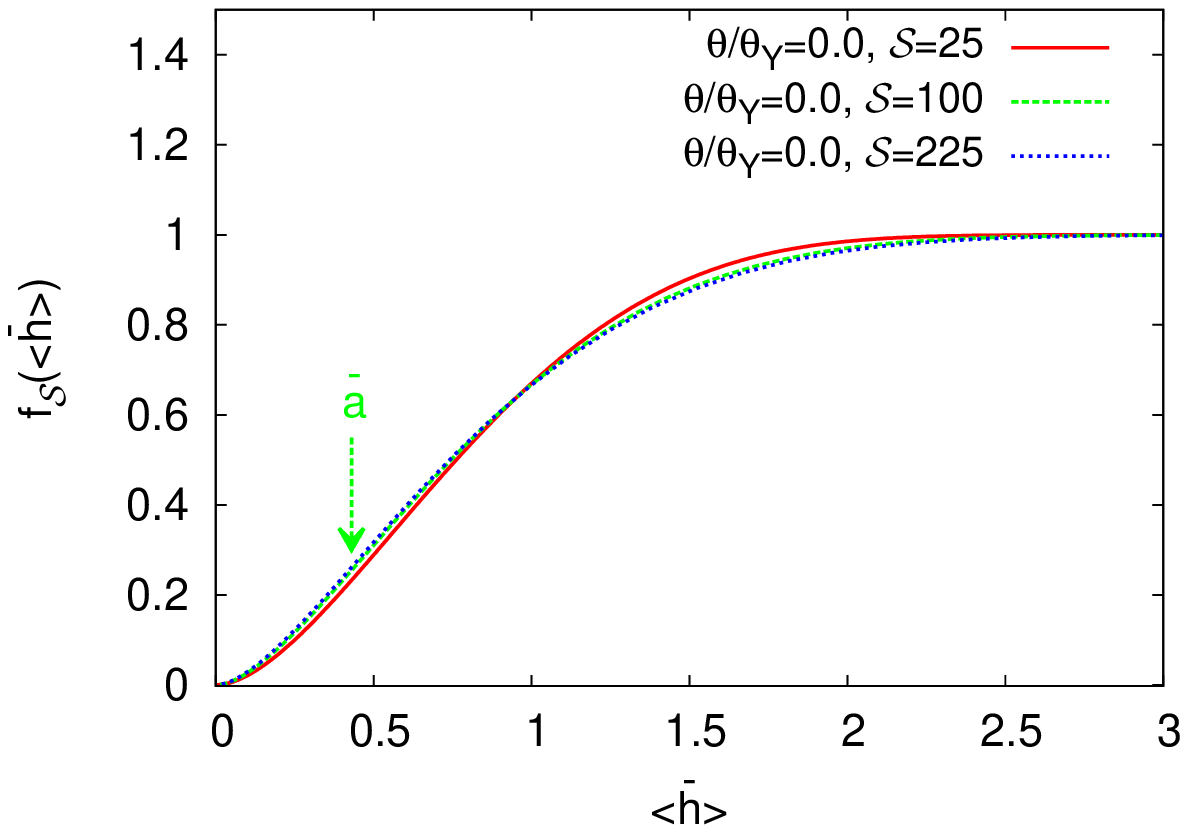}}
	\subfigure[]{\includegraphics[width=0.49\linewidth]{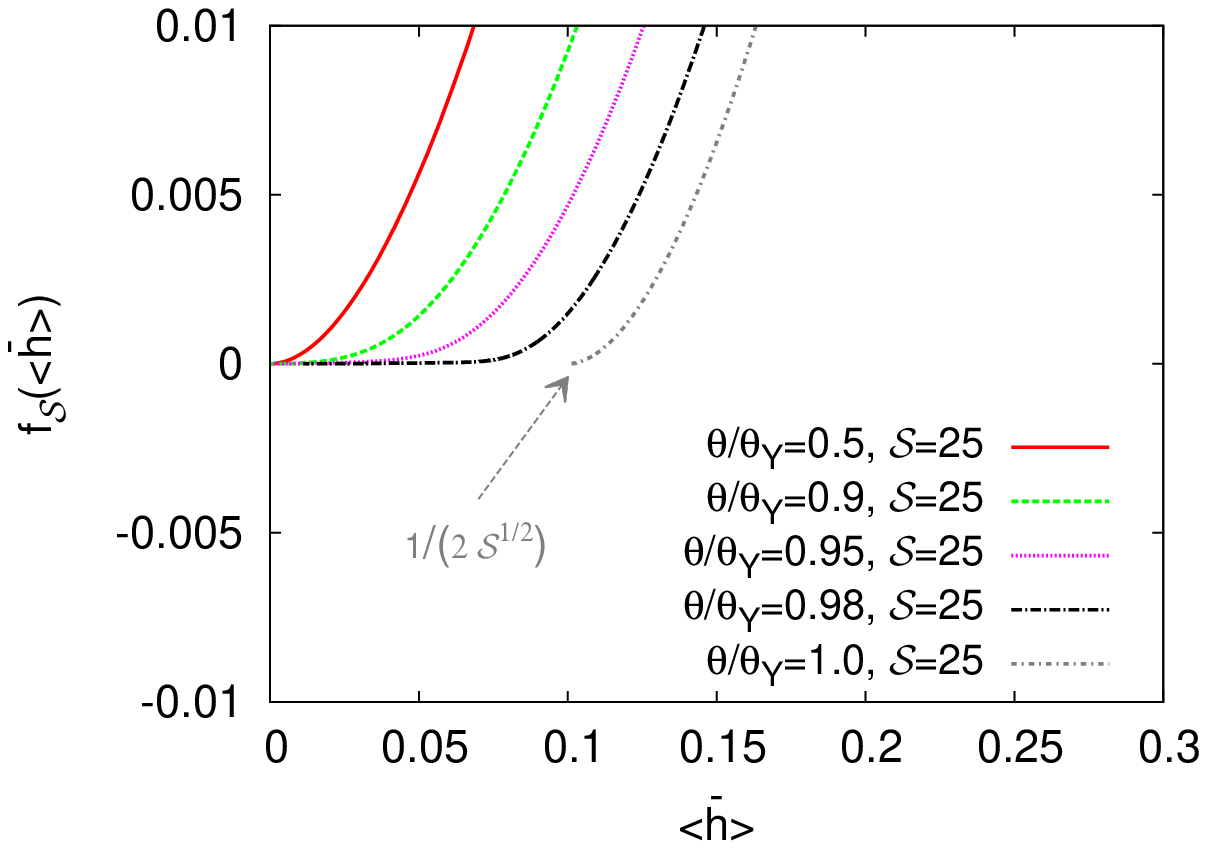}}
	\caption{ Effective disjoining pressure and binding potential [in dimensionless variables, see \cref{eq_hbar_Y,eq_xbar_Y}]. (a,b) Averaged Laplace pressure $\Angle{\bar{h}''(\bar{x})}$ [\cref{eq:secondder_adim}] as a function of the average height $\Angle{\bar{h}(\bar{x})}$ [\cref{eq:hav_theta<alpha_rescaled}], which can be identified with the (negative of the) effective disjoining pressure $f'_\TcalY$ defined by \cref{eq:average_eq_motion}. The maximum of $f'_\TcalY$ is controlled by $\TcalY$, see panel (a), where the behavior of $f'_\TcalY$ is illustrated for $\theta/\Young=0$ (reflecting wall; other values of $\theta/\Young\lesssim 0$ lead to a similar behavior). The parameter $\theta/\Young$, in contrast, influences the behavior of the disjoining pressure near the substrate [see panel (b)]. In particular, upon approaching the pseudo-partial wetting point ($\theta/\Young=1$), a prewetting film develops and the averaged Laplace pressure vanishes at a finite height [arrow in (b)]. (c,d) Effective  potential $f_{\TcalY}$ [\cref{eq:potential}] corresponding to the data in panels (a,b). $f_{\TcalY}$ saturates exponentially, governed by $\TcalY$, to its asymptotic value $1$ at large heights [\cref{eq:f_large_h}], whereas $f_{\TcalY}$ goes like $\Angle{\bar h}^2$ for small $\Angle{\bar{h}}$ [\cref{eq:f_small_h}]. The point of inflection of $f_{\TcalY}$ [panel (c)], corresponding to the maximum of the averaged Laplace pressure $\Angle{\bar h''}$ [panel (a)] is taken as the characteristic regularization scale induced by thermal fluctuations, $\bar{a}$ [\cref{eq:regularization}, indicated by the arrows in (a,c)]. Near $\Angle{\bar{h}}=0$, the boundary condition parameter $\theta/\Young$ regulates the curvature of the potentials [see panel (d) and \cref{eq:f_small_h}]. At the pseudo-partial wetting point ($\theta=\Young$), a minimum of $f_{\TcalY}$ at a finite height appears [arrow in (d)], corresponding to a uniform film thickness. \label{fig:rego}}
\end{figure}



\begin{figure}[h!]\centering 
        \vcenteredhbox{\includegraphics[width=0.49\linewidth]{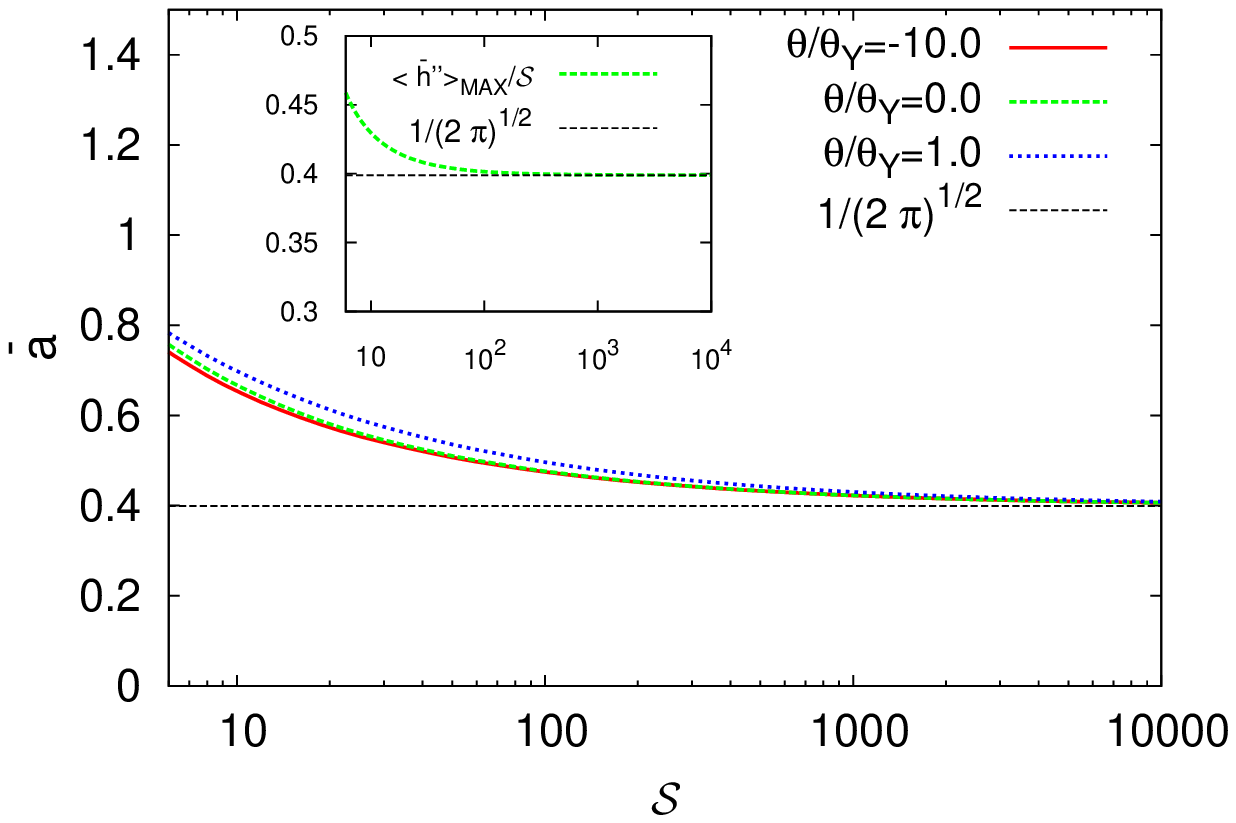}}
	\caption{Regularization length scale $\bar a$ [\cref{eq:regularization}] as a function of the scale separation $\TcalY$ for different values of $\theta/\Young$. Inset: maximum of the second order derivative (averaged Laplace pressure) $\Angle{\bar{h}''(\bar{x})}$ [cf.\ \cref{fig:rego}(a)] as a function of $\TcalY$, for $\theta/\Young=0$ (reflecting wall). The dashed lines represent the asymptotic values given in \cref{eq:asymptforce}. To highlight the linear nature of the asymptotics described by \cref{eq:asymptforce}, the maximum of the averaged Laplace pressure  is further normalized by $\TcalY$. \label{fig:rego2}}
\end{figure}


Since the present model contains \textit{a priori} only contact interactions, the difference of the profile $\Angle{h(x)}$ from a straight wedge can be understood as a consequence of the interaction of fluctuations with the impenetrable wall and of the averaging over the contact line position $\R$. 
Based on \cref{eq:eqnofmotion}, which expresses the balance between disjoining and Laplace pressure, we define the \emph{effective disjoining pressure} $-f'_{\TcalY}$ via \footnote{As may be easily checked by inserting the rescaled quantities defined in \cref{eq_hbar_Y,eq_xbar_Y,eq:singleentryparam}, \cref{eq:average_eq_motion} reduces to \cref{eq:eqnofmotion}.}
\beq\label{eq:average_eq_motion}
\Angle{\bar{h}''(\bar{x})} = \frac{\TcalY}{2} f_{\TcalY}'(\Angle{\bar{h}(\bar{x})}).
\eeq
The effective disjoining pressure is in fact the derivative with respect to $h$ of an effective binding potential $f_\TcalY(h)$, which is studied separately further below. Note that $f'_{\TcalY}$ is to be understood as a function of $\Angle{\bar h}$ and not of $\bar x$. An effective attraction (repulsion) of the interface towards (from) the wall corresponds to $-f_\TcalY'<0$ ($-f_\TcalY'>0$). The averaged curvature of the profile is defined as the second order derivative of \cref{eq:hav_theta<alpha_rescaled} with respect to $\bar{x}$:
\beq\label{eq:secondder_adim}
\begin{split}\Angle{\bar{h}''(\bar{x})}\equiv \frac{d^2\Angle{\bar h(\bar x)}}{d\bar x^2} &= \frac{\TcalY^{3/2}(\theta/\Young)(1-\theta/\Young)^2}{4}e^{\TcalY (1-\theta/\Young)[1-(1+\theta/\Young)\bar{x}/2]}\erfc\Round{\sTcalY\frac{1-(\theta/\Young)\bar{x}}{\sqrt{2\bar{x}}}}\\
&\quad+ \TcalY\frac{1+(\theta/\Young-1/\TcalY)\bar{x} + (1-\theta/\Young)^2\bar{x}^2}{4\bar{x}^2}\sqrt{\frac{2}{\pi\bar{x}}}e^{-\TcalY (\bar{x}-1)^2/2\bar{x}}. \end{split}
\eeq
The  effective disjoining pressure therefore simply follows by relating the curvature $\Angle{\bar{h}''}$ to the thickness $\Angle{\bar{h}}$, which can be straightforwardly done numerically, based on the implicit relation through $\bar x$.  In \cref{fig:rego}(a,b), $\Angle{\bar{h}''}$ is plotted as a function of $\Angle{\bar{h}}$ for different values of $\TcalY$ and $\theta/\Young$. These plots therefore represent $f_\TcalY'$, i.e., the negative of the effective disjoining pressure. For fixed $\theta/\Young$, \cref{fig:rego}(a) reveals that the quantity $\TcalY$ controls the distribution of the  effective disjoining pressure as a function of the distance from the wall. The pressure peaks for values of $\Angle{\bar{h}}$ that do not seem to strongly depend on $\TcalY$ and, in fact, turn out to be close to the average height at the classical contact line location $\Angle{\bar h(\bar x=1)}$ [cf.\ \cref{fig:fixedYoung}(a)]. This is consistent with the behavior in the classical limit ($\TcalY\to\infty$), where the disjoining pressure is strongly localized at the classical contact line location [see \cref{eq:normal}]. While in \cref{fig:rego}(a) we focus on values $\theta/\Young=0$  (reflecting boundary conditions), we remark that the behavior is similar for any value $\theta/\Young\lesssim 0$ (including the limit of purely absorbing boundary conditions, $\theta/\Young=-\infty$). This is a simple consequence of the similarity of the average profiles for different $\theta/\Young$, as illustrated in \cref{fig:fixedYoungb}. The effective disjoining pressure is, however, affected by the proximity to the pseudo-partial wetting transition, as controlled by the parameter $\theta/\Young$ [\cref{fig:rego}(b)]: for $\theta/\Young \nearrow 1$, a flat film emerges and the second order derivative must correspondingly vanish at a finite value of $\Angle{\bar{h}}=1/2\sTcalY$ (dotted arrow), in agreement with \cref{eq:filmRfixed_rescaled}.

The regularization length scale $\bar{a}$ associated with thermal fluctuations (see \cref{sec:disjoining}) is chosen as the height at which the dimensionless effective disjoining pressure $-f'_\TcalY$ is largest in magnitude: $\Angle{\bar{h}''}_{\st{MAX}} \equiv \Angle{\bar{h}''(\bar{x}_{\st{MAX}})}$, i.e.,
\beq\label{eq:regularization}
\bar{a} \equiv \Angle{\bar{h}(\bar{x}_{\st{MAX}})}.
\eeq
By definition, $\bar{a}$ corresponds to the inflection point of $f_{\TcalY}$ [see \cref{fig:rego}(c,d)]. It represents a reasonable approximation of the characteristic height above the wall at which the profiles starts to flatten and crosses over towards the outer wedge (cf.\ \cref{fig:fixedYoung} and \cref{fig:fixedYoungb}). The dependence of $\bar{a}$ and of $\Angle{\bar{h}''}_{\st{MAX}}$ on $\TcalY$ is displayed in \cref{fig:rego2}. We find that the regularization length scale only weakly depends on $\theta/\Young$. The asymptotic behaviors observed in the plot in the classical limit, $\TcalY \to \infty$, can be derived analytically. To this end, we determine the third order derivative of the dimensionless profile in \cref{eq:hav_theta<alpha_rescaled} with respect to $\bar{x}$, which for large $\TcalY$ reads:
\beq
\Angle{\bar{h}'''(\bar{x})} \equiv \frac{d^3\Angle{\bar h(\bar x)}}{d\bar x^3} \overset{\TcalY\to\infty}{\sim} \frac{\TcalY^2}{8} \frac{1-[2\theta/\Young+(1-2\theta/\Young)\bar{x}^2]\bar{x}^2}{[1-(\theta/\Young)\bar{x}]\bar{x}^2} \sqrt{\frac{2}{\pi\bar{x}}}e^{-\TcalY (\bar{x}-1)^2/2\bar{x}}.
\eeq
Note that the above asymptotic expression for $\Angle{\bar h'''(\bar{x})}$ vanishes at $\bar{x}=1$, implying that the largest force [see \cref{eq:average_eq_motion}] is localized at the classical contact line position in the limit of large $\TcalY$ (i.e., $\bar{x}_{\st{MAX}}\sim1$ for $\TcalY\to\infty$). Indeed, this is expected from the purely deterministic model, see \cref{eq:normal}. Correspondingly, from \cref{eq:secondder_adim,eq:hav_theta<alpha_rescaled} evaluated at $\bar{x}=1$, we obtain 
\rev{\begin{align}\label{eq:asymptforce}
&\Angle{\bar{h}''}_{\st{MAX}} \overset{\TcalY\to\infty}{\sim} \frac{\TcalY}{\sqrt{2\pi}}, &&\bar{a} \overset{\TcalY\to\infty}{\sim} \frac{1}{\sqrt{2\pi}}.
\end{align}}
For moderate $\TcalY$, the actual values of $\Angle{\bar{h}''}_{\st{MAX}}$ and $\bar{a}$ are slightly larger than the asymptotic values in \cref{eq:asymptforce}; however, the rescaled regularization length $\bar a$ is always of order of unity and depends only gently on $\theta/\Young$. This implies that the rescaling [\cref{eq_hbar_Y}] used for the vertical coordinates (i.e., the height fluctuations of standing capillary waves) is also the ``natural'' scale for the bare $a$, i.e., 
\beq
a \overset{\l/h_0\to0}{\sim} \sqrt{\frac{\l h_0}{2\pi\Young}}.
\label{eq_a_entr}
\eeq 

Equation \eqref{eq:average_eq_motion} also admits a study of the {\it effective binding potential} $f_{\TcalY}(\Angle{\bar{h}(\bar{x})})$ itself, which is the (negative of the) the integrated effective disjoining pressure.  This is possible if the profiles are strictly monotonous, which is fulfilled if $\TcalY\gtrsim 1$, i.e., if the outer scale $h_0$ is well separated from the thermal length $\l$ \footnote{This condition applies to all the cases shown in \cref{fig:rego}.}. In such a case, multiplying \cref{eq:average_eq_motion} by $\Angle{\bar{h}'(\bar{x})}$ and integrating yields
\beq\label{eq:potential}
\Angle{\bar{h}'(\bar{x})}^2=\TcalY f_{\TcalY}(\Angle{\bar{h}(\bar{x})}).
\eeq
We recall that the quantity $\TcalY$ is related to $\gamma(1-\cos \Young)$ in unscaled variables [see \cref{eq:singleentryparam_Y,freeenergy:disjoining}] and therefore reflects in \cref{eq:potential} the influence of the solid. Moreover, since according to  \cref{eq:asymptotics_x_large,eq:asympt_xsmall} the limits $\bar x\to\infty$ and $\bar x\to 0$ correspond to $\bar h\to 0$ and $\bar h\to \sTcalY$, respectively, we infer with the aid of \cref{eq:hprimeav_theta_rescaled} that \rev{(we will write the argument of $f_{\TcalY}$ simply as $\bar{h}$ instead of $\Angle{\bar{h}(\bar{x})}$, for shortness)}
\beq
f_{\TcalY}(\bar{h}) \overset{\bar{h}\to 0}{\sim} 0 \qquad\text{and}\qquad f_{\TcalY}(\bar{h}) \overset{\bar{h} \to \sTcalY} \sim 1.
\eeq
Consequently, \cref{eq:average_eq_motion} implies that the area below the curves displayed in \cref{fig:rego}(a) is equal to $\TcalY/2$.
The associated  effective potentials for the cases reported in \cref{fig:rego}(a,b) are displayed in the corresponding bottom panels (c) and (d).  In practice, the function $f_{\TcalY}$ is obtained here by relating $\Angle{\bar h'(\bar{x})}$ [\cref{eq:hprimeav_theta_rescaled}] to $\Angle{\bar h(\bar{x})}$ [\cref{eq:hav_theta<alpha_rescaled}] for each value of $\bar{x}$. 

Overall, $f_\TcalY$ exhibits an attractive character [\cref{fig:rego}(c)]. This is expected from the fact that we consider here a partially wetting liquid on a ``hydrophilic'' substrate. Indeed, the asymptotic value $f_\TcalY(\bar{h})\to 1$ in the limit $\bar h \to \sTcalY$ shows, recalling \cref{eq:potential}, that this attraction originates from the influence of the solid. Instead of Dirac's delta function localized at the wall obtained for the disjoining pressure in the limit of vanishing temperature [see \cref{eq:normal}], $f'_\TcalY$ is spread over a certain range in height. Specifically, the following effects contributing to $f_\TcalY$ can be identified: first, by construction, each stochastic realization of the profile is constrained to end precisely at the substrate, i.e., at the contact line position $\R$, with $\R$ being a fluctuating variable. Physically, this is a necessary condition in order to recover, after averaging over $\R$, Young's angle at $h_0$. Note further that, as evidenced in \cref{fig:fixedYoungb} and by \cref{eq:asympt_xsmall}, the shape of the profile at large scales is essentially independent of the value of the boundary condition parameter $\theta$. This insensitivity with respect to $\theta$ turns out to also apply to $f_\TcalY$ in the outer region [see \cref{eq:f_large_h}]. We therefore conclude that the profile constraint and the subsequent averaging over $\R$ are essentially responsible for the curvature the average profile and its elevation above the classical contact line position. As already anticipated, a second contribution to $f_\TcalY$ emerges from the direct interaction of the fluctuating profile with the substrate, expressed by the boundary condition in \cref{eq:prop_Robin}. For $\theta>0$, the profile is attracted to the wall, whereas for $\theta<0$, it is repelled (\cref{fig:P}). Due to the nature of the contact interaction, this effect is expected to be mainly significant for short-wavelength fluctuations near the substrate. 

These intuitive arguments are borne out by analytical calculations: for large $\TcalY$, the proper asymptotics of $f_{\TcalY}$ around the classical position of the contact line can be inferred by introducing the rescaling 
\beq \bar{x}=1+\kappa/\sTcalY,
\label{eq_x_kappa}\eeq 
with $\kappa$ being some arbitrary constant. By performing the limit $\TcalY\to\infty$ in \cref{eq:hav_theta<alpha_rescaled} keeping $\kappa=\sTcalY(\bar{x}-1)$ fixed, we get $\Angle{\bar{h}(\bar{x})}\to\bar{h}(\kappa)$, with
\beq
\bar{h}(\kappa) \equiv \frac{e^{-\kappa^2/2}}{\sqrt{2\pi}} - \frac{\kappa}{2} \erfc\Round{\frac{\kappa}{\sqrt{2}}}.
\label{eqas1}
\eeq
%
\rev{This profile does not depend anymore on $\theta/\Young$ and can be understood as the universal profile obtained in the limit of an infinite outer scale $h_0$ \footnote{\rev{The rationale behind \cref{eq_x_kappa} lies in the fact that, upon increasing $h_0$, the height $\Angle{h(\R_\st{cl})}$ at the effective contact point as well as the characteristic region in $x$ around $\R_\st{cl}$ both scale $\sim \sqrt{h_0}$ for sufficiently large $h_0$. The scaling of the average height is already taken into account in \cref{eq_hbar_Y}.}}.}
Similarly, from \cref{eq:hprimeav_theta_rescaled,eq:potential} one obtains $f_\TcalY\Round{\Angle{\bar{h}(\bar{x})}}\to f_\infty(\bar{h}(\kappa))$, with
\beq
f_\infty(\bar{h}(\kappa)) \equiv \frac{1}{4} \erfc^2\Round{\frac{\kappa}{\sqrt{2}}}.
\label{eqas2}
\eeq
In the limit of a large scale separation $\TcalY$, the binding potential is therefore given by the parametric \cref{eqas1,eqas2} and is independent of the boundary condition parameter $\theta$. 
From \cref{eq_x_kappa} we infer that, for $\TcalY\gg 1$, the outer region around $\bar x\simeq 0$ (where $\bar h\simeq \sTcalY$) corresponds to $\kappa\ll -1$. Accordingly, \cref{eqas1} reduces to $\bar h(\kappa)\sim -\kappa$ in this case and the effective potential in \cref{eqas2} therefore approaches its asymptotic value $1$ as
\beq\label{eq:f_large_h}
f_\infty(\bar{h}) \overset{\bar{h}\to\infty}{\sim} \frac{1}{4} \erfc^2\Round{-\frac{\bar h}{\sqrt{2}}}.
\eeq
This asymptotic relation describes the outer behavior of the binding potential, with a regularization length scale of order unity in scaled variables [cf. \cref{eq:asymptforce} and \cref{eq_a_entr}]. However, at small distances from the wall, $\theta$ has a strong influence: in particular, as shown in \cref{fig:rego}(d), the effective potential flattens as the pseudo-partial wetting point ($\theta/\Young=1$) is approached. This is expected because a thin film emerges and the associated potential must develop a minimum at the film thickness $\bar{h}=1/2\sTcalY$. This behavior is confirmed by the asymptotics of $f_{\TcalY}$ for small $\bar h$, which, making use of \cref{eq:asymptotics_x_large}, is found to be 
\beq\label{eq:f_small_h}
f_{\TcalY}(\bar{h}) \overset{\bar{h}\to0}{\sim} \frac{\TcalY\bar{h}^2}{4} \times \begin{cases}1 &\textup{if } \theta\leq0, \\
[1-(\theta/\Young)^2]^2 &\textup{if } 0< \theta<\Young.
\end{cases}
\eeq
The corresponding asymptotics for the effective disjoining pressure $-f'_\TcalY$ [see \cref{eq:average_eq_motion}] can be obtained from differentiation of \cref{eq:f_large_h,eq:f_small_h} with respect to $\bar{h}$. The inner solution [\cref{eq:f_small_h}] is valid for a range of $\bar h$ whose size decreases as $1/2\sTcalY$ in the limit of large $\TcalY$.

In conclusion, while our initial model [\cref{eq:Z,eq:prop_Robin}] includes only contact interactions, the interplay of the interfacial fluctuations with the wall gives rise to a \emph{finite-range} contribution to the disjoining pressure and its associated potential $f_\TcalY$. The singular contact force in the deterministic model [see \cref{eq:normal}] is therefore \emph{regularized} by fluctuations. The emergence of fluctuation contributions to a binding potential is, in principle, known and has previously been studied, for instance, in applications of the renormalization group to wetting phenomena \cite{brezin_critical_1983, kroll_universality_1983, fisher_wetting_1985, lipowsky_scaling_1987, huse_comment_1987, juelicher_functionalRG_1990, spohn_fixedpoints_1991,forgacs_review_1991, indekeu_thermal_2010}. Here, we have characterized this effect in the presence of a fluctuating contact line. 
The resulting potential decays much faster (exponentially) than for van der Waals interactions and is localized over a length $a$ given by \cref{eq_a_entr} for a one-dimensional interface. The form of the potential depends on the boundary condition parameter $\theta$ at small scales, but approaches its asymptotic value independently from this parameter.  Interestingly, while the average profile strongly depends on $\theta$ in the case of a pinned contact line (see \cref{sec:fixed} and, in particular, \cref{fig:fixedR}), once the contact line is allowed to fluctuate, the profile exhibits this strong dependence only near the substrate. The corresponding profile in the outer region ($h\gtrsim a$) varies instead only mildly with $\theta/\Young<1$.  These properties indicate that, in fact, the fluctuating nature of the contact line, rather than the direct interaction of the interfacial fluctuations with the wall, are responsible for the characteristic shape of the profile and the regularization of the binding potential at large scales. In particular, the entropic repulsion effect, caused by the impenetrability of the substrate, does not play a prominent role in the fluctuating contact line ensemble. The fluctuating contact line model predicts a pseudo-partial wetting transition to occur for $\theta/\Young=1$. 

\section{Conclusions}\label{sec:conclusions}

In this study, we have investigated the effect of thermal fluctuations on the morphology of an interface near a contact line in the presence of an impenetrable wall [see \cref{fig_wedge_sketch_fluct}]. To facilitate the analytical treatment, we have considered, within a path-integral approach, a one-dimensional profile $h(x)$ described by a capillary wave Hamiltonian. \rev{We have assumed a general (Robin-type) no-flux boundary condition for the interfacial fluctuations. This boundary condition can be equivalently represented by a contact potential in the Hamiltonian \cite{burkhardt_propagator_1989, upton_interfacemodel_1999, upton_interfacemodel_2002, Rajabpour_2009, walton_starquant_2005} and is characterized by a dimensionless parameter $\theta$ that encompasses fully absorbing ($\theta=-\infty$) and fully reflecting ($\theta=0$) types of walls. The quantity $\theta$ can be understood as a parametrization of the short-distance physics emerging from a finite-range binding potential.}
Values of $\theta \geq 0$ give rise to an interface that is microscopically bound to the wall, whereas $\theta<0$ represents an unbound interface [see \cref{fig:fixedR}]. Crucially, while we assume each stochastic realization of the profile $h(x)$ to touch the wall at some well-defined contact line position $\R$, we consider $\R$ to be a random variable. The fluctuation properties of $\R$ follow straightforwardly by promoting the capillary wave Hamiltonian to the action of a path integral [see \cref{eq:Z}]. The inclusion of a fluctuating contact line is a fundamental ingredient of the model, as it ensures that Young's angle $\Young$ is recovered as the slope of the mean profile far from the substrate [see \cref{eq:outerangle}]. In other words, Young's law, which derives from the homogeneity of the substrate \cite{snoeijer_microscopic_2008}, is robust in presence of thermal fluctuations, provided one chooses the ensemble in which both $h(x)$ and $R$ are fluctuating. Our model enables one to study the morphology of interfacial profiles for a fixed outer (Young's) angle $\Young$, while altering the ``micromechanics'' parametrized by the parameter $\theta$, which can be interpreted as a microscopic deviation from Young's angle.  This situation can be contrasted \rev{to the case where} the contact line is fixed [see \cref{sec:fixed}], which has previously been considered in the literature \cite{burkhardt_propagator_1989, WoodParry01, romero-enrique_interfacial_2004,upton_interfacemodel_1999,upton_interfacemodel_2002}: for such a pinned interface, the outer slope is governed by the boundary condition parameter $\theta$. Furthermore, for fixed but large $\R$, a precursor film emerges in this case, whose (asymptotically constant) thickness is controlled by the balance between thermal fluctuations and surface energy (i.e., $\theta$) [see \cref{fig:fixedR_rescaled}].

When the contact line is allowed to fluctuate, its ensemble-averaged mean  location $\Angle{\R}$ is found to be near the expected classical contact line position $\R_\st{cl}=h_0/\Young$ of a straight wedge, provided $\theta<\Young$ [see \cref{fig:Rfluct}]. The average profile in front of the wedge decays essentially exponentially with increasing distance from $\Angle{\R}$ [see \cref{eq:asymptotics_x_large}] -- in remarkable contrast to the case of fixed $\R$. When the boundary condition parameter $\theta$ approaches $\Young$ from below, we find a crossover to a pseudo-partial wetting state, characterized by a diverging mean contact line position $\Angle{\R}$ and the development of a flat film of constant thickness in front of the wedge. For $\theta=\Young$, the average profile finally becomes identical to the one in the pinned contact line case. \rev{For a reflective wall ($\theta=0$), the line tension is found to diverge logarithmically, $\tau\sim \ln \Young$ upon approaching the wetting transition at $\Young=0$.} 

The effect of the averaging over $\R$ as well as the direct interactions between the fluctuating interface and the wall can be captured in terms of an effective binding potential $f_{\TcalY}$ of finite range [see \cref{eq:potential} and \cref{fig:rego}(c,d)]. Physically, this potential accounts for the attractive character of the solid that is partially wet by the liquid: in unscaled variables ${\TcalY} f_{\TcalY}$ tends to $2 \gamma (1-\cos \Young) \simeq \gamma \Young^2$ at large distance as well as in the zero temperature limit. Due to fluctuations, this influence of the substrate is spread over a regularization length given by 
\beq \label{eq_a_1d}
 a \simeq \l \sqrt{\frac{h_0}{\Young \l}},
\eeq
where $\l$ is the characteristic thermal length [see \cref{eq:ell}]. Note that the scale $a$ simply corresponds to the square-root of the roughness of a free fluctuating interface. We emphasize, however, that the square-root dependence of $a$ on the scale $h_0/\Young$ is specific to a one-dimensional interface. One dimensional fluid interfaces can occur, for instance, in lipid bilayer films below their demixing transition \cite{honerkamp-smith_linetensions_2008, thiam_biophysics_nat2013}. Lipid bilayers are the building blocks of the membranes of biological cells and can be considered as a physical realization of a two-dimensional binary fluid \cite{honerkamp-smith_dynamic_prl2012}. In the case of a two-dimensional interface, we instead expect a logarithmic behavior \cite{rowlinson_molecular_1982, safran_statistical_1994, flekkoy_fluctuating_1996}:
\beq
 a \simeq\l \sqrt{\ln \frac{h_0}{\Young \ell}},
\eeq
where we used $h_0/\Young$ and $\l$ \cite{lekkerkerker_life_2008} as the small and large wavenumber cut-off, respectively. In $d=3$ dimensions,  the thermal length is defined as $\l=\sqrt{\frac{\kB T}{\gamma}}$. As before, the outer length $h_0$ is typically the capillary length. Considering first an ordinary molecular fluid, for which $h_0/\Young$ is of the order of millimeters, gives a scale separation ratio $h_0/\Young \l$ of around 7 decades, corresponding to $\sqrt{\ln(h_0/ \Young \l)} \simeq 4$. The regularization length is therefore typically a fraction of nanometer for ordinary fluids, as observed \cite{Israelachvili}. Note that \cref{eq_a_1d}, which pertains to the one-dimensional model, predicts a value of $a$ almost three orders of magnitude larger. We remark, however, that, if the interface binding potential is long-ranged -- which is the case when van der Waals interactions are present in the fluid --, the approximation of the large-scale physics by \rev{a contact potential} may seem questionable. Instead, the \rev{models considered here} are expected to be a more suitable effective description for systems with short-range binding potentials, such as colloidal fluids \cite{lekkerkerker_life_2008}. Indeed, in colloid-polymer mixtures van der Waals interactions are essentially absent and the surface tension is rather low, so that a value of $a$ in the micrometer regime seems realistic.

Interfacial and contact line fluctuations give rise to a regularization of the potential $f_{\TcalY}$ over a length $a$. 
Notably, it turned out that the large scale behavior of $f_\TcalY$ is essentially independent of the impenetrable character of the wall or the boundary condition for the fluctuation.
The latter instead only controls the scaling law obeyed by the effective potential in an inner boundary layer. For a bound interface ($\theta >0$), in the limit of small $h$ we have [see \cref{eq:f_small_h}]
\beq\label{f_small_h:real_variables}
f_{\TcalY}(h) \sim \frac{\Young^2h^2}{4\l^2} [1-(\theta/\Young)^2]^2.
\eeq
The influence of $\theta/\Young$ on the behavior of $f_{\TcalY}$ is felt over the characteristic length-scale 
$$
\eta \equiv \frac{\l}{\Young[1-(\theta/\Young)^2]},
$$
which according to \cref{eq:asymptotics_x_large} also governs the decay of the profile for $x \rightarrow \infty$. The length $\eta$ continuously diverges at the pseudo-partial wetting transition $\theta \nearrow \Young$. This divergence indicates that the boundary condition of the interface starts having an influence over a scale comparable to $a$ at the pseudo-partial wetting transition.

Except for the regularization scale $a$, a qualitative extension of \cref{f_small_h:real_variables} to a two-dimensional interface appears difficult. Indeed, \cref{f_small_h:real_variables} incorporates the decay in $x$ of the average profile, for which the impact of the dimensionality is \emph{a priori} unclear. Furthermore, inferring the influence of $\theta$ would require a better understanding of the nature of the parameter $\theta$ and its relation to the microscopic physics. This information is required to predict the conditions under which a pseudo-partial wetting transition can be observed for two dimensional interfaces. This is left for future studies.

The model introduced in the present study can be considered as the ``zeroth-order'' (equilibrium) problem in the dynamics of a thin film which advances on a planar wall via a fluctuating, but well-defined, contact line position \cite{gross_nanodrop_2014,colosqui_crossover_2013,colosqui_terraced_wetting_2015}. The identification of the boundary condition parameter $\theta$ by asymptotic matching with an inner layer description including the details of intermolecular interactions remains a challenging task for future studies. In a refined version of our model, also finite-range binding potentials may be considered. 
We have assumed the presence of a cutoff scale $h_0$ far above the substrate, at which interfacial fluctuations are suppressed and the Young's angle is exactly recovered as the slope of the average profile. In the presence of gravity, for instance, such a scale can be naturally identified with the capillary length \cite{rowlinson_molecular_1982}. Such a parameter may be dispensed with if, instead of a straight wedge, a droplet of fixed volume is considered.

\acknowledgements
We thank M.\ Popescu and L.\ Schimmele for useful discussions.
MS and DB kindly acknowledge funding from the European Research Council under the Europeans Community's Seventh Framework Programme (FP7/2007-2013) / ERC Grant Agreement N. 279004. 

\appendix

\section{Propagator}\label{app:av_prop}

Here we derive basic relations for the propagator. Although we focus in the main text exclusively on contact potentials, we consider in the following the more general case with a potential $u(h)$. The corresponding expressions that are valid for contact potentials can be recovered by setting $u(h)\equiv0$ (see also \cref{app:part_hav,app:cont_prop}).\\
Since the profile is restricted to the half-space $h\geq 0$, an appropriate definition for Dirac's delta function, which we denote by $\delta_q(h)$, has to be introduced. For any $q\geq0$, we set $\delta_{q}(h)=0$ if $h\neq q$, and require
\beq\label{eq:Dirac_h0}
\int_0^\infty dh\,\delta_{q}(h) = 1.
\eeq
We remark that the previous formula is understood to be valid also for $q=0$, while the usual definition of Dirac's delta function, denoted by $\delta(h)$, would give $1/2$ on the r.h.s.\ in that case. For $q\neq0$ one could equivalently write $\delta_q(h)=\delta(h-q)$ without a risk of confusion.\\
In the presence of a potential $u(h)$, the generalization of the diffusion equation~\eqref{eq:diff}, which has to be satisfied by $c(h_0,h;X)$, is given by \cite{feynman_book,chaichian_book, kleinert_book}
\beq\label{eq:diff_gen}
\l\frac{\pt c}{\pt X} = \frac{\l^2}{2}\frac{\pt^2c}{\pt h^2} - u(h) c, \hspace{.4in} c(h_0,h;0) = \delta_{h_0}(h),
\eeq
corresponding to the following path integral representation for the propagator, valid for any $X>0$:
\beq\label{eq:propagatorgeneral}
c(h_0,h;X) = \int_{q(0)=h_0}^{q(X)=h} \Dcal_\theta q\, e^{-\frac{1}{\l} \int_{0}^X ds\, \Square{\frac{1}{2}q'(s)^2 + u(q(s))}}.
\eeq
Recall that the notation $\Dcal_\theta h$ stands for the standard functional measure supplemented with the request of non-negativity of the profile (see, e.g., Refs.\ \cite{majumdar_brownian_2005,Rajabpour_2009,Farhi_1990,Clark_1980}), a property that will be further associated with a boundary condition  for the propagator \eqref{eq:prop_Robin} (see \cref{app:fixedR}). First, note that the propagator in \cref{eq:propagatorgeneral} is translation-invariant, in the sense that for $0<x<X$ we have
\beq\label{eq:prop_in}
c(h_0,h;X-x) = \int_{q(0)=h_0}^{q(X-x)=h} \Dcal_\theta q\, e^{-\frac{1}{\l} \int_{0}^{X-x} ds\, \Square{\frac{1}{2}q'(s)^2 + u(q(s))}} = \int_{Q(x)=h_0}^{Q(X)=h} \Dcal_\theta Q\, e^{-\frac{1}{\l} \int_{x}^X ds\, \Square{\frac{1}{2}Q'(s)^2 + u(Q(s))}},
\eeq
where $Q(s)=q(s-x)$ is a transformation that preserves both functional measure and non-negativity of the profiles. Furthermore, the definition~\eqref{eq:propagatorgeneral} is symmetric with respect to an exchange of $h_0$ and $h$, since
\beq\label{eq:h-h0_symm}
c(h,h_0;X) = \int_{q(0)=h}^{q(X)=h_0} \Dcal_\theta q\, e^{-\frac{1}{\l} \int_{0}^X ds\, \Square{\frac{1}{2}q'(s)^2 + u(q(s))}} = \int_{Q(0)=h_0}^{Q(X)=h} \Dcal_\theta Q\, e^{-\frac{1}{\l} \int_{0}^X ds\, \Square{\frac{1}{2}Q'(s)^2 + u(Q(s))}} = c(h_0,h;X)
\eeq
where the transformation $Q(s)=q(X-s)$ preserves both functional measure and non-negativity of the profiles.\\
Now, for $0<x<X$ we can use \cref{eq:Dirac_h0} 
\rev{with $q=q(x)$ inside the path integral in \cref{eq:propagatorgeneral}. Then,} due to the factor $\delta_{q(x)}(h)$, only those profiles such that $q(x)=h$ gives a non-vanishing contribution. The above path integral can then be factorized into the product of two path integrals, where one extends over all (non-negative) profiles with $q(0)=h_0$ and $q(x)=h$, and the other over all (non-negative) profiles with $q(x)=h$ and $q(X)=k$. Using \cref{eq:prop_in} directly leads to the {\it Chapman-Kolmogorov} equation \cite{chaichian_book}
\beq\label{eq:C-K_eq}
c(h_0,k;X) = \int_0^\infty dh\, c(h_0,h;x) c(h,k;X-x),
\eeq
valid for $0<x<X$.

\section{Pinned Contact Line Ensemble}\label{app:fixedR}

In the presence of a potential $u(h)$, the functional in \cref{eq:Ham} is 
\rev{easily generalized, in such a way} 
that the partition function~\eqref{eq:fixedRpart} becomes
\beq\label{eq_C0R_abbrev}
c(h_0,0;\R) = \int_{h(0)=h_0}^{h(\R)=0} \Dcal_\theta h\, e^{-\frac{1}{\l} \Hcal_\R[h]} = \int_{h(0)=h_0}^{h(\R)=0} \Dcal_\theta h\, e^{-\frac{1}{\l} \int_{0}^\R ds\, \Square{\frac{1}{2}h'(s)^2+u(h(s))}}.
\eeq
From \cref{eq:C-K_eq} it immediately follows that, for $x<\R$,
\beq\label{eq:C-K_eq_R}
c(h_0,0;\R) = \int_0^\infty dh\, c(h_0,h;x) c(h,0;\R-x).
\eeq
In the pinned contact line cases we define the average of a functional $\Fcal[h]$ over all the non-negative profiles $h=h(x)$ with $h(0)=h_0$ and $h(\R)=0$ as
\beq\label{eq:<F>_R}
\Angle{\Fcal}_\R \equiv \frac{1}{c(h_0,0;\R)}\int_{h(0)=h_0}^{h(\R)=0} \Dcal_\theta h\, \Fcal[h] e^{-\frac{1}{\l}\Hcal_\R[h]}.
\eeq
Now, let $F(h)$ be a function of $h$, and for any given position $x$ let
\beq\label{eq:Floc}
\Fcal[h](x)=F(h(x))
\eeq
be a functional depending on the profile only via its value $h(x)$ at that $x$. The average of $\Fcal[h](x)$ with pinned contact line [\cref{eq:<F>_R}] will be denoted by $\Angle{\Fcal(x)}_\R$. We remark that $x$ can be either smaller or larger than the contact line position $\R$. For $x<\R$, as in the derivation leading to the Chapman-Kolmogorov equation~\eqref{eq:C-K_eq}, we can use relation~\eqref{eq:Dirac_h0} to factorize the path integral defining $\Angle{\Fcal(x)}_\R$ and obtain
\beq\label{eq:Fav_x<R_loc}
\Angle{\Fcal(x)}_\R = \frac{1}{c(h_0,0;\R)}\int_0^\infty dh\, F(h) c(h_0,h;x) c(h,0;\R-x).
\eeq
While the choice $F(h)=1$ is in agreement with \cref{eq:C-K_eq_R}, the choice $F(h)=h$ yields the average profile in \cref{eq:profileR}. For $x\geq\R$, since $h(x)$ is assumed to be identically vanishing, we can simply write $\Fcal[h](x)=F(0)$, which is deterministic and thus equal to its average:
\beq\label{eq:Fav_R>x_loc}
\Angle{\Fcal(x)}_\R = F(0).
\eeq
\rev{Equations~\eqref{eq:Fav_x<R_loc} and~\eqref{eq:Fav_R>x_loc} will be used in the treatment of the fluctuating contact line ensemble in \cref{app:fluctuatingR}.}\\
We now characterize the probability density $\rho_\R(x,h)$ for finding the profile at position $x<\R$ with a height $h$ in the pinned contact line ensemble. This is the conditional probability density associated with the set of all the non-negative profiles $h=h(x)$ satisfying both the conditions $h(0)=h_0$ and $h(\R)=0$. By writing \cref{eq:Fav_x<R_loc} in the form
\beq\label{eq:FavR}
\Angle{\Fcal(x)}_\R = \int_0^\infty dh\, F(h) \rho_\R(x,h)
\eeq
we obtain
\beq\label{eq:prob_dens}
\rho_\R(x,h) \equiv \frac{c(h_0,h;x) c(0,h;\R-x)}{c(h_0,0;\R)}.
\eeq
where we used property~\eqref{eq:h-h0_symm} in order to write $c(h,0;\R-x)=c(0,h;\R-x)$. By computing the derivative with respect to $x$ of \cref{eq:prob_dens} and using \cref{eq:diff_gen}, we obtain a continuity equation of the form
\beq\label{eq:continuity}
\frac{\pt\rho_\R}{\pt x} + \frac{\pt j_\R}{\pt h} = 0,
\eeq
with the probability (density) flux $j_\R(x,h)$ given by
\beq \label{eq_cond_prob_flux}
j_\R(x,h) \equiv \frac{\l}{2}\rho_\R(x,h)\Square{\frac{\pt\ln c(0,h;\R-x)}{\pt h} - \frac{\pt\ln c(h_0,h;x)}{\pt h}}.
\eeq
Note that the associated initial condition for \cref{eq:continuity} is well posed because of the validity of relation \eqref{eq:Dirac_h0} for $q=0$. Requiring that $j_\R(x,0)$ vanishes for any $x<\R$ implies
\beq\label{eq:j=0}
\left.\frac{\partial\ln c(h_0,h;x)}{\partial h}\right|_{h=0} = \left.\frac{\partial\ln c(0,h;\R-x)}{\partial h}\right|_{h=0},
\eeq
which leads to the condition in \cref{eq:prop_Robin}.
Note that, in \cref{eq:j=0}, $h_0$ is a free parameter that does not appear on the r.h.s.\ Thus, for \cref{eq:j=0} to be generally valid, the l.h.s.\ must be independent of $h_0$ as well. Consequently, the parameter $\theta$ in \cref{eq:prop_Robin} must be independent of $h_0$. Analogously, a dependence of \cref{eq:j=0} on $x$ can be ruled out based on the manifest independence of the l.h.s.\ of $\R$. Therefore, $\theta$ in \cref{eq:prop_Robin} must also be independent of $X$.\\
We finally discuss the slope of the average profile in the pinned contact line cases. From \cref{eq:FavR} we get
\beqn
\frac{d\Angle{\Fcal(x)}_\R}{dx} = \int_0^\infty dh\, \frac{dF}{dh} j_\R(x,h),
\eeqn
where the continuity equation~\eqref{eq:continuity} has been used, together with an integration by parts \footnote{The latter requires suitable convergence conditions on $F(h)$ to ensure the vanishing of the boundary terms, which, however, are always satisfied here.}. 
In particular, for the choice $F(h)=h$ we obtain
\beqn
\frac{d\Angle{h(x)}_\R}{dx} = \frac{\l}{c(h_0,0;\R)}\int_0^\infty dh\,c(h_0,h;x)\frac{\pt c(h,0;\R-x)}{\pt h} + \frac{\l}{2}\rho_\R(x,0),
\eeqn
where the property~\eqref{eq:h-h0_symm} has again been used. Evaluating above expression at $x=0$, using $c(h_0,h;0)=\delta_{h_0}(h)$ (and hence $\rho_\R(0,0)=0$, being $h_0$ {\it strictly} positive), yields
\cref{eq:dh_R/dx}.

\section{Fluctuating Contact Line Ensemble}\label{app:fluctuatingR}

In the presence of a potential $u(h)$, the partition function of the fluctuating contact line ensemble is still given by \cref{eq:Z}, provided \cref{eq:fixedRpart} is replaced with \cref{eq_C0R_abbrev}. It should be stressed that, due to the integration over $\R$, the potential can no longer be shifted by an arbitrary constant: such a shift of the potential would produce an extra exponential prefactor in \cref{eq_C0R_abbrev}, resulting in a modification of the actual value of $\Young^2/2$ in \cref{eq:Z}. In order to avoid this, we henceforth fix the offset of the potential by requiring $u(h)\to0$ for $h\to\infty$, such that the expression valid for the contact potential is recovered by setting $u(h)\equiv0$.\\
In order to control a possible divergence of the outer integral in \cref{eq:Z} for large $\R$ (see also \cref{app:part_hav}), we consider the following ``infrared'' regularized quantity:
\begin{align}\label{eq:Z_h0}
&Z_X(h_0) = \int_0^X d\R\, e^{-\frac{\Young^2\R}{2\l}} c(h_0,0;\R), &&Z_\infty(h_0) \equiv Z.
\end{align}
For any $x>0$ and for sufficiently large $X$, we have
\beqn
\begin{split}Z_X(h_0) &= \int_x^X d\R\, e^{-\frac{\Young^2\R}{2\l}} c(h_0,0;\R) + \int_0^x d\R\, e^{-\frac{\Young^2\R}{2\l}} c(h_0,0;\R) = e^{-\frac{\Young^2x}{2\l}} \int_0^\infty dh\,c(h_0,h;x) Z_{X-x}(h) + Z_x(h_0),\end{split}
\eeqn
where \cref{eq:C-K_eq_R} has been used. 
Dividing by $Z_X(h_0)$ and taking the limit $X\to\infty$ we obtain
\rev{\begin{align}\label{eq:czeta_norm}
&\int_{0}^{\infty} dh\, c(h_0,h;x) \zeta(h,h_0;x) + \tfrac{Z_x(h_0)}{Z} = 1, &&\zeta(h,h_0;x) \equiv e^{-\frac{\Young^2x}{2\l}} \lim_{X\to\infty}\tfrac{Z_{X-x}(h)}{Z_X(h_0)}.
\end{align}}
The condition~\eqref{eq:czeta_norm} expresses the conservation of probability in the fluctuating contact line ensemble. This meaning of \cref{eq:czeta_norm} can be further understood by analyzing the associated probability density, as discussed in \cref{app:fluctuatingR} [see, in particular, \cref{eq:dens_comp}]. 
\rev{The r.h.s. in the definition of $\zeta(h,h_0;x)$ can be manipulated into}
\beq
e^{-\frac{\Young^2x}{2\l}} \lim_{X\to\infty}\tfrac{Z_{X-x}(h)}{Z_X(h_0)} = e^{-\frac{\Young^2y}{2\l}} e^{-\frac{\Young^2(x-y)}{2\l}} \lim_{X\to\infty}\frac{\frac{Z_{X-y}(k)}{Z_X(h_0)}}{\frac{Z_{X-y}(k)}{Z_{X-x}(h)}} = \frac{e^{-\frac{\Young^2y}{2\l}}\lim_{X\to\infty}\frac{Z_{X-y}(k)}{Z_X(h_0)}}{e^{-\frac{\Young^2(y-x)}{2\l}}\lim_{X\to\infty}\frac{Z_{X+x-y}(k)}{Z_X(h)}},
\eeq
enabling us to write, for any $k\geq0$ and $y>x$, the following useful relation:
\beq\label{eq:zeta_prop}
\zeta(k,h_0;y) = \zeta(k,h;y-x) \zeta(h,h_0;x).
\eeq
According to the definition~\eqref{eq:Z_h0}, the average of an arbitrary function $\varphi(\R)$ of the fluctuating variable $\R$ in the fluctuating contact line ensemble can now be defined as
\beqn
\Angle{\varphi} \equiv \lim_{X\to\infty}\tfrac{1}{Z_X} \int_0^X d\R\, \varphi(\R)e^{-\frac{\Young^2\R}{2\l}} c(h_0,0;\R).
\eeqn
In particular, by choosing $\varphi(\R)=\R$ we obtain (the infrared regularized version of) \cref{eq:Rav_general}. \rev{Furthermore, the function $\varphi(\R)=\Angle{\Fcal}_\R$ given in \cref{eq:<F>_R} can be averaged over $\R$ to obtain the corresponding average in the fluctuating contact line ensemble
.} In particular, as in the derivation of \cref{eq:czeta_norm}, for any fixed $x$ the average of the functional $\Fcal[h](x)$ in \cref{eq:Floc}, denoted by $\Angle{\Fcal(x)}$, can be written as
\beq
\begin{split}\Angle{\Fcal(x)} &= \lim_{X\to\infty}\tfrac{1}{Z_X} \int_x^X d\R\, \Angle{\Fcal(x)}_\R e^{-\frac{\Young^2\R}{2\l}} c(h_0,0;\R) + \tfrac{1}{Z} \int_0^x d\R\, \Angle{\Fcal(x)}_\R e^{-\frac{\Young^2\R}{2\l}} c(h_0,0;\R),\end{split}
\eeq
which, by making use of \cref{eq:Fav_x<R_loc,eq:Fav_R>x_loc,eq:czeta_norm}, becomes
\beq\label{eq:FLocav_dh}
\Angle{\Fcal(x)} = \int_{0}^{\infty} dh\, F(h) c(h_0,h;x) \zeta(h,h_0;x) + \tfrac{Z_x(h_0)}{Z}F(0).
\eeq
\rev{In particular, the choice $F(h)=1$ is in agreement with \cref{eq:czeta_norm}.} 
Equation \eqref{eq:FLocav_dh}, together with the expression 
\beqn
\Angle{\Fcal(x)} = \int_{0}^{\infty} dh\, F(h) \rho(x,h)
\eeqn
for the average of $\Fcal[h](x)$, allows us to define the probability density in the fluctuating contact line ensemble as
\beq\label{eq:dens_comp}
\rho(x,h) \equiv c(h_0,h;x) \zeta(h,h_0;x) + \tfrac{Z_x(h_0)}{Z}\delta_0(h).
\eeq
By integrating the previous equation over all the possible realizations of $h$ we correctly obtain $1$ due to the constraint~\eqref{eq:czeta_norm}. Next, we provide a number of relevant properties of the two-point correlation function. By following the same strategy leading to \cref{eq:FLocav_dh}, the correlations in the fluctuating contact line ensemble can be written, for $x<y$, as
\beq\label{eq:corr}
\Angle{h(x)h(y)} = \int_0^\infty dh\, h c(h_0,h;x)\int_{0}^{\infty} dk\, k c(h,k;y-x) \zeta(k,h_0;y) = \int_0^\infty dh\, h \Angle{h(y-x)}_{h_0=h} c(h_0,h;x) \zeta(h,h_0;x),
\eeq
\rev{where the last inequality is obtained using the property~\eqref{eq:zeta_prop}. Note that \cref{eq:corr} evaluated for $y=x$ and \cref{eq:FLocav_dh} with $F(h)=h^2$ coincide}. Although not necessary for the present study, we remark that correlations between $h(x)$ and $h'(x)$ can be defined from a suitable limit of the mixed derivative of the correlation function:
\begin{align}\label{eq:ordering}
&\Angle{h(x)h'(x)} \equiv \lim_{y\searrow x}\frac{\pt\Angle{h(x)h(y)}}{\pt y}, &&\Angle{h'(x)h(x)} \equiv \lim_{y\searrow x}\frac{\pt\Angle{h(x)h(y)}}{\pt x}.
\end{align}
The previous expressions are not equal, reflecting the ordering problem which is well known to come out from the path integral description \cite{Feynmanwiki}. We will discuss more in details on this issue in \cref{app:part_hav}, were explicit expressions for contact potentials are obtained.\\
Finally, it may be of interest to show how the previous results can be derived from a more general framework. To this aim, we consider functionals of the form
\beq\label{eq:F_x=intPhi}
\Fcal_x[h]=\int_0^x ds\, \Phi\Round{h(s),h'(s)},
\eeq
with $\Phi(h,h')$ being some regular function of its arguments $h$ and $h'$. We find
\rev{\beqn
\Angle{\Fcal_x} = \int_{0}^{\infty} dh\, \Angle{\Fcal_x}_x(h) c(h_0,h;x) \zeta(h,h_0;x) + \tfrac{1}{Z} \int_0^x d\R\, \Angle{\Fcal_\R}_\R e^{-\frac{\Young^2\R}{2\l}} c(h_0,0;\R)+ \tfrac{\Phi\Round{0,0}}{Z} \int_0^x d\R\, (x-\R)e^{-\frac{\Young^2\R}{2\l}} c(h_0,0;\R),
\eeqn}
where we introduced the notation
\begin{align}\label{eq:F[.,R](h)}
&\Angle{\Fcal}_x(k) \equiv \tfrac{1}{c(h_0,k;x)}\int_{h(0)=h_0}^{h(x)=k} \Dcal_\theta h\, \Fcal[h] e^{-\frac{1}{\l}\Hcal_x[h]}, &&\Angle{\Fcal}_x(0)\equiv\Angle{\Fcal}_x,
\end{align}
which is actually a generalization of the definition~\eqref{eq:<F>_R} with a non-vanishing ending height. \rev{In particular, for a functional of the form~\eqref{eq:Floc} we can write $\Angle{\Fcal(x)}_x(k) = F(k)$, as directly follows from the definition~\eqref{eq:F[.,R](h)}
.} Then, we note that \cref{eq:Floc} can be rewritten as
\begin{align}
&\Fcal[h](x)=F(h_0)+\int_0^x ds\, \Phi_F\Round{h(s),h'(s)}, &&\Phi_F(h,h')=h'\frac{dF}{dh},
\end{align}
showing that $\Fcal[h](x)-F(h_0)$ is a functional of the form~\eqref{eq:F_x=intPhi}. \rev{Thus, the results for~\eqref{eq:Floc} are only a particular case of the results for~\eqref{eq:F_x=intPhi}, and can be obtained by applying the latter to the difference $\Fcal_x[h]=\Fcal[h](x)-F(h_0)$.}

\section{Contact Potential: Pinned Contact Line Ensemble}\label{app:cont_prop}

In this appendix we derive the expression for the contact propagator reported in \cref{eq:propagator} by solving the associated eigenvalue equation~\eqref{eq:eigen}, subject to the Robin boundary condition~\eqref{eq:Robin} [see also \cite{burkhardt_propagator_1989}]. We furthermore derive here the results of section~\ref{sec:fixed} pertaining to the pinned contact line problem.\\
There exists no eigenstate for $\e=0$. The positive energy eigenvalues ($\e>0$) of \cref{eq:eigen} can be parametrized by $\e=\phi^2/2$ with a non-vanishing real parameter $\phi$. The corresponding eigenstates $\psi_\phi(h)\equiv\psi_{\e=\phi^2/2}(h)$ can then be written as \cite{ohya_path_2012}
\beqn
\psi_\phi(h) = \tfrac{1}{\sqrt{2\pi\l}}\Round{e^{-\frac{i\phi h}{\l}}+\tfrac{i\phi-\theta}{i\phi+\theta}e^{\frac{i\phi h}{\l}}}.
\eeqn
Since $\psi_{-\phi}(h)$ can be obtained from $\psi_\phi(h)$ by means of a unitary transformation \cite{ohya_path_2012}, the linearly independent eigenstates are those for $\phi>0$. Furthermore, if (and only if) $\theta>0$, there exists an eigenstate having a negative energy eigenvalue: the bound eigenstate corresponding to $\e=\e_0\equiv-\theta^2/2$ is given by \cite{ohya_path_2012}
\beqn
\psi_0(h) = \tfrac{1+\sgn\theta}{2} \sqrt{\tfrac{2\theta}{\l}} e^{-\frac{\theta h}{\l}},
\eeqn
$\sgn\theta$ being the sign of $\theta$. 
Note that the bound state is denoted by $\psi_0\equiv\psi_{\e=\e_0}$ and should not be confused with $\psi_{\phi=0}\equiv\psi_{\e=0}$, the latter being identically vanishing. This notation allows us to write the orthonormality relation in a compact form (a star $^*$ indicates complex conjugation):
\beqn
\int_0^\infty dh\, \psi_\varphi^*(h) \psi_\phi(h) = \delta_{\phi,0}\delta_{\varphi,0} + (1-\delta_{\phi,0}\delta_{\varphi,0})\delta(\phi-\varphi),
\eeqn
with $\delta_{\phi,\varphi}$ being Kronecker's delta. The above defined set of eigenstates is the largest possible, since the completeness relation
\beqn
\psi_0(h)\psi_0^*(k) + \int_0^\infty d\phi\, \psi_\phi(h)\psi_\phi^*(k) = \delta(h-k),
\eeqn
is satisfied. We remark that Dirac's delta $\delta(h-k)$ in the previous equation is understood to be $\delta_k(h)$, or $\delta_h(k)$, as explained in \cref{app:av_prop}. The propagator follows from \cref{eq_prop_eigen_decomp} as 
\rev{\beqn
\begin{split}c(h_0,h;X) &= \psi_0^*(h_0) \psi_0(h) e^{-\frac{\e_0 X}{\l}} + \int_0^\infty d\phi\, \psi_\phi^*(h_0) \psi_\phi(h) e^{-\frac{\phi^2 X}{2\l}}\\
&= \tfrac{1+\sgn\theta}{2} \tfrac{2\theta}{\l} e^{-\frac{\theta(h+h_0)}{\l}+\frac{\theta^2X}{2\l}} + \tfrac{1}{2\pi\l} \tfrac{1}{\sqrt{\pi}} \int_{-\infty}^{+\infty} du\, e^{-u^2} \int_{-\infty}^\infty d\phi\, \Square{e^{\frac{i\phi(h_0-h)}{\l}}+\tfrac{i\phi-\theta}{i\phi+\theta}e^{\frac{i\phi(h_0+h)}{\l}}}e^{-iu\phi\sqrt{\frac{2X}{\l}}},\end{split}
\eeqn}
which, when evaluated using the contour integration technique, results in \cref{eq:propagator}.\\
For the contact potential case, the probability density~\eqref{eq:prob_dens} 
\rev{can easily be written down by using for the propagators their explicit expression \eqref{eq:propagator}.} The case $\theta>0$ is extensively discussed in \cref{sec:fixed}, where it is shown that, for a sufficiently large $R$, a wedge-like structure emerges, and a universal rescaling of variables is then possible by using the wedge length as a universal cut-off for capillary waves in the limit $R \to \infty$. By definition, such a rescaling is independent of $R$. \rev{Explicitly, one gets
\beqn
\rho_\infty(x,h) = \tfrac{1+e^{-\frac{2hh_0}{\l x}}}{\sqrt{2\pi \l x}}e^{-\frac{(h-h_0+\theta x)^2}{2\l x}} + \tfrac{\theta}{\l} e^{-\frac{2\theta h}{\l}}\erfc\Round{\tfrac{h+h_0-\theta x}{\sqrt{2\l x}}}.
\eeqn
Integration of $h\rho_\infty(x,h)$ over $h\geq0$ then gives \cref{eq:hav_theta}.} The situation $\theta \le 0$ leads to a different morphology. We illustrate this for the special cases of reflecting ($\theta=0$, denoted by ``ref'') and absorbing ($\theta=-\infty$, denoted by ``abs'') boundary conditions, because these lend themselves to an analytical treatment. The probability densities are given by
\rev{\begin{align*}
&\rho_\R^{\st{ref}}(x,h) = \tfrac{e^{-\frac{(h-h_0)^2}{2\l x}}+e^{-\frac{(h+h_0)^2}{2\l x}}}{\sqrt{2\pi \l x(1-x/\R)}}e^{\frac{h_0^2}{2\l\R}-\frac{h^2}{2\l (\R-x)}}, &&\rho_\R^{\st{abs}}(x,h) = \tfrac{1}{1-x/\R}\tfrac{h}{h_0}\rho_\R^{\st{ref}}(x,h),
\end{align*}}
respectively, which, when inserted in \cref{eq:FavR} with $F(h)=h$, yield
\beq\label{eq:havref}
\Angle{h(x)}^{\st{ref}}_\R = h_{0}\Round{1-\tfrac{x}{\R}} \erf\Round{\sqrt{\tfrac{h_{0}^2}{2\l x}\Round{1-\tfrac{x}{\R}}}} + \sqrt{\tfrac{2\l x}{\pi}\Round{1-\tfrac{x}{\R}}}e^{-\frac{h_{0}^{2}}{2\l x}\Round{1-\frac{x}{\R}}}
\eeq
and
\beq\label{eq:havabs}
\Angle{h(x)}^{\st{abs}}_\R = \Angle{h(x)}^{\st{ref}}_\R + \tfrac{\l x}{h_0} \erf\Round{\sqrt{\tfrac{h_{0}^2}{2\l x}\Round{1-\tfrac{x}{\R}}}},
\eeq
respectively, where $\erf(z)=1-\erfc(z)$. As seen from \cref{eq:havref,eq:havabs}, in the limit of large $R$, no wedge structure survives. Hence the characteristic outer scale $\lambda_\st{M}$ used for the rescaling grows with $R$, yielding a rescaling that differs from the one adopted for $\theta>0$ in \cref{sec:fixed}. Following \cite{burkhardt_propagator_1989}, for a finite $R$ the appropriate rescaling of variables is given by
\beq
\bar{h}=\frac{h}{\sqrt{R \l}}, \hspace{.2in} \bar{x}= \frac{x}{R}, \hspace{.2in}  \bar{\theta}=\theta\sqrt{\frac{R}{\l}}.
\eeq
The profiles in \cref{eq:havref,eq:havabs} are then rescaled as
\beq\label{eq:havref_resc}
\Angle{\bar{h}(\bar{x})}^{\st{ref}}_1 = \bar{h}_{0}\Round{1-\bar{x}} \erf\Round{\sqrt{\tfrac{\bar{h}_{0}^{2}}{2 \bar{x}}(1-\bar{x})}} + \sqrt{\tfrac{2 \bar{x}}{\pi}(1-\bar{x})}e^{-\frac{\bar{h}_{0}^{2}}{2 \bar{x}}(1-\bar{x})}
\eeq
and
\beq\label{eq:havabs_resc}
\Angle{\bar{h}(\bar{x})}^{\st{abs}}_1 = \Angle{\bar{h}(\bar{x})}^{\st{ref}}_1+ \tfrac{\bar{x} }{\bar{h}_0} \erf\Round{\sqrt{\tfrac{\bar{h}_{0}^{2}}{2\bar{x}}(1-\bar{x})}},
\eeq
showing that for $\theta \le 0$ we have two dimensionless control parameters: $\bar{h}_0$ and $\bar{\theta}$, the latter being $\bar{\theta}=0$ for \cref{eq:havref_resc} and $\bar{\theta}=-\infty$ for \cref{eq:havabs_resc}. For a fixed $h_0$, in the limit $R \to \infty$, we find $\bar{h}_0 \to 0$ and $\bar{\theta}$ remains as the only dimensionless control parameter. The corresponding profiles are the ``bridge'' profiles already discussed in \cite{burkhardt_propagator_1989}.

\section{Contact Potential: Fluctuating Contact Line Ensemble}\label{app:part_hav}

Here, we derive expressions for the partition function [\cref{eq:Z}] as well as for the one- and two-point correlation functions in the fluctuating contact line ensemble, assuming the presence of a pure contact potential. We also pay attention to possible infrared ($\R\to\infty$) divergences. \\
We first determine the restrictions on the parameter $\theta$ that ensure a finite contact line position in the classical limit $\l \to 0$. To this end, we rewrite \cref{eq:Z} as
\rev{\begin{align}\label{eq:Znow}
&Z = \int_0^\infty d\R\, e^{-\frac{1}{\l}A(\R)}, &&A(\R)\equiv \tfrac{\Young^2}{2}\R-\l\ln c(h_0,0;\R).
\end{align}
In order to obtain a finite contact line position in the limit $\l \to 0$, the quantity 
$A(\R)$} must exhibit a minimum at a certain finite $\R$, at least for $\l$ sufficiently small. This is not the case if $A(\R)$ is a monotonically decreasing function. To further explore this possibility, notice that for $\R\to0$ we have $c(h_0,0;\R)\to \delta_{h_0}(0) = 0$ (with $h_0$ being {\it strictly} positive). Thus, $A(\R)$ turns out to diverge to $\infty$ for small $\R$ \footnote{This excludes a possible divergence of the integrand in \cref{eq:Znow} for $\R\to0$.}. On the other hand, the asymptotic behavior for large $\R$ depends on the model chosen. In our case it depends on the parameter $\theta$, and from \cref{eq:fixedRpart_expl} we get
\beq\label{eq:csim}
c(h_0,0;\R) \overset{\R\to\infty}{\sim} \begin{dcases} \tfrac{h_0}{|\theta| \R}\Round{1+\tfrac{\l}{|\theta| h_0}}\tfrac{2e^{-\frac{h_0^2}{2\l\R}}}{\sqrt{2\pi\l\R}} &\textup{if } \theta \leq0,\\ \tfrac{2\theta}{\l} e^{-\frac{\theta h_0}{\l}+\frac{\theta^2\R}{2\l}} &\textup{if } \theta > 0,\end{dcases}
\eeq
where the asymptotic expression in \cref{eq:erfc} has been used. As a consequence,
\beq
A(\R) \overset{\R\to\infty}{\sim} \begin{dcases} \tfrac{\Young^2}{2}\R &\textup{if } \theta \leq0,\\ \tfrac{\Young^2-\theta^2}{2}\R &\textup{if } \theta > 0.\end{dcases}
\eeq
Thus, if $\theta<\Young$, $A(\R)$ diverges to $\infty$ for large $\R$, implying that a certain finite $\R>0$ exists for which $A(\R)$ is minimum. In the remaining case $\theta\geq \Young$ we note that
\beq\label{eq:dlnc/dR}
\l \frac{\partial\ln c(h_0,0;\R)}{\partial \R} = \frac{\theta^2}{2} + \frac{h_0^2+(\theta h_0-\l)\R}{c(h_0,0;\R)\R^2}\frac{e^{-\frac{h_0^2}{2\l\R}}}{\sqrt{2\pi \l\R}} > \frac{\theta^2}{2},
\eeq
where the inequality holds for any $\R$ and $\l<\theta h_0$, i.e., if the thermal length is sufficiently small. In this case, we obtain
\beq
\frac{dA}{d\R} = \frac{\Young^2}{2}-\l \frac{\partial\ln c(h_0,0;\R)}{\partial \R} < \frac{\Young^2-\theta^2}{2},
\eeq
which shows that, if $\theta\geq\Young$ (and $\l<\theta h_0$), $dA/d\R$ is strictly negative and hence $A(\R)$ is strictly monotonically decreasing, thus excluding the existence of a minimum.\\
We now proceed to the calculation of the partition function. By using \cref{eq:fixedRpart_expl}, the regularized partition function in \cref{eq:Z_h0} can be cast into
\rev{\beq
Z_X(h_0) 
= \tfrac{2}{\Young}\int_0^\xi d\rho\, \tfrac{1}{\sqrt{\pi \rho}}e^{-\rho-\frac{\lambda^2}{\rho}} + \tfrac{2\tau}{\Young} \int_0^\xi d\rho\, e^{-2\lambda\tau+(\tau^2-1)\rho}\erfc\Round{\tfrac{\lambda-\tau \rho}{\sqrt{\rho}}},
\eeq}
where the integration variable has been non-dimensionalized by writing $\rho=\Young^2\R/2\l$, and, accordingly,
\begin{align}
&\xi=\tfrac{\Young^2X}{2\l}, &&\lambda=\tfrac{\Young h_0}{2\l}, &&\tau=\tfrac{\theta}{\Young}.
\end{align}
The first integral gives
\beq
\begin{split}\int_0^\xi d\rho\, \tfrac{1}{\sqrt{\pi \rho}}e^{-\rho-\frac{\lambda^2}{\rho}} &= \tfrac{1}{2}e^{-2\lambda}\erfc\Round{\tfrac{\lambda-\xi}{\sqrt{\xi}}}-\tfrac{1}{2}e^{2\lambda}\erfc\Round{\tfrac{\lambda+\xi}{\sqrt{\xi}}} \overset{\xi\to\infty}{\sim} e^{-2\lambda},\end{split}
\eeq
while the second integral gives
\beq\label{eq:int}
\begin{split}\int_0^\xi d\rho\, e^{-2\lambda\tau+(\tau^2-1)\rho}\erfc\Round{\tfrac{\lambda-\tau \rho}{\sqrt{\rho}}} &= \tfrac{1}{2}\tfrac{e^{-2\lambda}\erfc\Round{\frac{\lambda-\xi}{\sqrt{\xi}}}}{1-\tau}+\tfrac{1}{2}\tfrac{e^{2\lambda}\erfc\Round{\frac{\lambda+\xi}{\sqrt{\xi}}}}{1+\tau}+\tfrac{e^{-2\lambda\tau+(\tau^2-1)\xi}\erfc\Round{\frac{\lambda-\tau\xi}{\sqrt{\xi}}}}{\tau^2-1}\\
&\overset{\xi\to\infty}{\sim} \tfrac{e^{-2\lambda}}{1-\tau}+\tfrac{(1+\sgn\tau)e^{-2\lambda\tau+(\tau^2-1)\xi}}{\tau^2-1}.\end{split}
\eeq
The asymptotic result in \cref{eq:int} shows that, for a fixed $\tau$, the integral is finite for $\tau<1$, linearly diverging with $\xi$ for $\tau=1$, and exponentially diverging with $\xi$ for $\tau>1$. We then have
\beq
\begin{split}Z_X(h_0) &= \tfrac{\erfc\Round{\frac{h_0-\Young X}{\sqrt{2\l X}}}}{\Young-\theta}e^{-\frac{\Young h_0}{\l}}-\tfrac{\erfc\Round{\frac{h_0+\Young X}{\sqrt{2\l X}}}}{\Young+\theta}e^{\frac{\Young h_0}{\l}}+\tfrac{2\theta \erfc\Round{\frac{h_0-\theta X}{\sqrt{2\l X}}}}{\theta^2-\Young^2}e^{-\frac{\theta h_0}{\l}+\Round{\theta^2-\Young^2}\frac{X}{2\l}}\\
&\overset{X\to\infty}{\sim} \tfrac{2}{\Young-\theta} e^{-\frac{\Young h_0}{\l}} + \tfrac{2\theta(1+\sgn\theta)}{\theta^2-\Young^2}e^{-\frac{\theta h_0}{\l}+\Round{\theta^2-\Young^2}\frac{X}{2\l}}.\end{split}
\eeq
Three cases can be distinguished:
\rev{\begin{itemize}
\item $\theta<\Young$. Here, we obtain $Z_X(h_0) \overset{X\to\infty}{\sim} \tfrac{2}{\Young-\theta} e^{-\frac{\Young h_0}{\l}}$, resulting in the well defined partition function $Z\equiv Z_\infty(h_0)$ in \cref{eq:partitionfunctionTEXT}.
\item $\theta=\Young$. Here, we obtain $Z_X(h_0) \overset{X\to\infty}{\sim} \tfrac{2\Young}{\l}X e^{-\frac{\Young h_0}{\l}}$, resulting in a partition function that linearly diverges in the infrared, i.e., when the upper integration boundary approaches infinity.
\item $\theta>\Young$. Here, we obtain $Z_X(h_0) \overset{X\to\infty}{\sim} \tfrac{4\theta}{\theta^2-\Young^2}e^{-\frac{\theta h_0}{\l}+\Round{\theta^2-\Young^2}\frac{X}{2\l}}$, resulting in a partition function that exponentially diverges in the infrared.
\end{itemize}
In the limit $X\to\infty$, this results can be conveniently grouped together by writing the inverse of $Z\equiv Z_\infty(h_0)$ as:}
\begin{align}\label{eq:out}
& &&\tfrac{1}{Z} = \tfrac{\out-\theta}{2} e^{\frac{\out h_0}{\l}}, &&\out \equiv \max(\Young,\theta).
\end{align}
Equation~\eqref{eq:czeta_norm} takes the form
\begin{align}\label{eq:czeta_norm_mu}
&\int_{0}^{\infty} dh\, c(h_0,h;x) \zeta(h,h_0;x) = 1 - \tfrac{\out-\theta}{2} e^{\frac{\out h_0}{\l}}Z_x(h_0), &&\zeta(h,h_0;x) = e^{-\frac{\out^2x}{2\l}}e^{-\frac{\out (h-h_0)}{\l}},
\end{align}
and the rest of the computations can be straightforwardly carried out, recovering the three cases $\theta<\Young$, $\theta=\Young$ and $\theta>\Young$ at the end by choosing $\out$ according to \cref{eq:out}. Note that the propriety~\eqref{eq:zeta_prop} is satisfied by $\zeta(h,h_0;x)$ in \cref{eq:czeta_norm_mu}. 
\rev{While the integral in \cref{eq:FLocav_dh} can now be solved for $F(h)=h^p$ ($p>0$), we show the result for $p=1$ only. We obtain}
\beq\label{eq:hav_out}
\begin{split}\Angle{h(x)} &= \tfrac{\theta \l}{(\out+\theta)^2}e^{(\out-\theta)\frac{h_0}{\l}+(\theta^2-\out^2)\frac{x}{2\l}}\erfc\Round{\tfrac{h_0-\theta x}{\sqrt{2\l x}}}+\tfrac{h_0-\out x}{2}\erfc\Round{\tfrac{\out x-h_0}{\sqrt{2\l x}}}\\
&\quad- \Square{\tfrac{\theta \l}{(\out+\theta)^2}+\tfrac{\out-\theta}{\out+\theta}\tfrac{h_0+\out x}{2}}e^{\frac{2\out h_0}{\l}}\erfc\Round{\tfrac{h_0+\out x}{\sqrt{2\l x}}}+\tfrac{2\out}{\out+\theta}\sqrt{\tfrac{\l x}{2\pi}}e^{-\frac{(h_0-\out x)^2}{2\l x}},\end{split}
\eeq
resulting in the average profile in \cref{eq:hav_theta<alpha} if $\theta<\Young$, for which $\out\equiv\Young$. For $\theta\geq\Young$, instead, where $\out\equiv\theta$, \cref{eq:hav_out} reproduces the average profile $\Angle{h(x)}_\infty$ in \cref{eq:hav_theta}. This is not surprising, since for such a case the partition function diverges in the infrared and the resulting average contact line position goes to infinity (see \cref{sec:hxR}). Thus, for $\theta\geq\Young$, a flat precursor film is present and any information about $\Young$ is completely lost. 

The correlations between $h(x)$ and $h'(x)$ defined in \cref{eq:ordering} can be computed analytically [this is not the case for \cref{eq:corr}]:
\begin{align}
&\Angle{h(x)h'(x)} = -\out \Angle{h(x)}, &&\Angle{h'(x)h(x)} = -\out\Angle{h(x)} + \l \Round{1-\tfrac{\out-\theta}{2} e^{\frac{\out h_0}{\l}}Z_x(h_0)},
\end{align}
where \cref{eq:czeta_norm_mu} has been used. 
Note that, consequently,
\beq\label{eq:comm_h'h}
\Angle{h'(x)h(x)} - \Angle{h(x)h'(x)} = \l \Round{1-\tfrac{\out-\theta}{2} e^{\frac{\out h_0}{\l}}Z_x(h_0)}.
\eeq
The first contribution to the nonvanishing commutator is the constant $\l$, which reflects the usual ordering problem of the path integral description \cite{Feynmanwiki}. The second contribution in \cref{eq:comm_h'h} is related to the correction term $\tfrac{\out-\theta}{2} e^{\frac{\out h_0}{\l}}Z_x(h_0)$, which is due to the presence of the wall and the contact line. If $\theta < \Young$, and hence $\out\equiv\Young$, in the limit $x\to\infty$ this contribution cancels the first, hence we obtain $\Angle{h'(x)h(x)} - \Angle{h(x)h'(x)} \to 0$. This shows that the magnitude of the fluctuations of the interface diminishes upon increasing $x$ -- which is expected since the profile approaches the wall. If $\theta\geq\Young$, such a correction term is not present because $\out\equiv\theta$. This is an indication that in this case interfacial fluctuations remain of similar strength as $x\to\infty$. Indeed, $\theta=\Young$ marks the onset of the pseudo-partial-wetting regime, where a flat film covers the substrate.\\

\section{Basic Results for Square Well Potential}\label{app:squarewell}


In this section we recall a number of basic results for square well potentials in quantum mechanics \cite{shankar_QM_book, burkhardt_localisation_1981, fisher_review_membranebook}. Our main aim is to interpret the emergence of bound states as a competition between the thermal length $\l$ and the characteristic range $\sigma$ of the binding potential. We focus on square well potentials since these can be considered as a first-order approximation to general short ranged binding potentials (see, e.g., Ref.\ \cite{gopalakrishnan2006self} for more information on this idealization). 
The generalization of \cref{eq:eigen} to a square well potential (denoted by ``sw'') is given by
\beq\label{eq:eigen_u}
-\frac{\l^2}{2}\psi'' + u_\st{sw}(h/\sigma) \psi = \e \psi,
\eeq
with
\beq\label{eq:squarewell}
u_\st{sw}(h/\sigma) = \begin{dcases} \infty &\textup{if } h < 0, \\  -\Young^2/2 &\textup{if } 0<h<\sigma, \\   0  &\textup{if }  h > \sigma.\end{dcases}
\eeq
The (negative) energy $\e$ of a bound state can be parametrized as $\e=-\nu^2/2$, with $0<\nu<\Young$ (for $\nu=\Young$ the eigenstate identically vanishes). 
\rev{The admissible solutions, fulfilling $\psi(0)=0$ and $\psi(h)\to0$ for $h\to\infty$, are
\beqn
\begin{dcases}
\psi_\nu(h) = A_\nu\sin(h\sqrt{\Young^2 - \nu^2}/\l) &\textup{if } 0<h<\sigma,\\
\psi_\nu(h) = B_\nu e^{-\nu h/\l} &\textup{if } h>\sigma,
\end{dcases}
\eeqn
for some constants $A_\nu$ and $B_\nu$. Continuity of $\psi_\nu(h)$ and $\psi'_\nu(h)$ in $h=\sigma$ requires
\beq\label{eq:logder}
\l\frac{\psi_\nu'(\sigma)}{\psi_\nu(\sigma)} = \frac{\sqrt{\Young^2 - \nu^2}}{\tan(\sigma\sqrt{\Young^2 - \nu^2}/\l)} = - \nu.
\eeq
From the latter equation we see that the existence of $\nu$ is not ensured for any value of the ratio $\sigma/\l$. With the definitions $p\equiv2 \Young\sigma/\pi\l$ and $\xi\equiv\nu/\Young$, the second equality in \cref{eq:logder} then results in a relation for $\xi$:
\beq\label{eq:existence}
\sqrt{1 - \xi^2} = - \xi \tan(\pi p\sqrt{1 - \xi^2}/2),
\eeq
valid for any fixed (strictly) positive $p$. Note that, by construction, we consider only the range $0<\xi<1$. One can readily verify  that for $p\leq1$ no solution of \cref{eq:existence} exists, while for $p>1$ a number of solutions 
arise. 
These solutions progressively vanish as $p$ decreases from large values down to $1$, each of them reaching zero as $p$ reaches an odd number. For the last solution, denoted by $\xi_0$, to disappear, we infer from \cref{eq:existence} that $\xi_0 \sim \pi(p-1)/2$ as $p\searrow1$. Associated with each 
solution $\xi$ there is a bound state with energy 
$\e=-\Young^2\xi^2/2$. In particular, 
for the ground state we get $\e_0\equiv-\Young^2\xi_0^2/2\sim-\Young^2\Round{\Young\sigma/\l-\pi/2}^2/2$ as $\Young\sigma/\l\searrow\pi/2$.
In this limit the square well potential model can therefore be related to the contact potential model studied in the main text by choosing the (positive) parameter $\theta$ according to $\e_0=-\theta^2/2$, i.e.,
\beqn
\theta\,\overset{\Young\sigma/\l\,\searrow\,\pi/2}{\sim}\,\Young\Round{\frac{\Young\sigma}{\l}-\frac{\pi}{2}}.
\eeqn}




\bibliographystyle{ieeetr}
\bibliography{contact_potential}

\end{document}